\documentclass[12pt,a4paper]{article}

\usepackage{epsfig,graphicx,bbm}
\usepackage{amsmath,amsfonts,amssymb}
\usepackage{citesort}
 \usepackage[footnotesize]{caption}

\newcommand{\unit}{\leavevmode\hbox{\small1\kern-3.6pt\normalsize1}}

\allowdisplaybreaks

\begin{document}

\begin{flushright}
FTUAM-07/09\\
IFT-UAM/CSIC-07-27\\
IFIC/08-03\\

\vspace*{5mm}
\end{flushright}

\vspace*{10mm}
\begin{center}
{\Large \textbf{Lepton flavour violating semileptonic $\tau$ decays \\[3mm]
    in constrained MSSM-seesaw scenarios} }

\vspace{2cm} 
{\large
E.~Arganda\,$^{a}$, M.~J.~Herrero\,$^{a}$ and J.~Portol\'es\,$^{b}$}\\[0.4cm]

{$^{a}$\textit{Departamento de F\'{\i }sica Te\'{o}rica 
and Instituto de F\'{\i }sica Te\'{o}rica, IFT-UAM/CSIC \\[0.1cm]
Universidad Aut\'{o}noma de Madrid,
Cantoblanco, E-28049 Madrid, Spain}}\\[0.3cm]

{$^{b}$\textit {IFIC, Universitat de Val\`encia - CSIC, Apt. Correus 22085, E-46071 Val\`encia, Spain}}\\[0.1cm]

\vspace*{0.3cm} 

\begin{abstract}
In this work we study the Lepton Flavour Violating semileptonic $\tau$ decays:
1) $\tau \to \mu PP$ with $PP= \pi^+\pi^-, \pi^0\pi^0, K^+K^-, K^0 {\bar K}^0$;
2) $\tau \to \mu P$ with $P=\pi^0, \eta, \eta'$ and 3) $\tau \to \mu V$ with 
$V = \rho^0, \phi$. We work within the context of two constrained MSSM
scenarios: the CMSSM-seesaw and NUHM-seesaw, with a MSSM spectrum extended by three $\nu_R$ and their SUSY partners and where the seesaw mechanism for neutrino mass generation is implemented. A full SUSY one-loop computation is presented
and the importance of the various contributions, the $\gamma$-, $Z$-, and Higgs bosons
mediated ones, are analysed. The hadronisation of quark bilinears is performed within the
chiral framework. Some discrepancies in the predicted rates for 
BR($\tau \to \mu \eta$), BR($\tau \to \mu \eta'$) and  
BR($\tau \to \mu K^+K^-$) are found with respect to previous estimates,
which will be commented here.
These three channels will be shown to be the most competitive ones to
test simultaneously the Lepton Flavour Violation and the Higgs sector. 
We further present here a set of approximate formulas for all the
semileptonic channels which we believe can be useful for further comparison with
present and future data.  
\end{abstract}

\end{center}

\newpage 
\section{Introduction}\label{intro}
Lepton Flavour Violating (LFV) processes provide one of the most challenging tests 
of supersymmetric (SUSY)
extensions of the Standard Model (SM) of Particle
Physics~\cite{Hisano:1995nq,Hisano:1995cp,Hisano:1998fj,Kuno:1999jp}.
One of the most popular ones among these
extensions is the Minimal Supersymmetric Standard Model (MSSM) enlarged with three right handed 
neutrinos and their corresponding SUSY partners, and where the physical neutrino masses are generated
via a seesaw mechanism~\cite{seesaw:I,seesaw:II}.
Within this SUSY-seesaw context, the light neutrino masses and neutrino
mixing angles can be easily accommodated in agreement with present data~\cite{Yao:2006px} by
setting appropriate input values for the heavy right handed neutrino masses, 
within the range $M_R\sim (10^{10}-10^{15})$ GeV, and
appropriate Yukawa couplings, $Y_\nu$. The 
hypothesis 
of Majorana neutrinos is crucial in this concern, because it is only for them 
that large Yukawa couplings, say $Y_\nu \sim {\cal O}(1)$, can be set.
An interesting connection between neutrino and LFV physics then follows, because 
the large
Yukawa couplings induce, via loops of SUSY
particles~\cite{Borzumati:1986qx}, important contributions to the rare LFV processes. In fact, these contributions
are in some cases, already at the reach of their present experimental
sensitivity. So far, the most sensitive LFV process to the Yukawa couplings 
in the SUSY-seesaw context is $\mu \to e
\gamma$, where the present experimental sensitivity is at $1.2 \times 10^{-11}$~\cite{Brooks:1999pu,Ritt:2006cg}.
In the future, if 
the announced improvement in the sensitivity of $\mu-e$ conversion in
nuclei of up to $10^{-18}$ is finally
reached~\cite{PRIME}, this process will be by
far the most competitive one.
Regarding the tests of Lepton Flavour Violation (LFV) in the $\tau-\mu$ sector, the
most competitive one at present is $\tau \to \mu \gamma$, whose upper bound is now
set to $1.6\times
10^{-8}$~\cite{Aubert:2005ye,Abe:2006sf,Hayasaka:2007vc,Banerjee:2007rj}.
Furthermore, the
sensitivity to LFV in $\tau \to 3\mu$ has also improved notably in the last years. The
present upper bounds  
from BELLE and BABAR collaborations are $3.2 \times 10^{-8}$~\cite{Miyazaki:2007zw} and $5.3\times
10^{-8}$~\cite{Aubert:2007pw}, respectively. This leptonic channel has
the advantage over the radiative $\tau \to \mu \, \gamma$ decay that provides a test not 
only of SUSY but also of
the Higgs sector. It is remarkable that both $\tau \to 3\mu$
decay~\cite{Babu:2002et,Dedes:2002rh,Brignole:2003iv,Arganda:2005ji,Antusch:2006vw}
and $\mu-e$ conversion~\cite{Kitano:2003wn,Arganda:2007jw} in nuclei can get 
important contributions from Higgs mediated diagrams in SUSY scenarios with large 
$\tan\beta$ and light MSSM Higgs bosons.

In the present work, we study the LFV semileptonic tau decay channels which are also of 
interest because of the recently reported sensitivity by BELLE and BABAR
collaborations~\cite{Yusa:2006qq,Abe:2006qv,Aubert:2006cz,Miyazaki:2007jp}
that are, for some channels, already competitive 
with the LFV tau leptonic ones. In particular we analyse here the following
semileptonic tau decays:
1) $\tau \to \mu PP$ with $PP= \pi^+\pi^-, \pi^0\pi^0, K^+K^-, K^0 {\bar K}^0$; 
2) $\tau \to \mu P$ with $P=\pi^0, \eta, \eta'$ and 3) $\tau \to \mu V$ with 
$V= \rho^0, \phi$.
Their
present upper experimental bounds (90\% CL) are summarised 
in Table~\ref{LFVsemilep:bounds}. We perform a full
one-loop computation of the rates for all these processes within the context of 
two constrained SUSY-seesaw scenarios which are of particular interest: The
usual constrained MSSM-seesaw (CMSSM-seesaw)
scenario~\cite{Kane:1993td}, with universal soft SUSY 
masses at the gauge coupling
unification scale, and the so-called Non-Universal Higgs Mass (NUHM)
scenario~\cite{Ellis:2002iu},
with all those soft masses being universal except for the Higgs sector ones. In this later 
case the predicted Higgs particle masses can be low,
indeed close to their present experimental lower bounds (for the SM Higgs the present bound is $m_H > 114.4$
GeV $95 \%$ C.L.~\cite{Yao:2006px}), and  
the corresponding Higgs-mediated contribution to the previous LFV processes can be
relevant, even for large soft SUSY masses. 

In the previous related literature there are, to our knowledge, just a few theoretical 
computations of some of these LFV semileptonic $\tau$ decays induced by SUSY loops. 
In particular, $\tau \to \mu \eta$ was first computed in~\cite{Sher:2002ew}
within the context of the
unconstrained MSSM and in the approximation of large $\tan \beta$. A more refined
analysis of this channel, $\tau \to \mu \eta'$, $\tau \to \mu \pi$, and $\tau \to \mu
\rho$ was done in~\cite{Brignole:2004ah} for the unconstrained MSSM scenario
and large $\tan \beta$ approximation as well, but they used an effective lagrangian
framework for the LFV operators. An estimate of $\tau \to \mu \eta$ with the use of the mass insertion (MI)
approximation for the relevant lepton flavour mixing parameter between the $\tau$ and $\mu$ 
sectors, $\delta_{32}$, has been performed in~\cite{Paradisi:2005tk}. The decay mode 
$\tau \to \mu K^+K^-$ has been 
estimated in~\cite{Chen:2006hp} within the mass insertion and leading logarithmic (LLog)
approximations for $\delta_{32}$, and taking into account only the 
Higgs-mediated contribution in the large $\tan \beta$ limit. In all these previous works no
connection with the neutrino sector was considered and the
hadronisation of quark bilinears in the final state is simply parameterised
in terms of the proper meson decay constants and meson masses. Other estimates of some of these LFV semileptonic $\tau$ decays in different contexts, like $SO(10)$-SUSY-GUT model with universal soft breaking terms~\cite{Fukuyama:2005bh} and Littlest Higgs model~\cite{Blanke:2007db}, have also been performed in the literature.

Our analysis presented here is more complete in several aspects. First, we include 
both $Z$-boson and $A^0$-boson mediated contributions to $\tau \to \mu P$
($P=\eta,\eta',\pi^0$), and both $\gamma$ and $H^0,h^0$-bosons mediated
contributions to $\tau \to \mu K^+K^-$. The other
channels, $\tau \to \mu K^0 {\bar K}^0$ and $\tau \to \mu \pi^0 \pi^0$ have not
been estimated previously. We include $\gamma$ and $H^0,h^0$-bosons mediated
contributions in $\tau \to \mu K^0 {\bar K}^0$. The case $\tau \to \mu \pi^0 \pi^0$
can only be mediated by $H^0,h^0$-bosons. Second, we do not use either the mass
insertion nor the LLog approximation and our analytical computation 
is valid for all $\tan\beta$ values. Third, we make a connection with neutrino physics by
requiring compatibility through all this work with the neutrino data for masses 
and mixing angles. Fourth, we perform the hadronisation of quark bilinears
with close attention to the
chiral constraints, guided by the resonance chiral theory~\cite{Ecker:1988te} that has
proven to be a robust framework for the analyses of hadrodynamics when resonances are
involved. The $\gamma$ amplitude, due to its pole at $q^2=0$, is most sensitive to the 
hadronisation procedure. Hence the hadronisation of the electromagnetic current, 
that drives the $\gamma$ contributions, has been carried out by a careful construction
of the vector form factor that matches both the chiral low-energy limit and the asymptotic
smoothing at high $q^2$~\cite{Lepage:1980fj}. Those final states driven by heavy intermediate bosons
like the $Z^0$ or Higgses, on the other side, do not require such an involved scheme. 
In these cases we have used the leading chiral approximation of Chiral Perturbation Theory that we know, for sure, it
has to be fulfilled by the hadronisation. The advantage of our approach is that
it provides the most successful description up to date of the hadronic tau decays
and it can
be systematically improved by further developments of the appropriate form factors, whether
axial-vector, scalar or pseudoscalar cases.

The rest of this paper is organised as follows. The theoretical framework for the
computation of LFV semileptonic $\tau$ decays is described in Section 2. This includes
a short review of the SUSY-seesaw scenarios that we work within, CMSSM and NUHM, and a
description of our procedure for hadronisation of quark bilinears within the context
of Chiral Perturbation Theory ($\chi$PT) and Resonance Chiral Theory
(R$\chi$T).
In Section 3, the analytical results of the full one-loop 
branching ratios BR$(\tau \to \mu PP)$, BR$(\tau \to \mu P)$, BR$(\tau \to \mu \rho)$ and  
BR$(\tau \to \mu \phi)$ are presented.  Section 4 is
devoted to the numerical results and discussion. It includes, in addition a
comparison between the full one-loop and approximate results. A set of useful 
approximate 
formulas for the semileptonic
tau decay rates that are valid at large $\tan \beta$ are derived. A critical
comparison with previous predictions
in the literature is also included in this Section 4. Finally, Section 5 summarises the
conclusions.    

\begin{table}[h]
\begin{center}

\vspace{0.3cm}
\begin{tabular}{|c|c  c  c|}
\hline
LFV semilep. $\tau$ decays & BABAR  & Belle & BABAR \& Belle  \\
\hline
BR($\tau \to \mu \eta$) & $1.5 \times
10^{-7}$~\cite{Aubert:2006cz} & $6.5 \times 10^{-8}$~\cite{Abe:2006qv} & $5.1 \times 10^{-8}$~\cite{Banerjee:2007rj} \\
BR($\tau \to \mu \eta'$) & $1.4 \times 10^{-7}$~\cite{Aubert:2006cz} & $1.3 \times 10^{-7}$~\cite{Abe:2006qv} & $5.3 \times 10^{-8}$~\cite{Banerjee:2007rj} \\
BR($\tau \to \mu \pi$) & $1.1 \times 10^{-7}$~\cite{Aubert:2006cz} & $1.2 \times 10^{-7}$~\cite{Abe:2006qv} & $5.8 \times 10^{-8}$~\cite{Banerjee:2007rj} \\
BR($\tau \to \mu \rho$) & --- & $2.0 \times 10^{-7}$~\cite{Yusa:2006qq} & --- \\
BR($\tau \to \mu \phi$) & --- & $1.3 \times 10^{-7}$~\cite{Abe:2007exa} & --- \\
BR($\tau \to \mu \pi^+ \pi^-$) & --- & $4.8 \times 10^{-7}$~\cite{Yusa:2006qq} & --- \\
BR($\tau \to \mu \pi^0 \pi^0$) &  ---  &
--- & --- \\
BR($\tau \to \mu K^+ K^-$) & --- & $8.0 \times 10^{-7}$~\cite{Yusa:2006qq} & --- \\
BR($\tau \to \mu K^0 {\bar K}^0$) & ---  &
--- & --- \\\hline
\end{tabular}
\label{LFVsemilep:bounds}
\caption{Present upper bounds for LFV semileptonic $\tau$ decays.}
\end{center}
\end{table}

\section{Framework for LFV semileptonic $\tau$ decays}\label{th_framework}
In this section we describe the theoretical framework for the computation of
the LFV semileptonic $\tau$ decay rates. First we present the scenario for the generation of 
LFV in the $\tau$-$\mu$ sector, then we summarise the main ingredients to perform the hadronisation 
of quark bilinears within the context of $\chi$PT and R$\chi$T. 
\subsection{LFV in the SUSY-seesaw scenario}
The SUSY-seesaw scenario that we work within contains the full MSSM spectra 
and, in addition, the three right handed neutrinos and their SUSY partners.
It is defined in terms of both 
the SUSY and neutrino sector parameters which are summarised in the following. 

Regarding the SUSY sector we choose to work in two different constrained MSSM
scenarios: The usual Constrained MSSM (CMSSM) with similar input parameters as
in mSUGRA models, and the so-called Non-Universal Higgs Mass (NUHM) scenarios
with two additional parameters defining the non-universal soft Higgs masses. 
The corresponding sets of input parameters in these two scenarios are:
\begin{align}
\text{CMSSM :} \ & M_0\,, M_{1/2}\,,A_0\,, \tan \beta\,, 
\text{sign}(\mu)\,,
\nonumber \\
\text{NUHM :}\ & M_0\,, M_{1/2}\,,A_0 \,, \tan \beta\, ,
\text{sign}(\mu)\,,M_{H_1}, M_{H_2},
\end{align}
where $M_0$, $M_{1/2}$ and $A_0$ are the universal soft SUSY breaking scalar masses, gaugino
masses and trilinear couplings at the gauge
coupling unification scale, $M_X= 2 \times 10^{16}$ GeV. Notice that $M_0$ and
$A_0$ define also the soft parameters in the sneutrino sector. The other CMSSM
parameters are, as usual, the ratio of the two Higgs vacuum expectation values,
$\tan \beta=v_2/v_1$, and the sign of the $\mu$ parameter, $\text{sign}(\mu)$.
The departure from universality in the soft Higgs masses of the NUHM is parameterised 
in terms of two parameters $\delta_1$ and $\delta_2$ by, 
\begin{align}
  &
M^2_{H_1}\,=\,M^2_0\,(1+\delta_1)\,,\ \ M^2_{H_2}\,=\,M^2_0\,(1+\delta_2)\,.
\end{align}   
Notice that with the choice $\delta_{1,2}=0$ one recovers the universal case
defined by the CMSSM scenario.
For simplicity, and to further reduce the number of input parameters,
in all the numerical estimates of this work 
we will take $M_0= M_{1/2} \equiv M_{\rm SUSY}$, $A_0=0$ and $\text{sign}(\mu)=+1$.

To evaluate these two sets of parameters at low energies (taken here as 
the Z gauge boson mass $m_Z$) we solve the full one-loop Renormalisation Group Equations
(RGEs) including the extended neutrino and sneutrino sectors. For this and 
the computation of the full spectra at the low energy we use here
the public FORTRAN code SPheno~\cite{Porod:2003um}.

Regarding the neutrino sector, we use the seesaw mechanism for neutrino mass 
generation which is implemented here to the case of three right handed 
neutrinos. 
The usual input parameters in this case are 
the three right handed Majorana masses,
$M_{R_{1,2,3}}$ and the neutrino Yukawa coupling 3x3 matrix, $Y_\nu$. The Dirac
neutrino mass matrix is then related to the Yukawa couplings 
by $m_D=Y_\nu\, v_2$, where   
$v_{2}= \,v\,\sin \beta$ and $v=174$ GeV.  In this seesaw scenario, 
the physical Majorana neutrinos 
consist of
three light ones, $\nu_{1,2,3}$, with predicted masses being typically 
$m_{\nu_{1,2,3}} \sim {\cal O}(m_D^2/M_R)$, and three heavy ones, $N_{1,2,3}$, with masses 
$m_{N_{1,2,3}}\simeq M_{R_{1,2,3}}$. 
However, instead of this we will use another 
set of input parameters which are more convenient to accommodate the experimental
data on light neutrino masses and generational mixing angles. Within this
parameterisation, the Dirac mass matrix and, hence, the Yukawa coupling matrix, 
are derived in terms of the physical neutrino masses, neutrino mixings and a generic
complex orthogal $3 \times 3$ matrix, $R$, as follows~\cite{Casas:2001sr},
\begin{equation}\label{seesaw:casas}
m_D\,=Y_\nu\, v_2=\, i \sqrt{m^\text{diag}_N}\, R \,
\sqrt{m^\text{diag}_\nu}\,  U^\dagger_{\text{MNS}}\,,
\end{equation}
where,
\begin{equation}\label{def:Ndiag}
m_N^\text{diag}\,=\, \text{diag}\,(m_{N_1},m_{N_2},m_{N_3})\,,
\end{equation}
\begin{equation}\label{physicalmasses}
m_{\nu}^\text{diag}\,=\, \text{diag}\,(m_{\nu_1},m_{\nu_2},m_{\nu_3})\,,
\end{equation}
and we use the standard parameterisation for the unitary matrix 
$U_\mathrm{MNS}$~\cite{Maki:1962mu,Pontecorvo:1957cp}
containing the three generational mixing angles $\theta_{12}$,
$\theta_{13}$ and $\theta_{23}$ and the three CP violating phases, $\delta$,
$\phi_{1,2}$. In turn, the $R$
matrix is parameterised  
in terms of three complex angles, $\theta_i$ $(i=1,2,3)$ as~\cite{Casas:2001sr}
\begin{equation}\label{Rcasas}
R\, =\, 
\left( 
\begin{array}{ccc} 
c_{2}\, c_{3} & -c_{1}\, s_{3}\,-\,s_1\, s_2\, c_3
& s_{1}\, s_3\,-\, c_1\, s_2\, c_3 \\ 
c_{2}\, s_{3} & c_{1}\, c_{3}\,-\,s_{1}\,s_{2}\,s_{3} 
& -s_{1}\,c_{3}\,-\,c_1\, s_2\, s_3 \\ 
s_{2}  & s_{1}\, c_{2} & c_{1}\,c_{2}
\end{array} 
\right)\,,
\end{equation}
with $c_i\equiv \cos \theta_i$ and $s_i\equiv \sin\theta_i$. One interesting aspect
of this matrix is that it encodes 
the possible extra neutrino mixings (associated with the
right-handed sector) in addition to the ones in
$U_{\text{MNS}}$. Notice also that the previous Eq.~(\ref{seesaw:casas})
 is established at the 
right-handed neutrino mass scale $M_R$, so that the quantities appearing
in it are indeed the renormalised ones, namely, 
$m^\text{diag}_\nu\,(M_R)$ and $U_{\text{MNS}}\,(M_R)$. These latter 
are obtained here by means of the RGEs and by starting the running from their corresponding
renormalised values at $m_Z$, $m^\text{diag}_\nu\,(m_Z)$ and $U_{\text{MNS}}\,(m_Z)$
which are identified respectively with the physical $m^\text{diag}_\nu$ and $U_{\text{MNS}}$
from neutrino data.   

Concerning our choice for the size of the physical neutrino parameters, we shall focus
in this work on scenarios where both light and heavy neutrinos are hierarchical,
\begin{align}
&m_{\nu_1}\, \ll\, m_{\nu_2}\, \ll\, m_{\nu_3}\,,  \nonumber \\
&m_{N_1}\, \ll\, m_{N_2}\, \ll \,m_{N_3}\,, \nonumber 
\end{align}
and set the numerical values for the light neutrino parameters to the following 
ones which are compatible with present data~\cite{Yao:2006px,neutrinodata_fits}
\begin{align}
& m_{\nu_1}^2 \,\simeq\, 0 \,,\quad  \; \; \; m_{\nu_2}^2\,=\,
\Delta\, m^2_\text{sol} \,=\,8\,\times 10^{-5}\,\,\text{eV}^2\,, \; \; \; 
\quad  m_{\nu_3}^2\,=\,
\Delta \, m^2_\text{atm} \,=\,2.5\,\times 10^{-3}\,\,\text{eV}^2\,,
\nonumber \\[3mm]
& 
\theta_{12}\,=\,30^\circ\,, 
\quad \;\; \; 
\theta_{23}\,=\,45^\circ\,,
\quad \; \; \; 
\theta_{13}\,\simeq 0,
\quad \quad \; \; \; 
\delta\,=\,\phi_1\,=\,\phi_2\,=\,0\,.
\label{light}
\end{align}
Notice that, for simplicity, the three CP violating phases, 
have been set to zero. We have also set to zero the $\theta_{13}$ mixing angle and
the lightest neutrino mass in order to minimise as much as possible the LFV in the 
$\mu-e$ sector.  In fact, we have checked that for the explored parameters region in this
work, this $\mu-e$ LFV is below the sensitivity of the present data from 
$\mu \to e \gamma$, $\mu \to 3e$ and $\mu-e$ conversion in
nuclei.

In summary, the input parameters of the neutrino sector for the present work are:
\begin{align}
\text{Seesaw :}\,\,  & m_{N_{1,2,3}}\,\,\,\,,\,\,\,  \theta_{1,2,3} \,  .
\end{align}
 Regarding the generation of LFV in these constrained MSSM scenarios, we
remind 
that all lepton 
flavour mixing originates solely from the neutrino Yukawa couplings. These 
$Y_\nu$ first induce flavour violation 
in the slepton sector by the RGE running of the soft SUSY breaking parameters 
from
$M_X$ down to the electroweak scale $m_Z$. It is 
manifested in the non-vanishing values of the off-diagonal elements of the slepton squared
mass matrix at $m_Z$. We perform this running by solving the
full set of one-loop RGEs including the neutrino and sneutrino sectors. 
The resulting slepton mass matrices at $m_Z$ are then diagonalised and the
previous flavour mixing is then transmitted to the mass eigen-values and
eigen-states. Therefore, in this work where we deal with physical states, all
flavour mixing is implicit in the resulting physical charged slepton masses,  
$m_{\tilde l_1}^2,..,m_{\tilde l_6}^2$,  sneutrino masses, 
$m_{\tilde \nu_1}^2, \,
m_{\tilde \nu_2}^2, \,m_{\tilde \nu_3}^2$ and the corresponding 
matrices that rotate from the electroweak to the slepton and sneutrino mass 
bases, respectively, $R^{l}$ and $R^{\nu}$. 
The LFV in the physical processes, like $\l_j \to \l_i \gamma$, $\l_j \to 3
l_i$, $\mu-e$ conversion in nuclei, and the semileptonic $\tau$ decays studied
here, are then generated by the SUSY one-loop contributing diagrams that 
contain these slepton physical masses 
in the internal propagators, and also the previous rotation matrices in 
the interaction 
vertices, which connect between different lepton 
generations. A complete set of Feynman rules for the relevant LFV vertices can
be found in~\cite{Arganda:2005ji,Arganda:2007jw}.

Finally, in order to illustrate more quantitatively how important can be the 
size of the flavour mixing between the stau and smuon sectors, in the CMSSM-seesaw 
scenario, we include next the predictions of the mixing 
parameter $\delta_{32}$ that is defined in the LLog approximation as,
\begin{equation}
\delta_{32} \, =\, -\frac{1}{8\pi^2} \frac{(3\,M_0^2\,+\,A_0^2)}{M_{\rm SUSY}^2}
\left( Y_\nu^\dagger \,L\, Y_\nu \right)_{32}, 
\label{delta23}
\end{equation}
where $L$ is a $3 \times 3$ diagonal matrix whose elements are,
$L_{ii}=\log(M_X/m_{N_i})$ and
$M_{\rm SUSY}$ is an average SUSY mass. 
This phenomenological parameter $\delta_{32}$ measures the amount of 
flavour mixing between the second and third slepton generations in the left-handed sector (LL), which is by far the dominant  
one. The corresponding mixing in the right-handed slepton sector is extremely suppressed by
the smallness of the lepton masses which appear as global factors in the definitions of those (RR and RL) mixings (see, for instance,~\cite{Arganda:2005ji}).

One can estimate $\delta_{32}$ from the previous parameterisation of the seesaw model in
Eq.~(\ref{seesaw:casas}) by simply plugging in Eq.(\ref{delta23}) the value of 
$\left( Y_\nu^\dagger \,L\, Y_\nu \right)_{32}$ from the following expression,
\begin{eqnarray}\label{Y32:LLog}
v_2^2 \left( Y_\nu^\dagger \,L\, Y_\nu \right)_{32}&=& L_{33} \, m_{N_3}
\,\left[ \left( \sqrt{m_{\nu_3}} c_1^* c_2^* c_{13} c_{23} - \sqrt{m_{\nu_2}} s_1^* c_2^*  c_{12} s_{23} \right) \right. \nonumber \\
&& \left. \left( \sqrt{m_{\nu_3}} c_1 c_2 s_{23} + \sqrt{m_{\nu_2}} s_1 c_2 c_{12} c_{23} \right) \right] \nonumber \\
&+& L_{22} \, m_{N_2} \, \left[ \left( \sqrt{m_{\nu_3}} (-s_1^* c_3^* - c_1^* s_2^* s_3^*) c_{23} + \sqrt{m_{\nu_2}} (s_1^* s_2^* s_3^* - c_1^* c_3^*) c_{12} s_{23} \right) \right. \nonumber \\
&& \left. \left( \sqrt{m_{\nu_3}} (-s_1 c_3 - c_1 s_2 s_3) s_{23} + \sqrt{m_{\nu_2}} (c_1 c_3 - s_1 s_2 s_3) c_{12} \right) \right] \nonumber \\
&+& L_{11} \, m_{N_1} \, \left[ \left( \sqrt{m_{\nu_3}} (s_1^* s_3^* - c_1^* s_2^* c_3^*) c_{23} +
\sqrt{m_{\nu_2}} (s_1^* s_2^* c_3^* + c_1^* s_3^*)  c_{12} s_{23} \right) \right. \nonumber \\
&& \left. \left(\sqrt{m_{\nu_3}} (s_1 s_3 - c_1 s_2 c_3) s_{23} 
-\sqrt{m_{\nu_2}} (s_1 s_2 c_3 + c_1 s_3) c_{12} c_{23} \right) \right] \, , 
\end{eqnarray}
where, $s_{ij} \equiv \sin \theta_{ij}$ and  $c_{ij} \equiv \cos \theta_{ij}$, and we have already 
set $m_{\nu_1}=0$ and $\theta_{13}=0$. 
 \begin{figure}[t!]
   \begin{center} 
     \begin{tabular}{cc} \hspace*{-12mm}
         \psfig{file=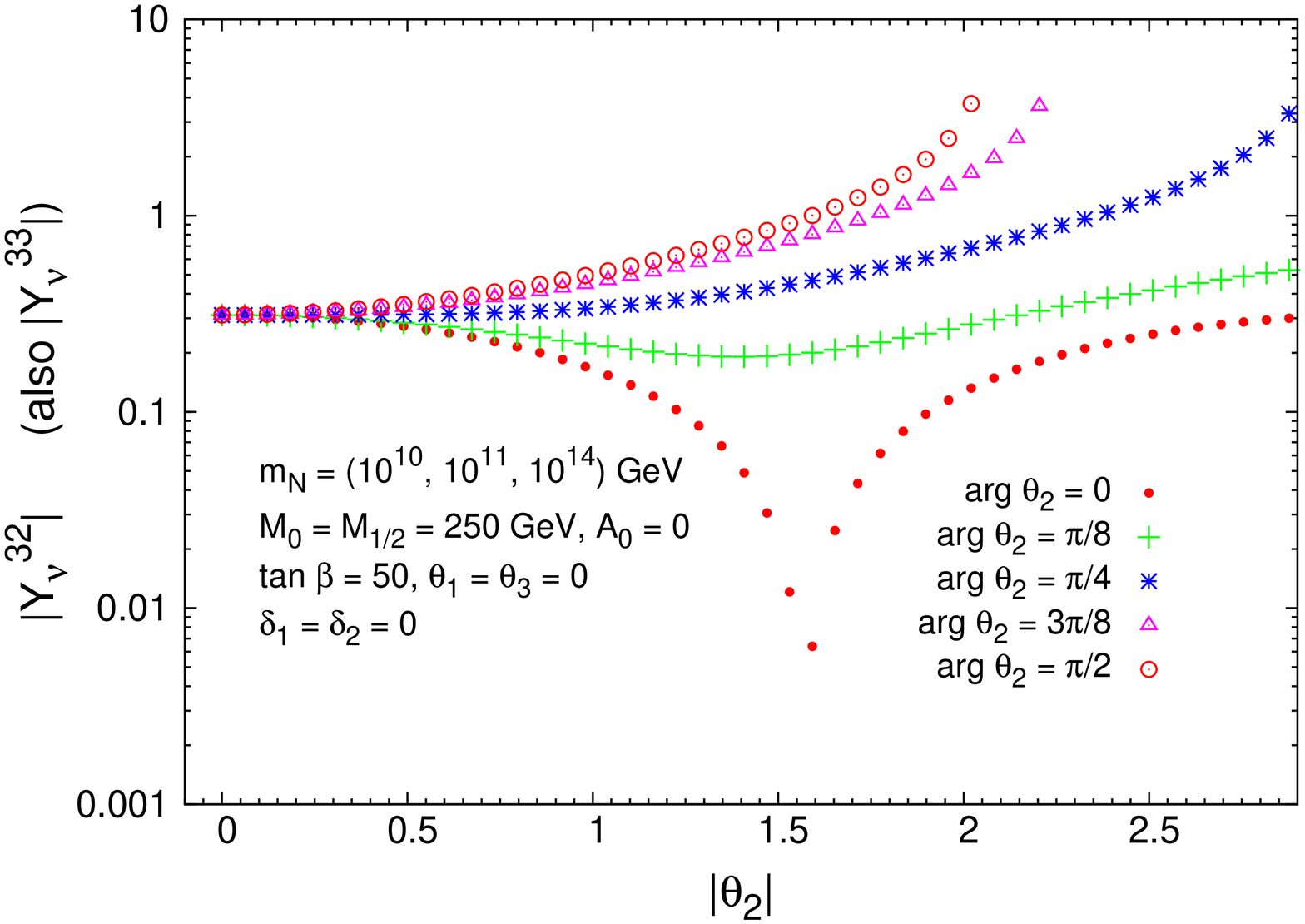,width=60mm,angle=270,clip=} 
 &
  	 \psfig{file=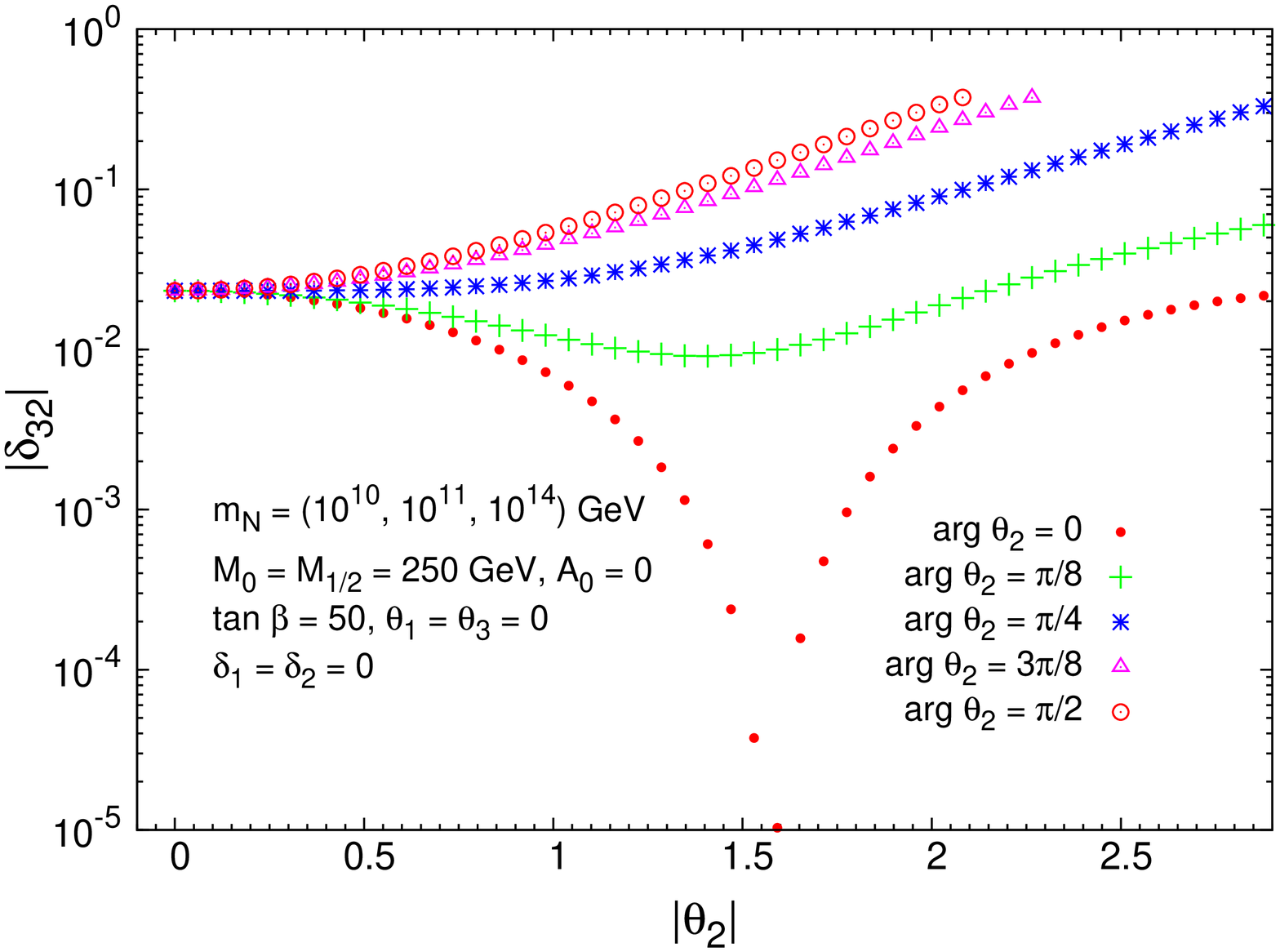,width=60mm,angle=270,clip=}   
     \end{tabular}
     \caption{$|Y_{\nu}^{32}|$ and $|\delta_{32}|$, in the CMSSM-seesaw scenario, 
     as a function of 
      $|\theta_2|$, for 
      $\arg \theta_2\,=\,\{0,\, \pi/8\,,\,\pi/4\,,\,3\pi/8, \,\pi/2
      \}$ (dots, crosses, asterisks, triangles and circles,
      respectively). Both $|\theta_2|$ and $\arg{\theta_2}$ are given in
      radians. The predictions for $|Y_{\nu}^{33}|$ are practically indistinguishable from those
      for $|Y_{\nu}^{32}|$.
     }\label{fig:theta2} 
   \end{center}
 \end{figure}

 \begin{figure}[ht]
   \begin{center} 
     \begin{tabular}{c} \hspace*{-12mm}
  	\psfig{file=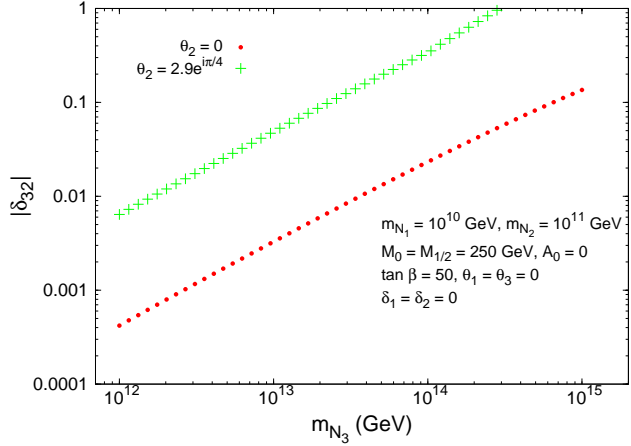,width=60mm,angle=270,clip=}   
     \end{tabular}
     \caption{$|\delta_{32}|$, in the CMSSM-seesaw
     scenario, as a function of $m_{N_3}$. 
     }\label{fig:mN3} 
   \end{center}
 \end{figure}
The numerical predictions for $|Y_\nu^{32}|$ and $|\delta_{32}|$ as a function of $\theta_2$ are shown in
Fig.~\ref{fig:theta2}. Here we have set $M_{\rm SUSY}=M_0=M_{1/2}= 250$ GeV, $A_0=0$, 
$(m_{N_1},m_{N_2},m_{N_3})=(10^{10},10^{11},10^{14})$ GeV, and the 
light neutrino parameters are those in Eq.~(\ref{light}). We see clearly 
that $|\delta_{32}|$ follows the same pattern as $|Y_\nu^{32}|$
(and $|Y_\nu^{33}|$) and can reach large values in the range 0.1-1 for several choices of $|\theta_2|$ 
and arg($\theta_2$). Notice also that the predictions for $|\delta_{32}|$ corresponding to 
Yukawa couplings larger than about 4 are not shown, because through all this work   
perturbativity in all the gauge and Yukawa couplings are imposed. This is set numerically in 
the Spheno program by the requirement $|Y_\nu|^2/(4\pi)< 1.5$ and corresponds to a maximal 
predicted value of about $|\delta_{32}|< 0.4$. The corresponding predictions with respect to $\theta_1$ 
are very
similar to those of $\theta_2$ and are not shown for brevity. The value of $|\delta_{32}|$ is
practically independent on $\theta_3$. For the rest of this work we will set $\theta_{1,3}=0$ and use
just $\theta_2$ as input parameter.
 
The numerical predictions for $|\delta_{32}|$ as a function of the heaviest neutrino mass, $m_{N_3}$ are 
shown in Fig.\ref{fig:mN3}. $|\delta_{32}|$ values within the range 0.1-1 are obtained for large 
$m_{N_3}$ values, say within the interval $10^{13}-10^{15}$ GeV. Notice that the predictions enter into 
the above commented 
non-perturbative region for values larger than $m_{N_3}=10^{14}$ GeV, and 
for the particular choice of $\theta_2=2.9\, e^{i\frac{\pi}{4}}$. Concretely, the value 
$|\delta_{32}|=1$, which is interesting for later discussion and comparison with other works,
corresponds to $m_{N_3}=3 \times 10^{14}$ GeV and lies clearly in the non-perturbative region. Finally,
just to mention that $|\delta_{32}|$ is not much dependent on $\tan \beta$ nor on $m_{N_{1,2}}$. The
values of these two heavy neutrino masses will be set in the following to the reference values 
$m_{N_{1,2}}=10^{10},10^{11}$ GeV.
\subsection{Hadronisation of quark bilinear currents}\label{hadronisation}

Semileptonic decays of the tau lepton are a relatively clean scenario from the strong interaction point of 
view. Hadrons in the final state stem from the hadronisation of quark bilinears, namely 
$\overline{\psi} \,\Gamma \psi$, where $\psi$ is a vector in the $SU(3)_F$ flavour space and $\Gamma$ is, in general, a matrix both in the spinor and the flavour space.
\par
An appropriate framework to handle the procedure of hadronisation is provided by the large-$N_C$ expansion
of $SU(N_C)$ QCD~\cite{'tHooft:1973jz}, being $N_C$ the number of colours. In short it stays that in the
$N_C \rightarrow \infty$ limit any Green function is given by meromorphic expressions provided by the tree level diagrams of a Lagrangian theory with an infinite spectrum of zero-width states. Though we do not know
how to implement fully this limit, a fruitful~\cite{Peris:1998nj} if debatable~\cite{Bijnens:2003rc}
approach lies in cutting the spectrum, keeping only the lightest multiplets of resonances. We will attach
to this tenet as a guiding principle.
\par
A suitable tool to realise the $1/N_C$ expansion is provided by chiral Lagrangians. In those processes
where hadron resonances do not play a dynamical role,
$\chi$PT~\cite{Weinberg:1978kz,Gasser:1984gg}
is the appropriate scheme to describe the strong interaction of Goldstone bosons ($\pi$, $K$ and 
$\eta$). This is the case, for instance, of $\tau \rightarrow \mu P$ (being $P$ short for a pseudoscalar
meson). When resonances participate in the dynamics of the process, as in $\tau \rightarrow \mu PP$,
it is necessary to include them as active degrees of freedom into the Lagrangian as it is properly done
in the R$\chi$T frame~\cite{Ecker:1988te}. Hence we will make use of 
R$\chi$T, that naturally includes $\chi$PT, to hadronise the relevant currents that appear in the
processes under study here.
\par
We consider bilinear light quark operators coupled to external sources and added to the massles QCD Lagrangian~:
\begin{equation}
\label{eq:currents}
 {\cal L}_\text{QCD} \, = \, {\cal L}_\text{QCD}^0 + \overline{q} \left[ \gamma_{\mu} \, \left( v^{\mu} \, + \,  
\gamma_5 \, a^{\mu} \right) \, - \, \left( \, s \, - \, i \, p \, \gamma_5 \right) \right] q \, ,
\end{equation}
where vector ($v^{\mu}= v^{\mu}_i \lambda^i/2$), axial-vector ($a^{\mu} = a^{\mu}_i \lambda^i/2$), scalar ($s= s_i \lambda^i$) and pseudoscalar ($p= p_i \lambda^i$) fields are 
matrices in the flavour space, and ${\cal L}_\text{QCD}^0$ is the massless QCD Lagrangian~\footnote{The Gell-Mann
matrices $\lambda^i$ are normalised as $\langle \lambda_i 
\lambda_j \rangle = 2 \delta_{ij}$ and the gluons are denoted here by $G_\mu$.}.
This Lagrangian density gives the QCD generating functional
${\cal Z}_\text{QCD} \left[v,a,s,p \right]$ as
\begin{equation}
 e^{i \, {\cal Z}_\text{QCD}[v,a,s,p]} \, = \, \int \, [\, D G_{\mu} \,] [\, D q\, ] [\, D \overline{q}\,] \, 
e^{i \, \int d^4 x \, {\cal L}_\text{QCD}[q,\overline{q},G,v,a,s,p]} \,.
\end{equation}
\par
In order to construct the corresponding Lagrangian theory
in terms of the lightest hadron modes we need to specify them. The lightest $U(3)$ nonet of pseudoscalar mesons~:
\begin{eqnarray}
\label{eq:phi}
 \phi(x) & = & \sum_{a=0}^{8} \, \frac{\lambda_a}{\sqrt{2}} \, \varphi_a \\ 
&=&\left(
\begin{array}{ccc}
 \displaystyle\frac{1}{\sqrt 2}\,\pi^0 + \displaystyle\frac{1}{\sqrt
 6}\,\eta_8 
 + {1\over\sqrt 3}\eta_0 
& \pi^+ & K^+ \\
\pi^- & - \displaystyle\frac{1}{\sqrt 2}\,\pi^0 + 
\displaystyle\frac{1}{\sqrt 6}\,\eta_8 
 + {1\over\sqrt 3}\eta_0  
& K^0 \\
 K^- & \bar{K}^0 & - \displaystyle\frac{2}{\sqrt 6}\,\eta_8 
+ {1\over\sqrt 3}\eta_0  
\end{array}
\right)
\ , \nonumber
\end{eqnarray}
is realised nonlinearly into the unitary matrix in the flavour space~:
\begin{equation}
 u(\varphi) = \exp \left[ i \, \frac{\Phi}{\sqrt{2} F} \right].
\end{equation}
Hence the leading ${\cal O}(p^2)$ $\chi$PT $SU(3)_L \otimes SU(3)_R$ chiral Lagrangian is 
\footnote{Notice that though we include a $U(3)$ nonet we are not relying on the 
$U(3)_L \otimes U(3)_R$ chiral Lagrangian~\cite{HerreraSiklody:1996pm} on grounds of 
predictability, as the latter introduces unknown functions.} ~:
\begin{equation}
 {\cal L}_{\chi}^{(2)} \, = \, \frac{F^2}{4} \, \langle u_{\mu} \, u^{\mu} \, + \, \chi_{+} \rangle \, ,
\end{equation}
where 
\begin{eqnarray}
\label{eq:uchi}
 u_{\mu} & = & i [ u^{\dagger}(\partial_{\mu}-i r_{\mu})u-
u(\partial_{\mu}-i \ell_{\mu})u^{\dagger} ] \ , \nonumber \\ 
\chi_{+} & = & u^{\dagger}\chi u^{\dagger} + u\chi^{\dagger} u\ \ 
\ \ , \ \ \ \ 
\chi=2B_0(s+ip) \; \; ,
\end{eqnarray}
and $\langle \ldots \rangle$ is short for a trace in the flavour space. Interactions with 
electroweak bosons can be accommodated through the vector $v_{\mu} = (r_{\mu} + \ell_{\mu})/2$
and axial-vector $a_{\mu} = (r_{\mu} - \ell_{\mu})/2$ external fields. The scalar field $s$ incorporates
explicit chiral symmetry breaking through the quark masses $s = {\cal M} + ...$ and, finally, 
$F \simeq F_{\pi} \simeq 92.4 \, \mbox{MeV}$ is the pion decay constant and 
$B_0 F^2 = - \langle 0 | \overline{\psi} \psi |0 \rangle_0$ in the chiral limit.
The chiral tensor $\chi$ provides masses to the Goldstone bosons through the external scalar field,
as can be seen in Eq.~(\ref{eq:uchi}). Indeed in the isospin limit we have~:
\begin{equation}
 \chi \, = \, 2 \, B_0 \, {\cal M} \, + ... \, = \, \left( \begin{array}{ccc}
                                                           m_{\pi}^2 &&  \\
                                                           & m_{\pi}^2 & \\
                                                           && 2 m_K^2 - m_{\pi}^2 \\
                                                           \end{array}
                                                     \right) \, + \, .... \, .
\end{equation}
Hence we identify~:
\begin{eqnarray}
 B_0 \, m_u & = & B_0 \, m_d \; = \; \frac{1}{2} \, m_{\pi}^2 \; , \nonumber \\
B_0 \, m_s & = & m_K^2 - \frac{1}{2} \, m_{\pi}^2  \, ,
\label{mq-mP}
\end{eqnarray}
that will be useful when considering the Higgs contributions.
The mass eigenstates $\eta$ and $\eta'$ are defined from the octet $\eta_8$ and singlet $\eta_0$ 
states through the rotation~:
\begin{equation}
\left(  \begin{array}{c}
  \eta \\
  \eta' 
 \end{array} \right)\, = \,  \left(
\begin{array}{cc}
\cos \theta & - \sin \theta \\
\sin \theta &  \cos \theta 
\end{array}  \right) \;  \left(
\begin{array}{c}
 \eta_8 \\
\eta_0
\end{array} \right)\, ,
\end{equation}
and we input\footnote{The values of $\theta$ in the literature range between
$\theta \sim -12^{\circ}$ up to $\theta \sim -20^{\circ}$~\cite{Kaiser:1998ds}.} a value of $\theta \simeq -18^{\circ}$.
\par
The hadronisation of a final state of two pseudoscalars is driven by vector and scalar resonances
though the latter, because their higher masses, play a lesser role and we will not include them
in the following. We will introduce the vector resonances in the antisymmetric formalism; hence
the nonet of resonance fields $V_{\mu \nu}$~\cite{Ecker:1988te} is defined by analogy
with Eq.~(\ref{eq:phi}) with the same flavour structure. By demanding the chiral symmetry 
invariance the resonance Lagrangian reads~:
\begin{equation}
 {\cal L}_V \, = \,  {\cal L}^{V}_{\mbox{\tiny kin}} \, + \, {\cal L}^{V}_{(2)} \; , 
\end{equation}
where
\begin{eqnarray}
\label{eq:vectors}
 {\cal L}^{V}_{\mbox{\tiny kin}} & = & -\frac{1}{2} \langle \,
\nabla^\lambda V_{\lambda\mu} \nabla_\nu V^{\nu\mu} \, \rangle + \frac{M_V^2}{4} \, \langle \, 
V_{\mu\nu} V^{\mu\nu}\,  \rangle \, , \nonumber \\[3.5mm]
{\cal L}^{V}_{(2)}  & = & \frac{F_V}{2\sqrt{2}} \langle V_{\mu\nu}
f_+^{\mu\nu}\rangle + i\,\frac{G_V}{\sqrt{2}} \langle V_{\mu\nu} u^\mu
u^\nu\rangle  \; \, , 
\end{eqnarray}
and in the latter the subscript $(2)$ indicates the chiral order of the tensor accompanying
$V_{\mu\nu}$. In Eq.~(\ref{eq:vectors}) we have used the definitions~:
\begin{eqnarray}
\nabla_\mu X &	\equiv &	 \partial_{\mu} X	 + [\Gamma_{\mu}, X] \; \; , \\
\Gamma_\mu & = & \frac{1}{2} \, [ \,
u^\dagger (\partial_\mu - i r_{\mu}) u +
u (\partial_\mu - i \ell_{\mu}) u^\dagger \,] \; \; ,\nonumber \\
f_+^{\mu\nu} & = &  u F_L^{\mu\nu} u^\dagger + u^\dagger F_R^{\mu\nu} u \, , \nonumber
\end{eqnarray}
being $F_{L,R}^{\mu \nu}$ the field strength tensors associated with the external right and
left fields. The couplings $F_V$ and $G_V$ are real.
\par
Accordingly our R$\chi$T framework is provided by~:
\begin{equation}
 {\cal L}_\text{R$\chi$T} \, = \,  {\cal L}_{\chi}^{(2)}  \, + \, {\cal L}_V \, ,
\end{equation}
and the contribution of the low modes to the QCD functional is formally given by~:
\begin{equation}
 e^{i \, {\cal Z}_\text{QCD}[v,a,s,p]} \, \Bigg|_{\mbox{\tiny low modes}} \, = \, \int \, [D u] [D V]  \, 
e^{i \, \int d^4 x \, {\cal L}_\text{R$\chi$T}[u,V,v,a,s,p]} \,.
\end{equation}
With this identification we can already carry out the hadronisation of the bilinear quark currents
included in Eq.~(\ref{eq:currents}) by taking
the appropriate partial derivatives, with respect to the external
auxiliary fields, of the functional action,
\begin{eqnarray}
 V_{\mu}^i \,  = \,  \overline{q} \, \gamma_{\mu} \, \frac{\lambda^i}{2} \, q \,  \, = \, \, \frac{\partial \, 
{\cal L}_\text{R$\chi$T}}{\partial \, v^{\mu}_i} \, \Bigg|_{j=0} , && \quad
A_{\mu}^i  \, = \,  \overline{q} \, \gamma_{\mu} \, \gamma_5 \, \frac{\lambda^i}{2} \, q \,  \, = \, \, \frac{\partial \, 
{\cal L}_\text{R$\chi$T}}{\partial \, a^{\mu}_i} \, \Bigg|_{j=0} , \nonumber \\ [3.5mm]
S^i \,   = \,  - \, \overline{q} \,  \lambda^i \, q \,  \, = \, \, \frac{\partial \, 
{\cal L}_\text{R$\chi$T}}{\partial \, s_i} \, \Bigg|_{j=0} , && \quad
P^i  \,  = \,  \overline{q} \, i \gamma_5  \lambda^i \, q \,  \, = \, \, \frac{\partial \, 
{\cal L}_\text{R$\chi$T}}{\partial \, p_i} \, \Bigg|_{j=0} ,
\end{eqnarray}
where $j = 0$ indicates that all external currents are set to zero.
This gives~:
\begin{eqnarray}
\label{eq:expcur}
V_{\mu}^i & = & \frac{F^2}{4} \, \langle \,\lambda^i \, \left( u \, u_{\mu} \, 
u^{\dagger} \, -  \, u^{\dagger} \, u_{\mu} \, u \right) \, \rangle - \,\frac{F_{V}}{2 \sqrt{2}} \, 
\langle \, \lambda^i \, \partial^{\nu}  \left( u^{\dagger} \, V_{ \nu \mu} \, u 
\, + \, u \, V_{\nu \mu} \, u^{\dagger} \, \right) \, \rangle \, , \nonumber \\
A_{\mu}^i & = & \frac{F^2}{4} \, \langle \,\lambda^i \, \left( u \, u_{\mu} \, 
u^{\dagger} \, +  \, u^{\dagger} \, u_{\mu} \, u \right) \, \rangle \; , \nonumber \\
 S^i & = & \frac{1}{2} \, B_0 F^2 \, \langle \, \lambda^i \, \left( u^{\dagger}u^{\dagger}
\, + \, uu \right) \rangle \; , \nonumber \\
P^i & = & \frac{i}{2} \, B_0 F^2 \, \langle \, \lambda^i \, \left( u^{\dagger}u^{\dagger} 
\, - \,uu \right) \rangle \; .
\end{eqnarray}
With these expressions we are able to hadronise the final states in 
$\tau \rightarrow \mu P P$ and $\tau \rightarrow \mu P$
processes as we explain now~:
\vspace*{0.5cm} \\
{\bf $\gamma$ contribution}
\vspace*{0.5cm} \\
The photon contribution to the decay into two pseudoscalar mesons is driven by the electromagnetic
current~:
\begin{equation}
 V_{\mu}^{\mbox{\tiny em}} \, = \, \sum_{q}^{u,d,s} \, Q_q \, \overline{q} \, 
\gamma_{\mu} \, q  \, = \,  V_{\mu}^3 \, + \,  \frac{1}{\sqrt{3}} \,V_{\mu}^8 \, , 
\label{emcurrent}
\end{equation}
where $Q_q$ is the electric charge of the $q$ quark in units of the positron charge $e$.
The electromagnetic form factor is then defined as~:
\begin{equation} 
\label{eq:ffem}
\langle \, P_1 (p_1) P_2(p_2)\,|\, V_{\mu}^{em} \, | \, 0 \, \rangle \,
= \, (p_1-p_2)_{\mu} \, F_V^{\tiny P_1 P_2}(s) \, ,
\end{equation}
where  
$F_V^{\tiny P_1 P_2}(s)$ is steered by both $I=1$ and $I=0$ vector resonances, in particular the $\rho(770)$ that is the lightest of resonances. Due to the $q^2=0$ pole of the photon propagator this is, by far, the dominant contribution to this hadronic final state. Hence the result is more sensitive to the construction of this
form factor. Accordingly we will elaborate a more complete expression than the one provided by the
vector current in Eq.~(\ref{eq:expcur}) though it will reduce to this one in the $N_C \rightarrow \infty$
limit, including only one multiplet of resonances and at $q^2 \ll M_{\rho}^2$.
A proper construction of $F_V^{\tiny P_1 P_2}(s)$ is given in Appendix~\ref{ap:2}.
\vspace*{0.5cm} \\
{\bf $Z^0$ contribution}
\vspace*{0.5cm} \\
Here both vector and axial-vector currents do contribute. In terms of the quark fields these are~:
\begin{eqnarray}
\label{eq:z0current}
J_{\mu}^Z & = & V_{\mu}^Z \, + \, A_{\mu}^Z \, , \nonumber \\
 V_{\mu}^Z & = & \frac{g}{2 \,\cos \theta_W} \, \overline{q} \, \gamma_{\mu} \,
\left[ 2 \sin^2 \theta_W Q - T_3^{(q)} \right] \, q \; , \nonumber \\
A_{\mu}^Z & = & \frac{g}{2 \,\cos \theta_W} \, \overline{q} \, \gamma_{\mu} \,
\gamma_5 \, T_3^{(q)} \, q \, ,
\end{eqnarray}
with $Q = \mbox{diag}(2,-1,-1)/3$ and $T_3^{(q)} = \mbox{diag}(1,-1,-1)/2$ the electric charge and weak hypercharges, respectively, $g$ is the $SU(2)$ gauge coupling and $\theta_W$ is the weak angle.
\par
In order to proceed to the hadronisation of these currents one has to write the currents in 
Eq.~(\ref{eq:z0current}) in terms of $V_{\mu}^i$ and $A_{\mu}^i$ defined in Eq.~(\ref{eq:expcur}).
This gives
\begin{eqnarray}
V_{\mu}^Z & = & \frac{g}{2 \, \cos \theta_W} \, \frac{F^2}{2} \, \left[
2 \, \sin^2 \theta_W \, \langle \, Q \, \left( u u_{\mu} u^{\dagger} - u^{\dagger} u_{\mu} u \right) \rangle
\, - \, \langle \, T_3^{(q)} \,  \left( u u_{\mu} u^{\dagger} - u^{\dagger} u_{\mu} u \right) \rangle \right] \, 
\, , \nonumber \\
A_{\mu}^Z & = & \frac{g}{2 \, \cos \theta_W} \, \frac{F^2}{2} 
\langle \, T_3^{(q)} \,  \left( u u_{\mu} u^{\dagger} + u^{\dagger} u_{\mu} u \right) \rangle \, .
\label{ZPPcurrent} 
\end{eqnarray}
Notice that the vector current contributes to an even number of pseudoscalar mesons while the axial-vector
current provides 1,3,... mesons.
\vspace*{0.5cm} \\
{\bf Higgs bosons contribution}
\vspace*{0.5cm} \\
Hadronisation of scalar Higgs bosons like $h^0$ and $H^0$ into two pseudoscalar mesons 
proceeds through the scalar current while the pseudoscalar $A^0$ Higgs boson hadronises 
through the pseudoscalar current into one pseudoscalar meson. As Higgses are rather massive
the hadronisation is not so sensitive to resonances as in the case of the photon contribution. Hence we will not
elaborate on scalar of pseudoscalar form factors (analogous to the vector case defined by 
Eq.~(\ref{eq:ffem})) that, moreover, are not so well known. We will rely in the following 
scalar and pseudoscalar
currents,
\begin{eqnarray}
 \overline{u}  \, \Gamma \,  u & = &  \frac{1}{2} J^3 + \frac{1}{2 \sqrt{3}} J^8 + \frac{1}{\sqrt{6}} J^0  \, , \nonumber \\
  \overline{d} \, \Gamma \, d & = & 
- \frac{1}{2} J^3 + \frac{1}{2 \sqrt{3}} J^8 + \frac{1}{\sqrt{6}} J^0  \, , \nonumber \\
  \overline{s}  \, \Gamma \, s & = & 
 -\frac{1}{\sqrt{3}} J^8 + \frac{1}{\sqrt{6}} J^0 \, , 
\label{scalarbilinear}
\end{eqnarray}
where $\Gamma = -1$ for $J^i \equiv S^i$,  $\Gamma = i \gamma_5$ for $J^i \equiv P^i$ and
the $S^i$ and $P^i$ currents are given in Eq.~(\ref{eq:expcur}).

\section{Analytical results of the LFV semileptonic $\tau$ decays}\label{analytical}
 In this section we present the analytical results of the branching ratios for the 
 LFV semileptonic $\tau$ decays: $\tau \to \mu P P$, with $PP$ = $\pi^+\pi^-$ ,$\pi^0\pi^0$, $K^+K^-$, 
 $K^0 \bar{K}^0$ and $\tau \to \mu P$, with $P$ = $\pi$, $\eta$ and $\eta'$. The
 predictions for the $\tau \to \mu \rho^0$ and $\tau \to \mu \phi$ channels, which are related 
 to $\tau \to \mu \pi^+ \pi^-$ and $\tau \to \mu K^+K^-,\mu K^0 \bar{K}^0$ respectively, will also be included. 
 
 \begin{figure}[t!]
   \begin{center} 
     \begin{tabular}{cc} \hspace*{-12mm}
         \psfig{file=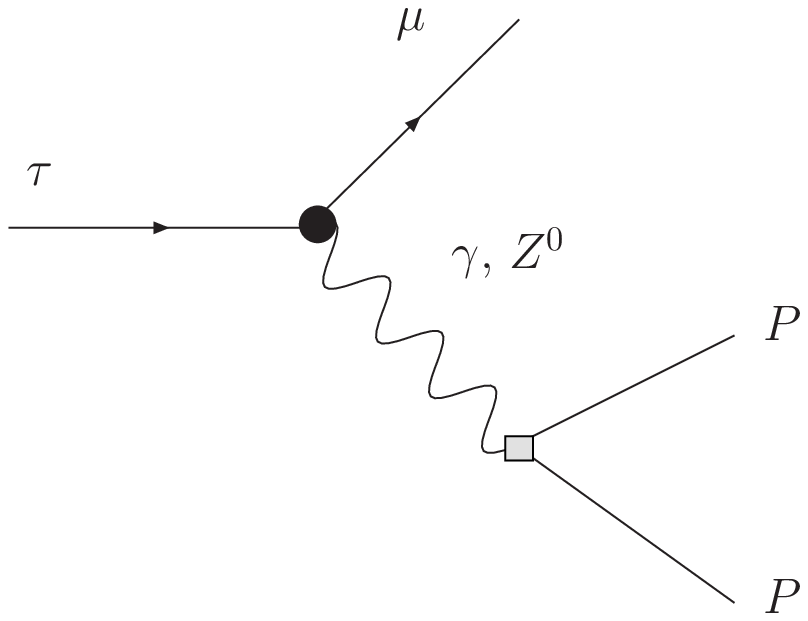,width=50mm,clip=} 
 &
  	\psfig{file=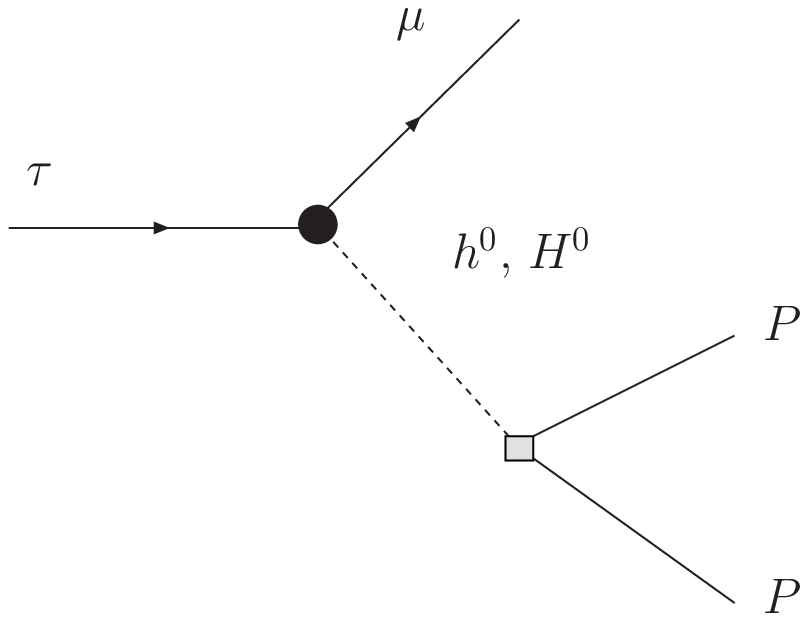,width=50mm,clip=} \\
         \psfig{file=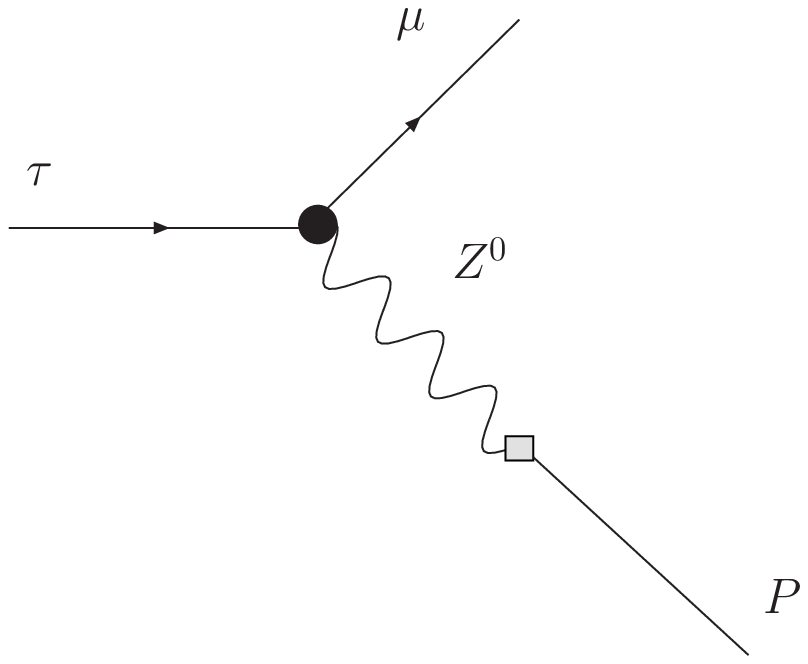,width=50mm,clip=} 
 &
  	\psfig{file=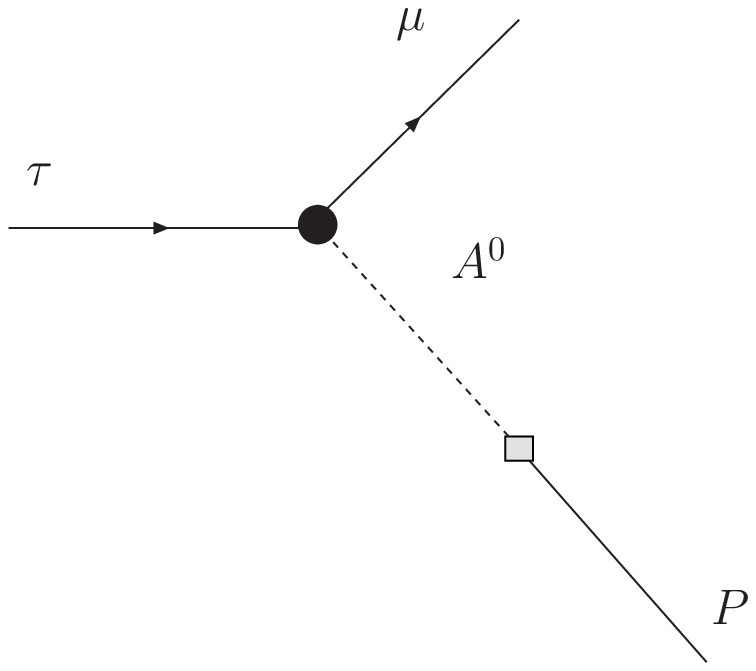,width=50mm,clip=}
     \end{tabular}
     \caption{Contributions to LFV semileptonic $\tau$
       decays into one and two pseudoscalar mesons.
     }\label{fig:diagrams} 
   \end{center}
 \end{figure}

\subsection{Predictions for $\tau \to \mu PP$}
The semileptonic $\tau \to \mu PP$ channels can be mediated by a photon, a $Z$ gauge boson
and a CP even Higgs boson, $h^0$ and $H^0$. The various contributing diagrams are depicted
in Fig.~\ref{fig:diagrams}. In these diagrams, the LFV vertex is represented by a black
circle and the hadronic vertex by a white box. The $Z$-mediated contribution is expected to
be much smaller than the $\gamma$-mediated contribution due to the 
${\cal O}(1/m_Z^2)$
suppression factor in the amplitude from the $Z$ propagator. This has been shown to happen in the
leptonic channels like $\tau \to 3 \mu$, where the $Z$-mediated contribution to its branching
ratio has been estimated to be a factor $10^{-3}-10^{-5}$ smaller than the 
$\gamma$-mediated contribution, for $\tan \beta=5-50$~\cite{Arganda:2005ji}. Consequently, 
we have neglected here 
the $Z$ contribution to the $\tau \to \mu PP$ decays. By using again this comparison with 
$\tau \to 3 \mu$, the $\gamma$ contribution to  $\tau \to \mu PP$ is expected to be the
dominant one, and the $h^0$ and $H^0$-mediated 
contributions are expected to be relevant only at large $\tan \beta$. Therefore, we have
included these three $\gamma$, $h^0$ and $H^0$ contributions in the computation.

The total amplitude for the $\tau \to \mu PP$ process can  be written as,
\begin{equation}
 T \, = \, T_{\gamma} \, + \, T_H \; ,
\end{equation}
where $T_{\gamma}$ and $T_H=T_{h^0}+T_{H^0}$ are the amplitudes of the $\gamma$-mediated and $H$-mediated
contributions respectively. 
First we present the result of $T_{\gamma}$ and $T_H$ in terms of the final state quarks, that is for 
$\tau \to \mu \overline{q} q$, and in terms of the corresponding $\tau$-$\mu$ 
LFV form factors:
\begin{eqnarray}
 T_{\gamma} &=& \overline{\mu} \left[ k^2 \gamma_{\mu} \left(A_1^L P_L +A_1^R P_R \right)
 + i m_{\tau}
\sigma_{\mu \nu} k^{\nu} \left( A_2^L P_L + A_2^R P_R \right) \right] \tau \nonumber \\
&& \times \, \frac{e^2 Q_q}{k^2} \,  \overline{q} \gamma^{\mu} q \, , 
\label{Tgamma}\\[2.5mm]
 T_H & = & \sum_{h^0, H^0} \frac{1}{m_{H_p}^2} \left\{ H_L^{(p)} S_{L,q}^{(p)} 
 \left[ \overline{\mu} P_L \tau \right] 
\left[ \overline{q} P_L q\right]  \, + \, 
H_R^{(p)} S_{R,q}^{(p)} 
\left[ \overline{\mu} P_R \tau   \right] 
\left[\overline{q} P_R q \right] \right. \nonumber \\
& &  \; \; \; \; \; \; \; \; \left.  + \; H_L^{(p)} S_{R,q}^{(p)} 
\left[  \overline{\mu} P_L  \tau \right] 
\left[\overline{q} P_R q \right]  \, + \, 
H_R^{(p)} S_{L,q}^{(p)} 
\left[  \overline{\mu} P_R \tau  \right] 
\left[ \overline{q}  P_L q \right] \right\} \; ,
\label{TH}
\end{eqnarray}
where, $k$ is the photon momentum, $Q_q$ the electric charge of the quark $q$ in
units of the positron charge $e$, $P_{L,R}=(1\mp \gamma_5)/2$, $m_\tau$ is the
$\tau$ 
lepton mass, and $m_{h^0}$,
$m_{H^0}$ are the Higgs boson masses. Notice that the momentum of the Higgs
propagators has been neglected against the Higgs boson mass.  The Higgs boson
couplings to quarks are correspondingly given by,
\begin{equation}
S_{L,u}^{(p)}   =   \frac{g}{2 m_W} \, \left(\frac{-\sigma_2^{(p)*}}{\sin
 \beta}\right) m_u , \,\, \; \; \quad \; 
S_{L,(d,s)}^{(p)}  =  \frac{g}{2 m_W} \, \left(\frac{\sigma_1^{(p)*}}{\cos \beta}
\right) m_{d,s}, \,\;  \; \; \quad 
 S_{R,q}^{(p)}   =   S_{L,q}^{(p)*} , 
\label{Hqqcouplings} 
\end{equation} 
where $m_q$ is the q quark mass, $m_W$ the W gauge boson mass, $g$ the $SU(2)$
gauge coupling,  and 
\begin{eqnarray}
\sigma_1^{(p)} &=& \left( \begin{array}{c} \sin{\alpha} \\ -\cos{\alpha} \\ i
    \sin{\beta} \end{array} \right) \,,\,\,\,\,  \quad \; \quad  \quad 
\sigma_2^{(p)}\, =\, \left( \begin{array}{c} \cos{\alpha} \\ \sin{\alpha} \\ -i
    \cos{\beta} \end{array} \right) \,. 
\label{sigmas}    
\end{eqnarray} 
The three entries for the index (p) in the previous expressions and in the 
following ones
correspond to $H_p =h^0, H^0, A^0$, respectively. The angle $\alpha$ rotates, as usual, from the electroweak neutral Higgs basis to the mass eigenstate basis.

The LFV form factors $A_{1,2}^{L,R}$ in Eq.~(\ref{Tgamma}) describe the
effective 
$\gamma \tau \mu$ vertex and get contributions from the SUSY one-loop diagrams depicted in  
Fig.~\ref{Photon_diagrams}. The full results for these form factors can be found in
the literature~\cite{Hisano:1995cp,Arganda:2005ji} and are collected 
in Appendix A.1 for completeness. Notice that we are presenting all the results in the physical
mass eigenstate basis for all the particles involved. Therefore the LFV is encoded
in the physical slepton and sneutrino masses and in the corresponding slepton and
sneutrino rotation matrices. The later appear in the chargino-sneutrino-lepton and
neutralino-slepton-lepton couplings. Similarly, the LFV form factors $H_{L,R}^{(p)}$
in Eq.~(\ref{TH})
describe the effective $H_p \tau \mu$ vertex and get contributions from the SUSY
one-loop diagrams shown in Fig.~\ref{H_diagrams}. These set of diagrams where 
computed in~\cite{Arganda:2004bz} and the results are collected  
in Appendix A.3. Again the LFV is encoded in the slepton and sneutrino masses and
in the rotation matrices.

The next step is to hadronise the quark bilinears appearing in Eqs.~(\ref{Tgamma}) and
(\ref{TH}). For this, we proceed as explained in Sec.~\ref{hadronisation}.
The quark bilinears in $T_\gamma$, $\left[\overline{q} \gamma_\mu q
\right]$, are hadronised via the electromagnetic
current in Eq.~(\ref{emcurrent}) which, for the final state with two mesons
$P_1(p_1)P_2(p_2)$, is
then written in
terms of the corresponding electromagnetic form factor, $F_V^{P_1P_2}$, by
means of Eq.~(\ref{eq:ffem}). Thus, one gets the photon amplitude in terms
of the final state hadrons:
\begin{equation}\label{Tgamma_hadron}
T_{\gamma} \,=\, \frac{e^2}{k^2} \, F_V^{P_1P_2}(k^2) \,
\overline{\mu}\left[ k^2 (p_1 \!\!\!\!\! \! / \, \, - p_2 \! \! \! \! \! \! / \, \, )
 \left(A_1^L P_L +A_1^R P_R \right) + 2 \, i \,  m_\tau \, p_1^{\mu}
\sigma_{\mu \nu} \,  p_2^{\nu} \left( A_2^L P_L + A_2^R P_R \right) \right]
\tau .
\end{equation}
The expressions of the $F_V^{P_1P_2}$ form factors for each 
of the final states, $P_1P_2$ =$\pi^+\pi^-$, $K^+K^-$ and $K^0
\bar{K}^0$ are collected in Appendix B. Obviously, the $\pi^0\pi^0$
final state does not get photon-mediated  contributions since $\gamma$ does
not couple to $\pi^0\pi^0$. Hence we set  $F_V^{\pi^0\pi^0}=0$.

The quark bilinears in $T_H$, $\left[ \overline{q} P_{L,R} q \right]$, when
hadronised in a final state of two mesons, get contributions just from scalar
currents, $S^i$, but not from pseudoscalar currents, $P^i$. Then, one
substitutes $\left[ \overline{q} P_{L,R} q \right]$ by 
$\left[(-1/2)(-\overline{q} q) \right]$, where $(-\overline{q} q)$ is given in
Eq.~(\ref{scalarbilinear}), and the relevant scalar currents, $S^0$, $S^3$
and $S^8$, are written in terms of two mesons by using
Eq.~(\ref{eq:expcur}). This gives:
 \begin{eqnarray}
S^3 & = & - B_0  \, \left[\frac{2}{\sqrt{3}} \left(\cos \theta - \sqrt{2} \sin \theta \right) \pi^0 \eta 
\, + \, \frac{2}{\sqrt{3}} \left( \sqrt{2} \cos \theta + \sin \theta \right) \pi^0 \eta' \,
+ \, K^+ K^- \, -  \, K^0 \bar K^0 \right]  \, , \nonumber \\
S^8 & = & \frac{B_0}{\sqrt{3}} \left[ K^+ K^- \, + \, K^0 \bar K^0 \, - \, 2 \pi^+ \pi^- \, - \, \pi^0 \pi^0\, 
+ \, \left( \cos^2 \theta \, + \, 2 \sqrt{2} \sin \theta \cos \theta \right) \, \eta \eta  \right. \nonumber \\
& & \; \; \; \; \; \; \; \left. + \, 2 \left( \sqrt{2} \sin^2 \theta + \sin \theta \cos \theta - \sqrt{2} \cos^2 \theta \right) \, \eta \eta' \, \right] \; , \nonumber \\
S^0 & = & - B_0 \sqrt{\frac{2}{3}} \left[ \, 2 \pi^+ \pi^- \, + \,  2 K^+ K^- \, + \,  2 K^0 \bar K^0 \, + \, 
\pi^0 \pi^0 \, + \, \eta \eta \right] \, .
\end{eqnarray}
Thus, one gets the Higgs boson amplitude in terms of the final state
hadrons:
\begin{equation}
 T_H \, = \, \sum_{p= h^0, H^0} \, \overline{\mu} \left[ \, c^{(p)}_{PP} \,
 + \, d^{(p)}_{PP} \, \gamma_5 \right] \, \tau
\, ,
\end{equation}
where
\begin{eqnarray} \label{eq:cdhiggs}
 c^{(p)}_{PP} & = & \frac{g}{2 m_W} \frac{1}{2 m_{H_p}^2} \, 
 \left( {J_L^{(p)}(PP)} + {J_R^{(p)}(PP)} \right) 
 \left( H_R^{(p)} + H_L^{(p)} \right) \, , \nonumber \\
d^{(p)}_{PP} & = &\frac{g}{2 m_W} \frac{1}{2 m_{H_p}^2} \, 
\left(  {J_L^{(p)}(PP)} +  {J_R^{(p)}(PP)} \right) 
 \left( H_R^{(p)} - H_L^{(p)} \right) \, ,
\end{eqnarray}
and
\begin{eqnarray}
 J_L^{(p)}(\pi^+ \pi^-) & = & J_L^{(p)}(\pi^0 \pi^0) \; = \; \frac{1}{4} \left( \left( \frac{-\sigma_2^{(p)*}}{\sin \beta} \right) m_{\pi}^2 + 
\left( \frac{\sigma_1^{(p)*}}{\cos
\beta} \right) m_{\pi}^2 \right) \, ,
\nonumber \\[2mm]
J_L^{(p)}(K^+ K^-) & = &  \frac{1}{4} \left( \left( \frac{-\sigma_2^{(p)*}}{\sin \beta} \right) m_{\pi}^2 +
\left( \frac{\sigma_1^{(p)*}}{\cos \beta}\right) (2m_K^2 - m_{\pi}^2) 
\right) \, ,
\nonumber \\[2mm]
J_L^{(p)}(K^0 \bar K^0) & = & \frac{1}{2} \left( \frac{\sigma_1^{(p)*}}{\cos \beta}
\right) m_K^2 \, ,\nonumber \\[2mm]
J_R^{(p)}(PP) & = & J_L^{(p)*}(PP) \, .
\label{Jotas}
\end{eqnarray}
Notice that in Eq.~(\ref{Jotas}) we have already used the relations between the 
quark and the meson masses of $\chi$PT given in Eq.~(\ref{mq-mP}). 
 
Finally, we get the following result for the branching ratio:
\begin{equation}
 {\rm BR}(\tau \to \mu PP)\, = \,  \frac{\kappa_{PP}}{64\, \pi^3 \, m_\tau^2 \, \Gamma_\tau}  \, 
\int_{s_{\rm min}}^{s_{\rm max}} ds \,
 \int_{t_{\rm min}}^{t_{\rm max}} dt \,
 \, \frac{1}{2} \sum_{i,f} |T|^2 \,,
\end{equation}
where $\Gamma_\tau$ is the total $\tau$ decay width, and the coefficient
$\kappa_{PP}$ is 1 for $PP=\pi^+\pi^-,K^+K^-,K^0 \bar{K}^0$ and 
1/2 for
$PP=\pi^0\pi^0$. In addition
\begin{eqnarray}
t_{\rm min}^{\rm max} &= &  \frac{1}{4 s} \left[ \left( m_\tau^2 - m_\mu^2 \right)^2 -
\left( \lambda^{1/2}\left( s,m_P^2, m_P^2 \right) \mp \lambda^{1/2} \left(
m_\tau^2, s, m_\mu^2 \right) 
\right)^2 \right] , \nonumber \\
 s_{\rm min}  &=&  4 m_P^2 \,\,,\,\, 
 s_{\rm max} =  \left( m_\tau - m_\mu \right)^2 \, \,, 
\, \, \lambda(x,y,z)
 = (x+y-z)^2-4xy \,.
\label{lambda}
\end{eqnarray}
The averaged squared amplitude is, 
\begin{equation}\label{eq:gammaH}
 \frac{1}{2} \sum_{i,f} |T|^2 \, = \, \frac{1}{8 \, m_\tau} \, 
\left[ g_1(s) \, + \,  g_2(s) \, t \, + \, g_3(s) \, t^2 \right] \, .
\end{equation}
where
\begin{eqnarray}
 g_1(s) & = & h_0 \, + \, h_1 \, s \, + \, h_2 \, s^2 \, + \, h_3 \, s^3 \; , \nonumber \\
 g_2(s) & = & j_1 \, s \, + \, j_2 \, s^2 \, + \, j_3 \, s^3 \; , \nonumber \\
 g_3(s) & = & k_1 \, s \, + \, k_2 \, s^2 \;,
\end{eqnarray}
with
\begin{eqnarray} 
 h_0  & = & -8 M_P^2 m_\tau^2 \left(  m_\mu^2 - m_\tau^2 \right)^2  \left( A_2^- A_2^{-*} + 
A_2^+ A_2^{+*} \right) \, \nonumber \\
& & + \, 2 \left( m_\mu + m_\tau \right)^2 \, c_H c_H^* \, 
+ \, 2 \left( m_\mu - m_\tau \right)^2 \, d_H d_H^* \,,  \nonumber \\
h_1 & = & - 8 m_\tau^2 \left( m_\tau m_\mu + M_P^2 \right)^2 \, A_2^- A_2^{-*} \, 
- 8 m_\tau^2 \left( m_\tau m_\mu - M_P^2 \right)^2 \, A_2^+ A_2^{+*} \, \nonumber \\
& & + 8 \left( m_\mu- m_\tau \right) m_\tau \left( m_\mu + m_\tau \right)^2 M_P^2 
\, \left( A_1^{-*}A_2^- + A_1^- A_2^{-*} \right) \, \nonumber \\
& & + 8 \left( m_\mu- m_\tau \right)^2 m_\tau \left( m_\mu + m_\tau \right) M_P^2 
\, \left( A_1^{+*}A_2^+ + A_1^+ A_2^{+*} \right) \, \nonumber \\
&& -2 \left( m_\mu + m_\tau \right) \, \left( m_\mu^2 + m_\tau^2 + 2 M_P^2 \right) \left( c_H A_1^{+*} + 
c_H^* A_1^+ \right) \nonumber \\
& & + 2 m_\tau \left( m_\mu^2 + m_\tau^2 + 2 M_P^2 \right) \, \left( c_H A_2^{+*} + 
c_H^* A_2^+ \right) \nonumber \\
& &   -2 \left( m_\mu - m_\tau \right) \, \left( m_\mu^2 + m_\tau^2 + 2 M_P^2 \right) \left( d_H A_1^{-*} + 
d_H^* A_1^- \right) \nonumber \\
& & + 2 m_\tau \left( m_\mu^2 + m_\tau^2 + 2 M_P^2 \right) \, \left( d_H A_2^{-*} + 
d_H^* A_2^- \right) \nonumber \\
&& - 2 c_H c_H^* - 2 d_H d_H^* \, , \nonumber \\
h_2 &=& 2 \left[\left( m_\mu^2 + m_\tau^2 \right)^2 + 4 M_P^4 + 8 m_\mu m_\tau M_P^2 
\right] \, A_1^+ A_1^{+*} \nonumber \\
& & + 2 \left[\left( m_\mu^2 + m_\tau^2 \right)^2 + 4 M_P^4 - 8 m_\mu m_\tau M_P^2 
\right] \, A_1^- A_1^{-*} \, \nonumber \\
& & 
+ \, 2 m_\tau^2 \left[ \left( m_\mu - m_\tau \right)^2 + 4 M_P^2 \right] \, A_2^+ A_2^{+*} \, 
+ \, 2 m_\tau^2 \left[ \left( m_\mu + m_\tau \right)^2 + 4 M_P^2 \right] \, A_2^- A_2^{-*}
\nonumber \\
& & 
- \, 2 m_\tau \left( m_\mu - m_\tau \right) \left[ \left( m_\mu + m_\tau \right)^2 + 4 M_P^2 
\right] \left( A_1^{-*} A_2^- + A_1^- A_2^{-*} \right) \nonumber \\
&&
- \, 2 m_\tau \left( m_\mu + m_\tau \right) \left[ \left( m_\mu - m_\tau \right)^2 + 4 M_P^2 
\right] \left( A_1^{+*} A_2^+ + A_1^+ A_2^{+*} \right) \nonumber \\
&&
+ \, 2 \left( m_\mu + m_\tau \right) \left( c_H A_1^{+*} + c_H^* A_1^+ \right) \,
+ \, 2 \left( m_\mu - m_\tau \right) \left( d_H A_1^{-*} + d_H^* A_1^- \right) \nonumber \\
&&
- \, 2 m_\tau \left[ c_H A_2^{+*} + c_H^* A_2^+ + d_H A_2^{-*} + d_H^* A_2^- \right] \, , \nonumber \\
h_3 & = & - 2 \left( m_\mu - m_\tau \right)^2 \, A_1^- A_1^{-*} \, 
- \,  2 \left( m_\mu + m_\tau \right)^2 \, A_1^+ A_1^{+*} \, 
-\, 2 m_\tau^2 \, \left[ A_2^- A_2^{-*} + A_2^+ A_2^{+*} \right] \nonumber \\
&&
+ \, 2 m_\tau \left( m_\mu - m_\tau \right)  \, \left[ A_1^- A_2^{-*} + A_1^{-*} A_2^- \right]\,
+ \, 2 m_\tau \left( m_\mu + m_\tau \right)  \, \left[ A_1^+ A_2^{+*} +
A_1^{+*} A_2^+ \right]\, , \nonumber \\
j_1 &= & 8 m_\tau^2 \left( m_\mu^2 + m_\tau^2 + 2 M_P^2 \right) 
\left( A_2^- A_2^{-*} + A_2^+ A_2^{+*} \right) \nonumber \\
&&
- \, 4 m_\tau \left[ c_H A_2^{+*} + c_H^* A_2^+ + d_H A_2^{-*} + d_H^* A_2^- \right] \nonumber \\
&&
+ \, 4 \left( m_\mu + m_\tau \right)  \left( c_H A_1^{+*} + c_H^* A_1^+ \right) \, 
+ \, 4 \left( m_\mu - m_\tau \right)  \left( d_H A_1^{-*} + d_H^* A_1^- \right) \,, \nonumber \\
j_2 &=& - 8 \left( m_\mu^2 + m_\tau^2 + 2 M_P^2 \right) \, 
\left[ A_1^+ A_1^{+*} + A_1^- A_1^{-*} \right] 
\, - \, 8 m_\tau^2 \, \left[ A_2^+ A_2^{+*} + A_2^- A_2^{-*} \right] \, , \nonumber \\
j_3 &=& 8 \, \left( A_1^- A_1^{-*} + A_1^+ A_1^{+*} \right) \; ,  \nonumber \\
k_1 &=& - 8 m_\tau^2 \, \left( A_2^- A_2^{-*} + A_2^+ A_2^{+*} \right) \, , \nonumber \\
k_2 &=& 8 \, \left( A_1^- A_1^{-*} +  A_1^+ A_1^{+*} \right) \, ,
\end{eqnarray}
and
\begin{eqnarray}
 A_i^{\pm} & = & \frac{e^2}{2 \, s} \, F_V^{PP} (s) (A_i^R \pm A_i^L)
 \,\,,\,\,
c_H  =  c^{(h^0)}_{PP} \, + \, c^{(H^0)}_{PP} \,\,,\,\, 
d_H  =  d^{(h^0)}_{PP} \, + \, d^{(H^0)}_{PP} \, .
\end{eqnarray}

\subsection{Predictions for $\tau \to \mu P$}
The semileptonic  $\tau \to \mu P$ channel can be mediated by a $Z$ gauge boson
and a CP odd $A^0$ Higgs boson, as represented in Fig.~\ref{fig:diagrams}. 
Both contributions are included here. The total amplitude for this 
$\tau \to \mu P$ decay can then be written as, 
\begin{equation}
 T \, = \, T_{Z} \, + \, T_{A^0} \;,
\end{equation}
where $T_Z$ and $T_{A^0}$ are the $Z$ and $A^0$ mediated amplitudes respectively.
As in the previous case, these are first evaluated in terms of the final state quarks, that is for $\tau
\to \mu \overline{q} q$, and in terms of the corresponding $\tau-\mu$ LFV form
factors: 
\begin{eqnarray}
T_{Z} &=& \frac{1}{m_Z^2} \overline{\mu} \left[ \gamma_{\mu} 
( F_L P_L + F_R P_R) \right] \tau \,\,.\,\,
\overline{q} \left[ \gamma^{\mu} 
\left( Z_L^{(q)} P_L + Z_R^{(q)} P_R \right) \right]q \, , 
\label{TZ} \\[2.5mm]
T_{A^0} & = & \frac{1}{m_{A^0}^2} \left\{ H_L^{(A^0)} S_{L,q}^{(A^0)} 
\left[ \overline{\mu}  P_L \tau  \right] \left[  \overline{q} P_L q \right]  \, + \, 
H_R^{(A^0)} S_{R,q}^{(A^0)} 
\left[ \overline{\mu} P_R \tau  \right] \left[ \overline{q}  P_R q \right] \right. \nonumber \\
& &  \; \; \; \; \; \; \; \; \left.  + \; H_L^{(A^0)} S_{R,q}^{(A^0)} 
\left[ \overline{\mu}  P_L \tau  \right] \left[ \overline{q}  P_R q \right]  \, + \, 
H_R^{(A^0)} S_{L,q}^{(A^0)} 
\left[ \overline{\mu}  P_R  \tau \right] \left[ \overline{q}  P_L q \right]
\right\} \;,
\label{TA}
\end{eqnarray}
where $Z_L^{(q)}=(-g/\cos\theta_W)(T_3^{(q)}-Q_q\sin^2\theta_W)$ and 
$Z_R^{(q)}=(g/\cos\theta_W) Q_q \sin^2\theta_W$ are the $Z$ couplings to  
quarks, and $S_{L,q}^{(A^0)}$ and  $S_{R,q}^{(A^0)}$ are the $A^0$ couplings to 
quarks, which are given by the third entry 
in Eqs.~(\ref{Hqqcouplings}) and (\ref{sigmas}). Notice that, as in the previous cases of $h^0$ and $H^0$, we have neglected the $k^2$ in the $Z$ and $A^0$ propagators.

The LFV form factors $F_{L,R}$ in Eq.~(\ref{TZ}) describe the effective 
$Z\tau\mu$
vertex and receive contributions from the SUSY one-loop diagrams depicted in 
Fig.~\ref{Z_diagrams}. The results for these form factors where found
in~\cite{Hisano:1995cp}
and corrected in~\cite{Arganda:2005ji}. We collect them
in Appendix A.2, for completeness. The LFV form factors $H_{L,R}^{(A^0)}$ in 
Eq.~(\ref{TA}) describe
the effective $A^0 \tau \mu$ vertex and, as in the previous $H \tau \mu$ vertices
with $H=h^0,H^0$, receive contributions from the one-loop diagrams in 
Fig.~\ref{H_diagrams}. The corresponding results are collected in Appendix A.3.

The hadronisation of the quark bilinears in $T_Z$ proceeds by means of the vector
and axial-vector currents in Eq.~(\ref{eq:z0current}), which in turn are written
in terms of one $P$ meson
by means of Eq.~(\ref{ZPPcurrent}). This leads to:   
\begin{eqnarray}
V_{\mu}^Z &=& 0 , \\
 A_{\mu}^Z &=& - \frac{g}{2 \cos{\theta_W}} F \left\{ C(\pi^0) \, \partial_{\mu}
 \pi^0 + C(\eta) \, \partial_{\mu} \eta + C(\eta^\prime) \, 
 \partial_{\mu} \eta^\prime \right\},
\end{eqnarray}
where the $C(P)$ functions are given by,
\begin{eqnarray}
C(\pi^0) &=& 1 , \nonumber \\
C(\eta) &=& \frac{1}{\sqrt{6}} \left( \sin \theta + \sqrt{2} \cos \theta \right) , \nonumber \\
C(\eta^\prime) &=& \frac{1}{\sqrt{6}} \left( \sqrt{2} \sin \theta - \cos \theta \right) \, .
\end{eqnarray}
The hadronisation into one pseudoscalar meson $P$ 
of the quark bilinears in $T_{A^0}$  
proceed via the pseudoscalar currents $P^i$. Concretely, $P^0$, $P^3$ and $P^8$,
whose expressions in terms of one $P$ meson can be obtained from
Eq.~(\ref{eq:expcur}). This leads to: 
\begin{eqnarray}
P^3 & = & 2 B_0 F \, \pi^0 \, , \nonumber \\
P^8 & = & 2 B_0 F \left( \, \cos \theta \, \eta + \sin \theta \, \eta' \, \right) \, , \nonumber \\
P^0 & = & 2 B_0 F \left( \, - \sin \theta \, \eta + \cos \theta \, \eta' \, \right) \, .
\end{eqnarray}  
Finally, by putting all together, we get the following result for the branching ratio: 
\begin{equation}
  {\rm BR}(\tau \to \mu P) \, = \, \frac{1}{4 \pi} \,
  \frac{\lambda^{1/2}(m_\tau^2,m_\mu^2,m_P^2)}{m_\tau^2 \,\,\Gamma_\tau} \,
\frac{1}{2} \,  \sum_{i,f} |T|^2 \, ,
\end{equation}
where the $\lambda(x,y,z)$ function is defined in Eq.~(\ref{lambda}) and again $\Gamma_\tau$
is the total decay width of the $\tau$ lepton. 
The averaged squared amplitude is given by,    
\begin{equation}
 \frac{1}{2} \sum_{i,f} |T|^2  =  \frac{1}{4 \, m_\tau} 
\sum_{k,m} \left[ 2 m_\mu m_\tau \left( a^k_P a^{m\,*}_P - b^k_P b^{m\,*}_P \right) 
+ (m_\tau^2 + m_\mu^2 - m_P^2 ) \left( a^k_P a^{m\,*}_P  + b^k_P b^{m\,*}_P \right)
\right],
\end{equation}
with $k,m = Z,A^0$, and
\begin{eqnarray}
 a^Z_P & = & - \frac{g}{2 \cos\theta_W} \frac{F}{2} \frac{C(P)}{m_Z^2} 
 \left( m_\tau-m_\mu \right) 
\left( F_L + F_R \right) \, ,\nonumber \\
b^Z_P & = &  \frac{g}{2 \cos\theta_W} \frac{F}{2} \frac{C(P)}{m_Z^2} 
\left( m_\tau+ m_\mu \right) 
\left( F_R - F_L \right) \, ,\nonumber \\
a^{A^0}_P & = &  \frac{g}{2 m_W} \frac{F}{2 m_{A^0}^2} 
 \left(  {B_L^{(A^0)}(P)} -  {B_R^{(A^0)}(P)} \right)
\left( H_L^{(A^0)} + H_R^{(A^0)} \right) \, , \nonumber \\
b^{A^0}_P & = & \frac{g}{2 m_W} \frac{F}{2 m_{A^0}^2} 
 \left(  {B_L^{(A^0)}(P)} -  {B_R^{(A^0)}(P)} \right)
\left( H_R^{(A^0)} - H_L^{(A^0)} \right) \, .
\end{eqnarray}
The $B_{L,R}^{(A^0)}(P)$ functions are given, correspondingly, by the third entry of: 
\begin{eqnarray}
 B_L^{(p)}(\pi) & = & \frac{m_{\pi}^2}{4} 
 \left( \frac{-\sigma_2^{(p)*}}{\sin \beta} - 
 \frac{\sigma_1^{(p)*}}{\cos \beta} \right) \, , \nonumber \\[2mm]
B_L^{(p)}(\eta) & = & \frac{1}{4 \sqrt{3}} 
\left[ \frac{-\sigma_2^{(p)*}}{\sin \beta}m_{\pi}^2 \left( \cos \theta - \sqrt{2}
\sin \theta \right) + 
\frac{\sigma_1^{(p)*}}{\cos \beta} \left[ \left( 3 m_{\pi}^2 - 4 m_K^2 \right) \cos \theta -
2 \sqrt{2} m_K^2 \sin \theta \right] \right] \, , \nonumber \\[2mm]
B_L^{(p)}(\eta') & = & \frac{1}{4 \sqrt{3}} 
\left[ \frac{ -\sigma_2^{(p)*}}{\sin \beta} m_{\pi}^2 \left( \sin \theta + \sqrt{2}
\cos \theta \right) + 
\frac{\sigma_1^{(p)*}}{\cos \beta} \left[ \left( 3 m_{\pi}^2 - 4 m_K^2 \right) \sin \theta +
2 \sqrt{2} m_K^2 \cos \theta \right] \right] \,, \nonumber \\[2mm]
B_R^{(p)}(P) & = & B_L^{(p)*}(P) \, ,
\label{Bs}
\end{eqnarray}
where the $\sigma_{1,2}^{(p)}$ functions are defined in Eq.~(\ref{sigmas}).
Notice that in this Eq.~(\ref{Bs}) the relations between the quark and meson masses of 
Eq.~(\ref{mq-mP}) have been used again.  

\subsection{Predictions for $\tau \to \mu \rho$ and $\tau \to \mu \phi$}
The $\tau \to \mu \rho^0$ decay is related to the $\tau \to \mu \pi^+ \pi^-$
channel since the $\rho$ decay proceeds mainly to $\pi^+\pi^-$. Indeed a $\rho^0$ is not
an asymptotic state~: the experiment reconstructs its structure from the two observed
pions. In addition, from the chiral point of view, two pions in a $J=I=1$ state are
indistinguishable from a $\rho$. Therefore one has to 
define the branching ratio of $\tau \to \mu \rho^0$ in close
relation to that of $\tau \to \mu \pi^+\pi^-$ as follows:
\begin{eqnarray}
 {\rm BR}(\tau \to \mu \rho^0)\, &=& \frac{1}{64\, \pi^3 \, m_\tau^2 \, \Gamma_\tau} 
\, \int_{s_{\rm min}}^{s_{\rm max}} ds \,
\left[ \int_{t_{\rm min}}^{t_{\rm max}} dt \,
 \frac{1}{2} \sum_{i,f} |T_\gamma|^2 \, \right]_{\pi^+\pi^-},
\end{eqnarray}
where $T_\gamma$ is defined in Eq. (\ref{Tgamma_hadron}) and all functions and
form factors involved are as 
those of $\tau \to \mu \pi^+\pi^-$ decay, with the exception of the
integration limits in $s$ which are now:
\begin{eqnarray} \label{eq:smnax}
s_{\rm min} &= &M_\rho^2- \frac{1}{2} M_\rho \Gamma_\rho\,\,,\,\, \quad  \quad 
s_{\rm max} = M_\rho^2 + \frac{1}{2} M_\rho \Gamma_\rho \,.
\end{eqnarray}
Similarly, the $\tau \to \mu \phi$ decay is related to the 
$\tau \to \mu K^+K^-$ and $\tau \to \mu K^0 \bar{K}^0$ decays since 
the $\phi$ decays proceeds mainly to  $K^+K^-$ and to $K^0 \bar{K}^0$.
Therefore, we define:
\begin{eqnarray}
 {\rm BR}(\tau \to \mu \phi)\,& = &\frac{1}{64\, \pi^3 \, m_\tau^2 \, \Gamma_\tau} \, 
\left\{ \int_{s_{\rm min}}^{s_{\rm max}} ds 
 \left[\int_{t_{\rm min}}^{t_{\rm max}} dt \,
 \frac{1}{2} \sum_{i,f} |T_\gamma|^2 \right]_{K^+K^-} 
\right. \nonumber \\[2.5mm]  &  &\,\left. \quad \quad \quad \quad \quad \; \; + \, 
 \int_{s_{\rm min}}^{s_{\rm max}} ds 
 \left[ \int_{t_{\rm min}}^{t_{\rm max}} dt \,
 \frac{1}{2} \sum_{i,f} |T_\gamma|^2 \right]_{K^0 \bar{K}^0} \right\} \, , 
\end{eqnarray}
where again $T_\gamma$ is defined in Eq. (\ref{Tgamma_hadron}) and all functions and
form factors involved are as those of $\tau \to \mu K^+ K^-$ and $\tau \to \mu
K^0 \bar{K}^0$ correspondingly, except for the
integration limits in $s$ which are now:
\begin{eqnarray} \label{eq:smnox}
s_{\rm min} &= &M_\phi^2- \frac{1}{2} M_\phi \Gamma_\phi\,\,,\quad \quad \quad \; \,\,
s_{\rm max} = M_\phi^2 + \frac{1}{2} M_\phi \Gamma_\phi \, .
\end{eqnarray}
In Eqs.~(\ref{eq:smnax},\ref{eq:smnox}), $\Gamma_{\rho}= \Gamma_{\rho}(M_{\rho}^2)$
and $\Gamma_{\phi}= \Gamma_{\phi}(M_{\phi}^2)$.

\section{Numerical results and discussion}\label{results}

In this section we present the numerical results of the LFV
semileptonic $\tau \to \mu PP$ and $\tau \to \mu P$ decay rates within
the constrained MSSM-seesaw scenarios described in Section~\ref{th_framework}.
Since our main goal is to explore if the predicted rates can or cannot
reach the present experimental sensitivities we will focus mainly on
choices of the input parameter values that lead to large $\delta_{32}$
and therefore to large LFV semileptonic $\tau$ decay rates. As we have
seen in the previous Section~\ref{th_framework}, within the scenario
with hierarchical heavy neutrinos and for $\theta_{1, 3} = 0$, this
means large values of $\theta_2$ and large values of $m_{N_3}$. On the
other hand, since all these rates grow with $\tan \beta$, in the
following numerical analysis we will focus mainly on large $\tan
\beta$ values. In the first subsection we will present the numerical results,
from our full computation of the LFV semileptonic tau decay rates and will
explore the dependence with the most relevant parameters in the constrained
MSSM scenarios.  In the second subsection we will include a comparison between
our full and some approximate results in the large $\tan \beta$ region. 
Moreover, we will also analyse to what extent the Higgs dominance
hypothesis holds for these LFV semileptonic $\tau$ decays and compare our 
predictions with other
results in the literature.
We will conclude this
section by showing that for some particular choices of the input
parameters, the rates for some channels indeed reach the present
experimental sensitivity.  
\subsection{LFV semileptonic tau decay rates} 
Firstly, we present the results for the simplest case of
$\theta_2 = 0$ and study the relative importance of the various contributions
to the decay rates that have been presented in the previous
section. Then we explore the increase in the rates for larger values of $\theta_2$. Since we are setting in the whole
numerical analysis $A_0 = 0$ and sign $(\mu) = +1$, the relevant SUSY
parameter will be $M_\text{SUSY} = M_0 = M_{1/2}$. In the study of the
behaviour of the rates with $M_\text{SUSY}$ we pay
special attention to the decoupling or non-decoupling behaviour of the SUSY particles at
large $M_\text{SUSY}$. 

\begin{figure}[t!]
   \begin{center} 
     \begin{tabular}{cc} \hspace*{-12mm}
         \psfig{file=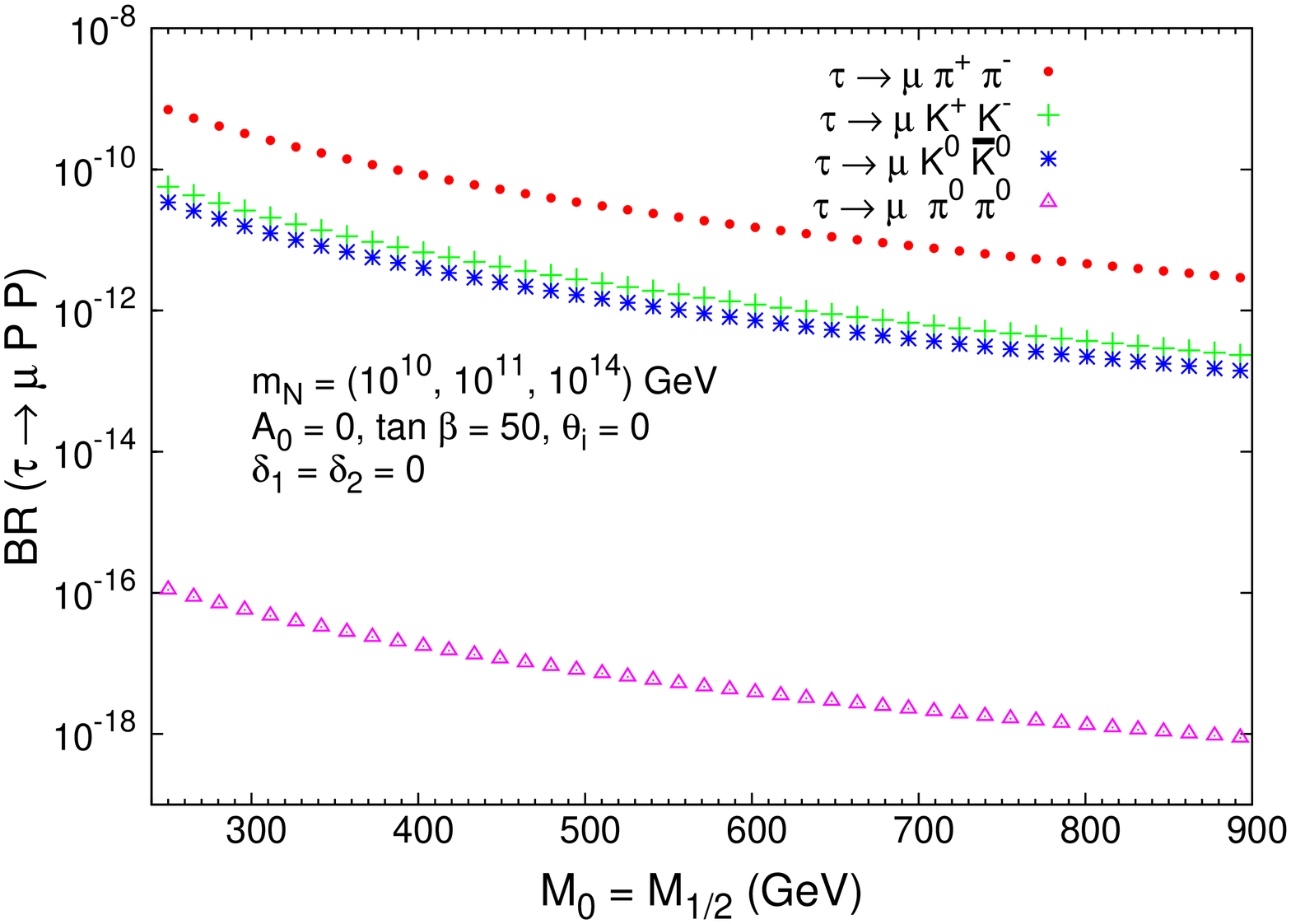,width=60mm,angle=270,clip=} 
 &
  	 \psfig{file=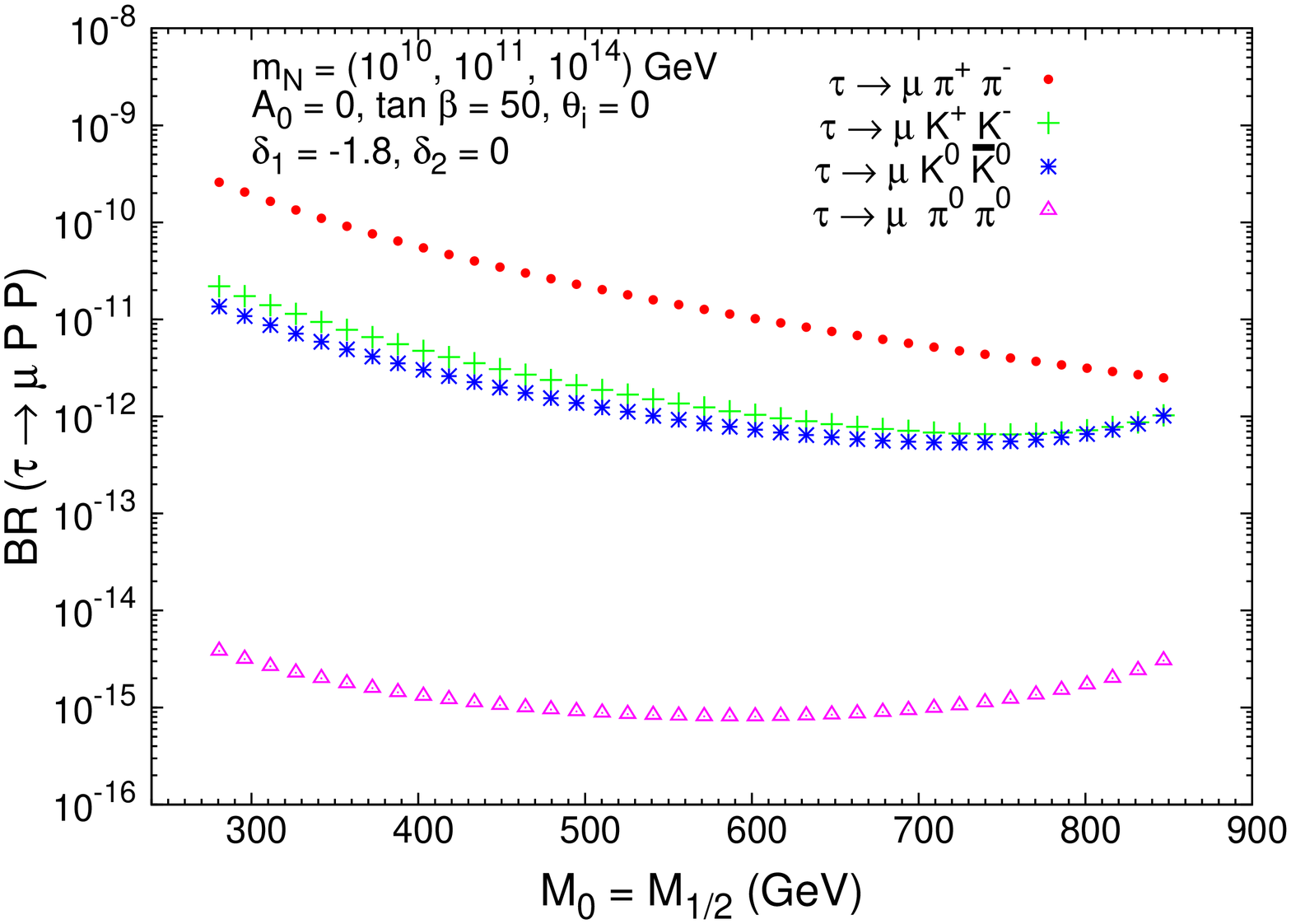,width=60mm,angle=270,clip=}   
     \end{tabular}
     \caption{BR($\tau \to \mu PP$) for $PP = \pi^+ \pi^-, K^+ K^-,
     K^0 \bar{K}^0, \pi^0 \pi^0$ as a function of $M_\text{SUSY} = M_0
     = M_{1/2}$ in the contrained MSSM-seesaw scenarios: CMSSM (left
     panel) and NUHM (right panel).
     }\label{fig:tau_muPP_R1} 
   \end{center}
 \end{figure}
 
In Fig.~\ref{fig:tau_muPP_R1} we display the prediction of BR$(\tau
\to \mu PP)$, with $PP =$ $\pi^+ \pi^-$, $K^+ K^-$, $K^0
\bar{K}^0$, $\pi^0 \pi^0$, as a function of $M_\text{SUSY}$ and
for the particular choice of $\theta_i = 0$ ($i = 1, 2, 3$). We
consider both CMSSM (left panel) and NUHM (right panel)
scenarios. We set here $\tan \beta=50$ and our ``reference'' values of 
$m_{N_{1,2,3}} =(10^{10},10^{11},10^{14})$ GeV.
For the NUHM case we set in addition $\delta_1 = -1.8$ and $\delta_2 =
0$, which have been shown in~\cite{Arganda:2007jw} to lead to low
Higgs boson mass values. Concretely, for $\theta_i = 0$ and 250 GeV $<
M_\text{SUSY} <$ 900 GeV the predicted masses are within the range 110
GeV $< m_{A^0}, m_{H^0} <$ 180 GeV, which are indeed very close to their
present experimental bounds.

 \begin{figure}[t!]
   \begin{center} 
     \begin{tabular}{cc} \hspace*{-12mm}
         \psfig{file=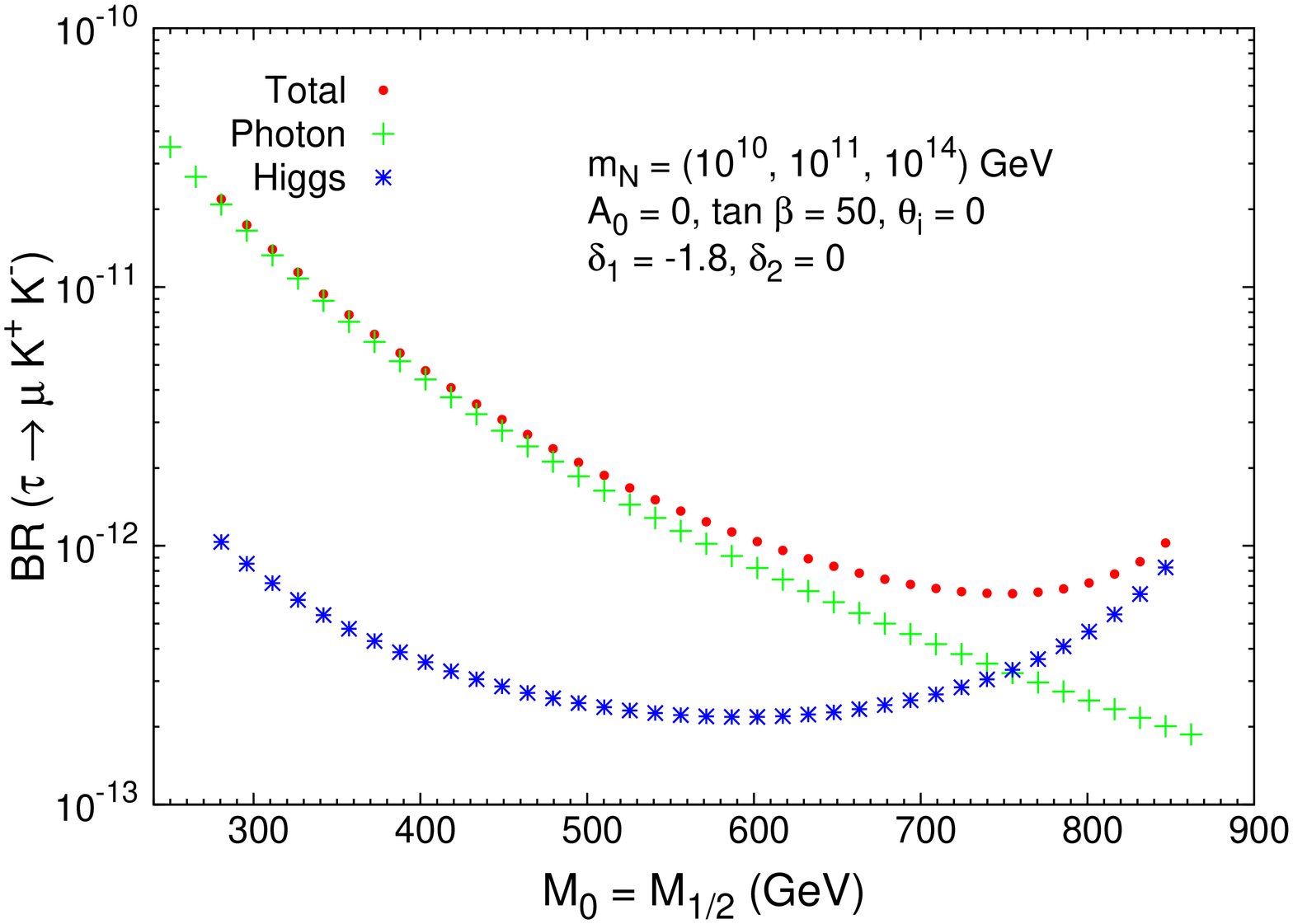,width=60mm,angle=270,clip=} 
 &
  	 \psfig{file=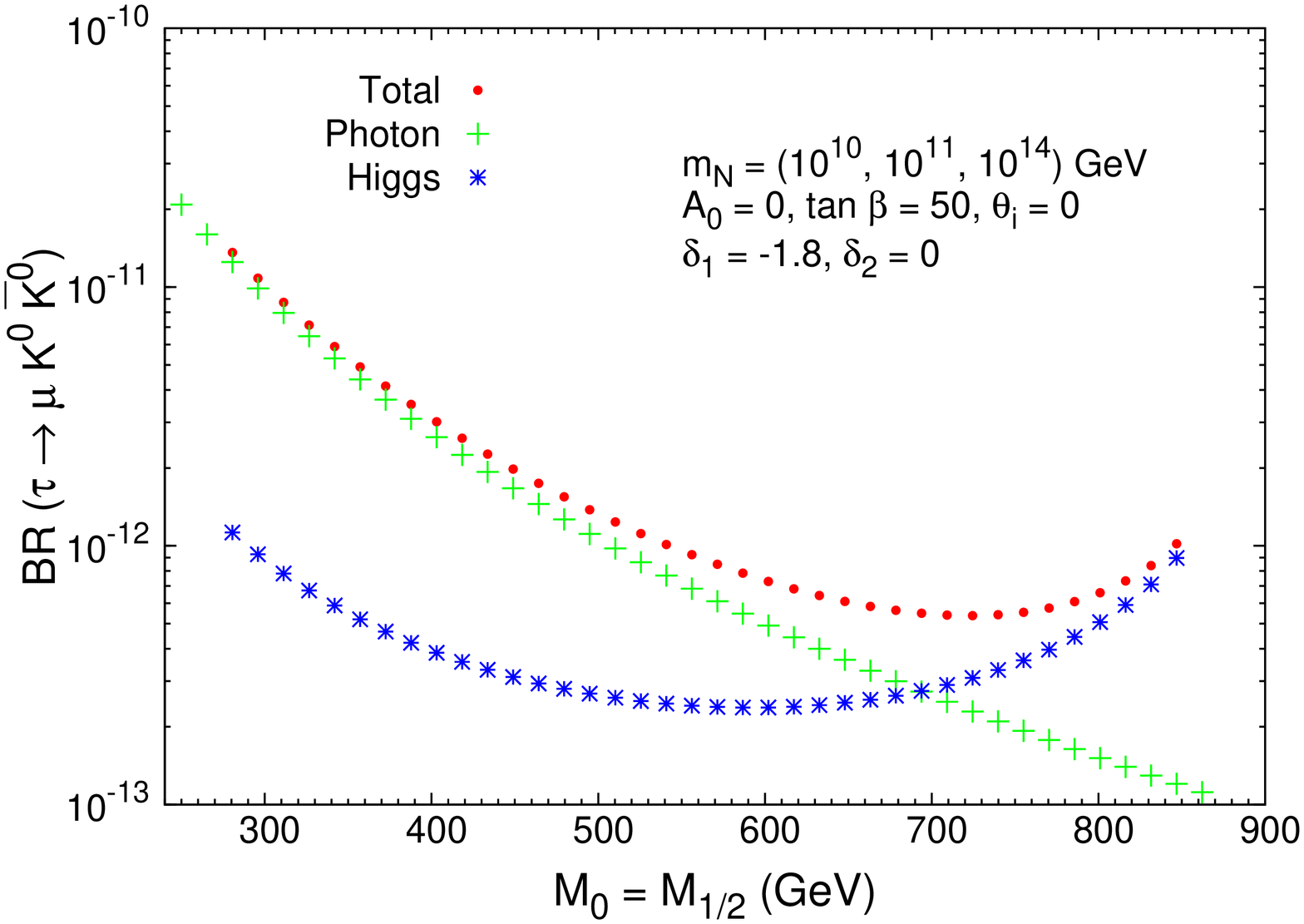,width=60mm,angle=270,clip=} \\
         \psfig{file=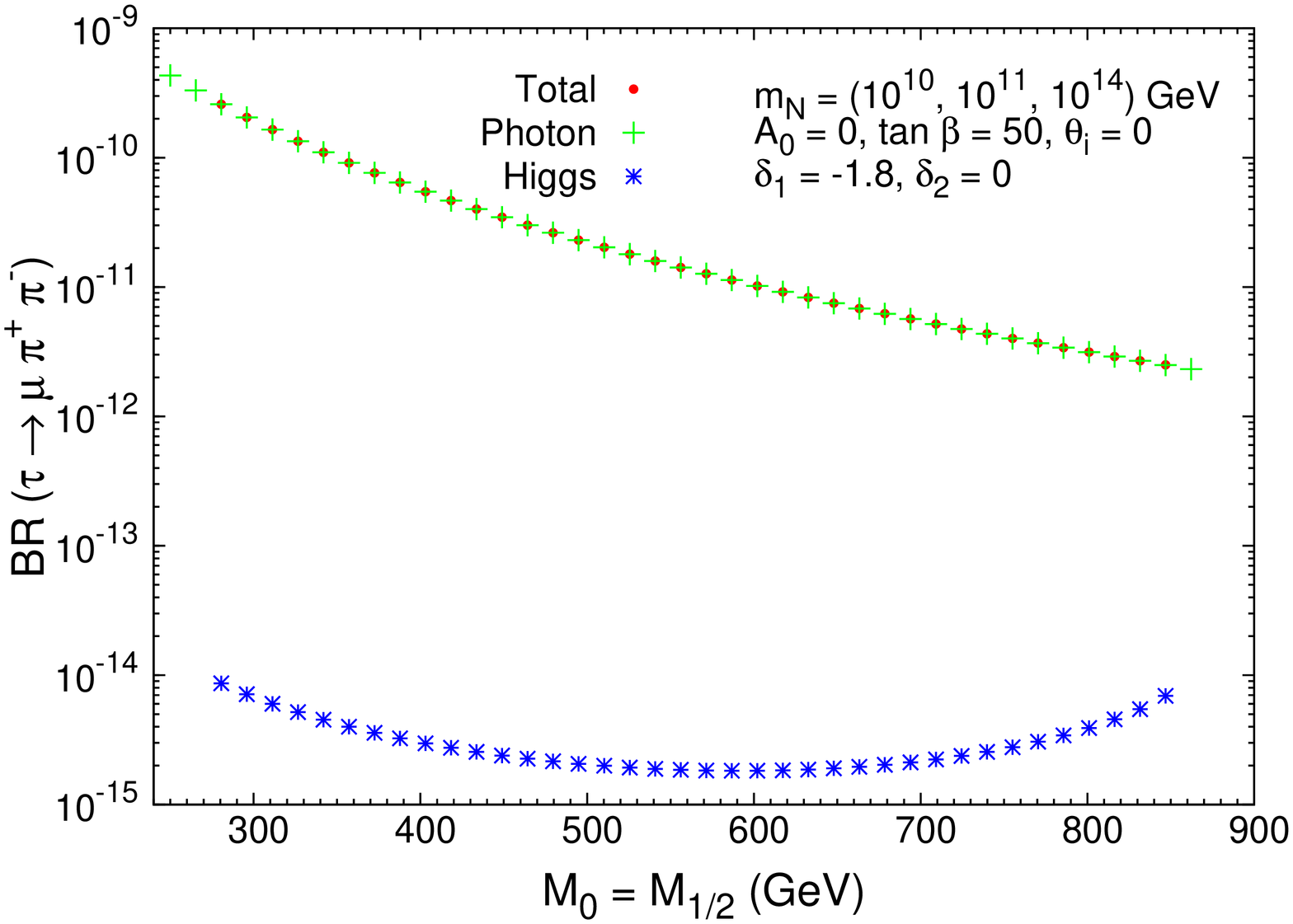,width=60mm,angle=270,clip=} 
 &
  	 \psfig{file=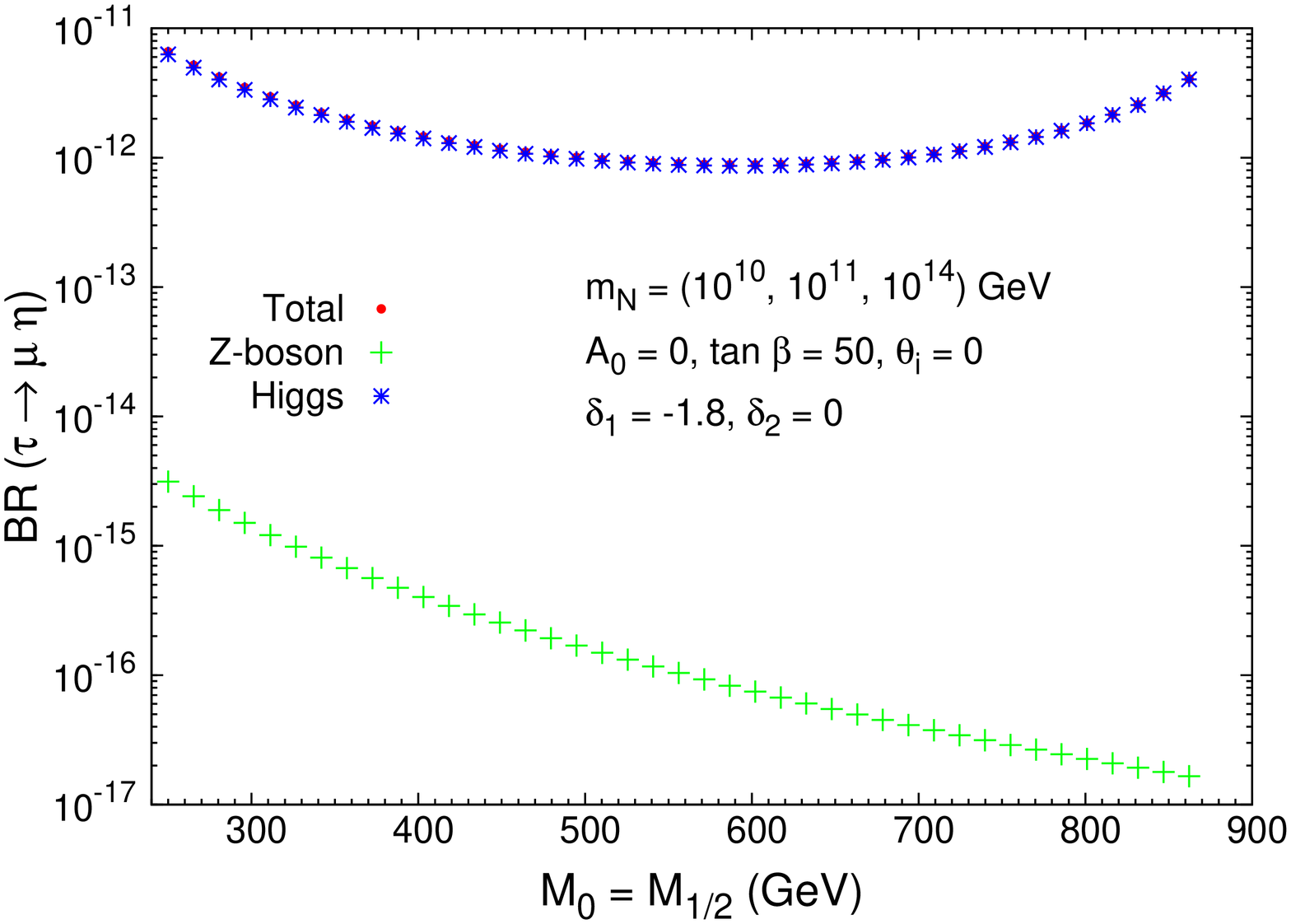,width=60mm,angle=270,clip=}
     \end{tabular}
     \caption{Rates of the various contributions to BR$(\tau
       \rightarrow \mu K^+ K^-)$ (upper left panel), BR$(\tau
       \rightarrow \mu K^0 \bar{K}^0)$ (upper right panel), BR$(\tau
       \rightarrow \mu \pi^+ \pi^-)$ (lower left panel) and BR$(\tau
       \rightarrow \mu \eta)$ (lower right panel) as a function of
       $M_\text{SUSY} = M_0 = M_{1/2}$ in the NUHM scenario.
     }\label{fig:tau_muhadrons_contributions}
   \end{center}
 \end{figure}

The first obvious conclusion from Fig.~\ref{fig:tau_muPP_R1} is that
the rates of the different channels exhibit the following hierarchy,
BR$(\tau \to \mu \pi^+ \pi^-)$ $>$ BR$(\tau \to \mu K^+ K^-)$
$\gtrsim$ BR$(\tau \to \mu K^0 \bar{K}^0)$ $\gg$ BR$(\tau \to \mu
\pi^0 \pi^0)$. This hierarchy can be understood in terms of the
dominant electromagnetic contribution and the relative phase space suppression.
We also see that the decoupling behaviour for large $M_\text{SUSY}$ is
clearly manifest in the universal case, where all the rates decrease
as $M_\text{SUSY}$ grows. In contrast, it turns out that, in the NUHM
case, the decoupling behaviour is only manifest
in the $\tau \to \mu \pi^+ \pi^-$ channel, whereas the
$\tau \to \mu K^+ K^-$, $\tau \to \mu K^0 \bar{K}^0$
and $\tau \to \mu \pi^0 \pi^0$ rates do not decrease with
$M_\text{SUSY}$ in the large $M_\text{SUSY}$ region.
This behaviour can be better comprehended by analysing separately the
different contributions to these channels, as shown in
Fig.~\ref{fig:tau_muhadrons_contributions}.

The results displayed in Fig~\ref{fig:tau_muhadrons_contributions}
for the $\tau \to \mu \pi^+ \pi^-$ channel show the dominance of the
photon-mediated contribution in this case, which is in fact
indistinguishable from the total rate in this plot,
for all the explored parameter values. The Higgs-mediated contribution
is subdominant by far due to the highly suppressed couplings of the
Higgs to the light $u$ and $d$ quarks, which after the hadronisation
of the corresponding quark bilinears  
result in $H \pi^+ \pi^-$ couplings proportional to $m_\pi^2$ (see
Eq.~(\ref{Jotas})). This plot also exhibits the non-decoupling behaviour of the
SUSY particles in the Higgs-mediated contribution. The particular
pattern of this contribution as a function of $M_\text{SUSY}$ is a
consequence of two facts. First, the well known constant behaviour with
$M_\text{SUSY}$ of
the LFV $H \tau \mu$ form factor at large $M_\text{SUSY}$. Second, the
encountered Higgs mass behaviour with $M_\text{SUSY}$, analysed
in~\cite{Arganda:2007jw}, which, for this choice of $\delta_{1,2}$ and
for the studied $M_\text{SUSY}$ interval, first
grows softly, reaches a maximum and then decreases softly.

The $\tau \to \mu \pi^0 \pi^0$ channel is only mediated by the
Higgs bosons and a similar suppression of $H \pi^0 \pi^0$ couplings as
in the $H \pi^+ \pi^-$ case occurs, leading to very low predicted
rates. These low rates and the non-decoupling behaviour of this
channel can be clearly
seen in Fig.~\ref{fig:tau_muPP_R1}.

The results for the $\tau \to \mu K^+ K^-$  channel that are depicted in
Fig~\ref{fig:tau_muhadrons_contributions} are interesting because the photon- and
the Higgs-mediated contributions compete in this decay. In fact the Higgs-mediated contribution can
equalise, or even exceed that of the photon,
dominating the total rate in the large $M_\text{SUSY}$
region. Both photon- and Higgs-mediated contributions are similar around
$M_\text{SUSY}=750$ GeV. The reason for this larger Higgs
contributions than in the previously studied $\pi \pi$ case is because
of the larger Higgs couplings to the strange quarks which result in $H
K K$ couplings proportional to $m_K^2$ (see Eq.~(\ref{Jotas})).

The results for the $\tau \to \mu K^0 \bar{K}^0$ channel in Fig~\ref{fig:tau_muhadrons_contributions}
are very similar to those for $\tau \to \mu K^+ K^-$. One difference is the
point where the Higgs-mediated contribution crosses the photon one, which for 
$\tau \to \mu K^0 \bar{K}^0$ is around
$M_\text{SUSY}=700$ GeV. Another interesting difference is that this
rate is always slightly smaller than $\tau \to \mu K^+ K^-$ due to the fact that the
photon-mediated contribution to $\tau \to \mu K^0 \bar{K}^0$
occurs just by the meson resonances, whereas the $\tau \to \mu K^+
K^-$ channel can also be mediated via pure electromagnetic
interaction. This difference is clearly summarised in the several
contributions to the  $F_V^{K^+ K^-}$ and $F_V^{K^0 \bar{K}^0}$
form factors in Eq.~(\ref{eq:pp}) of Appendix~\ref{ap:2}.

 \begin{figure}[t!]
   \begin{center} 
     \begin{tabular}{cc} \hspace*{-12mm}
         \psfig{file=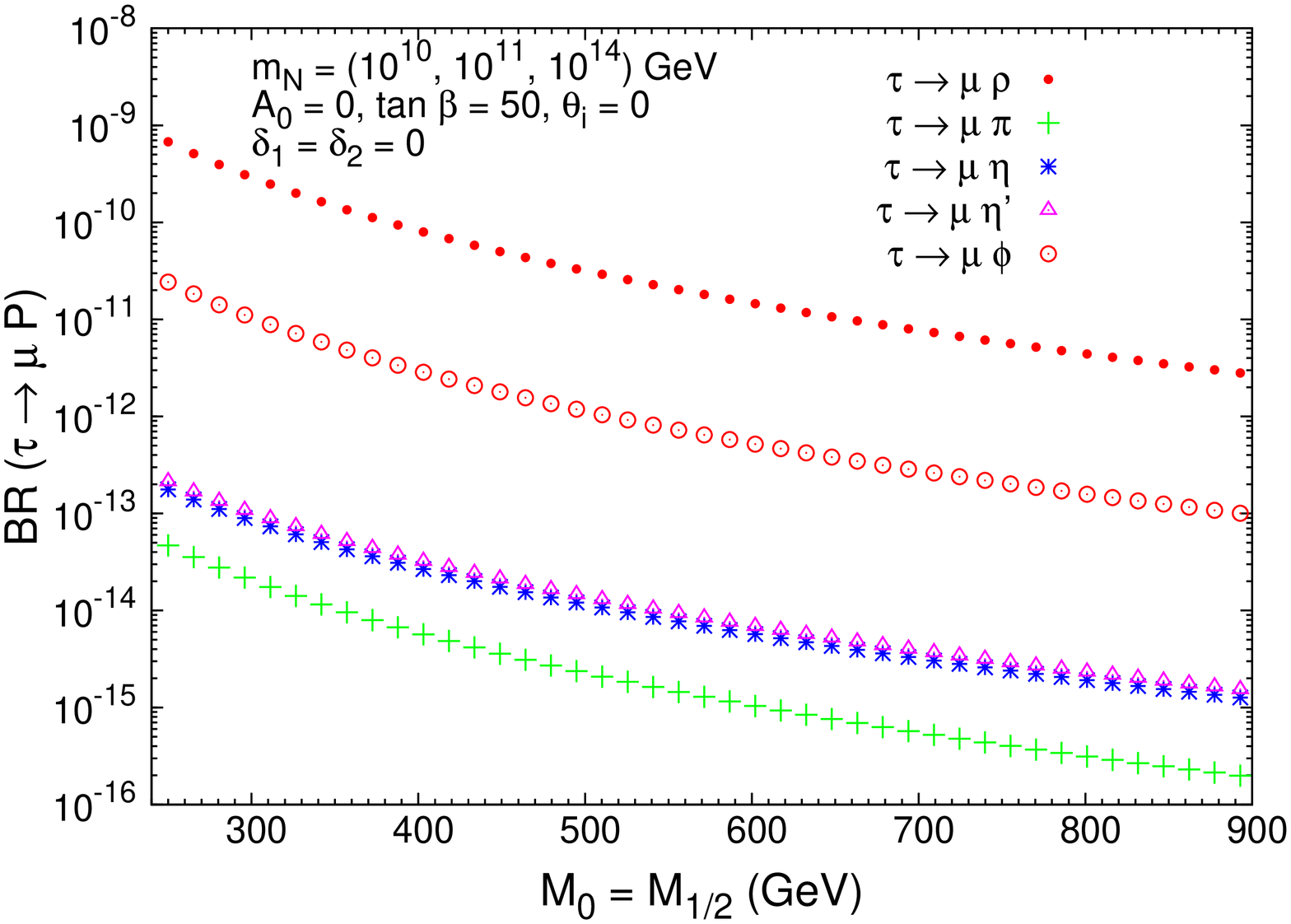,width=60mm,angle=270,clip=} 
 &
  	 \psfig{file=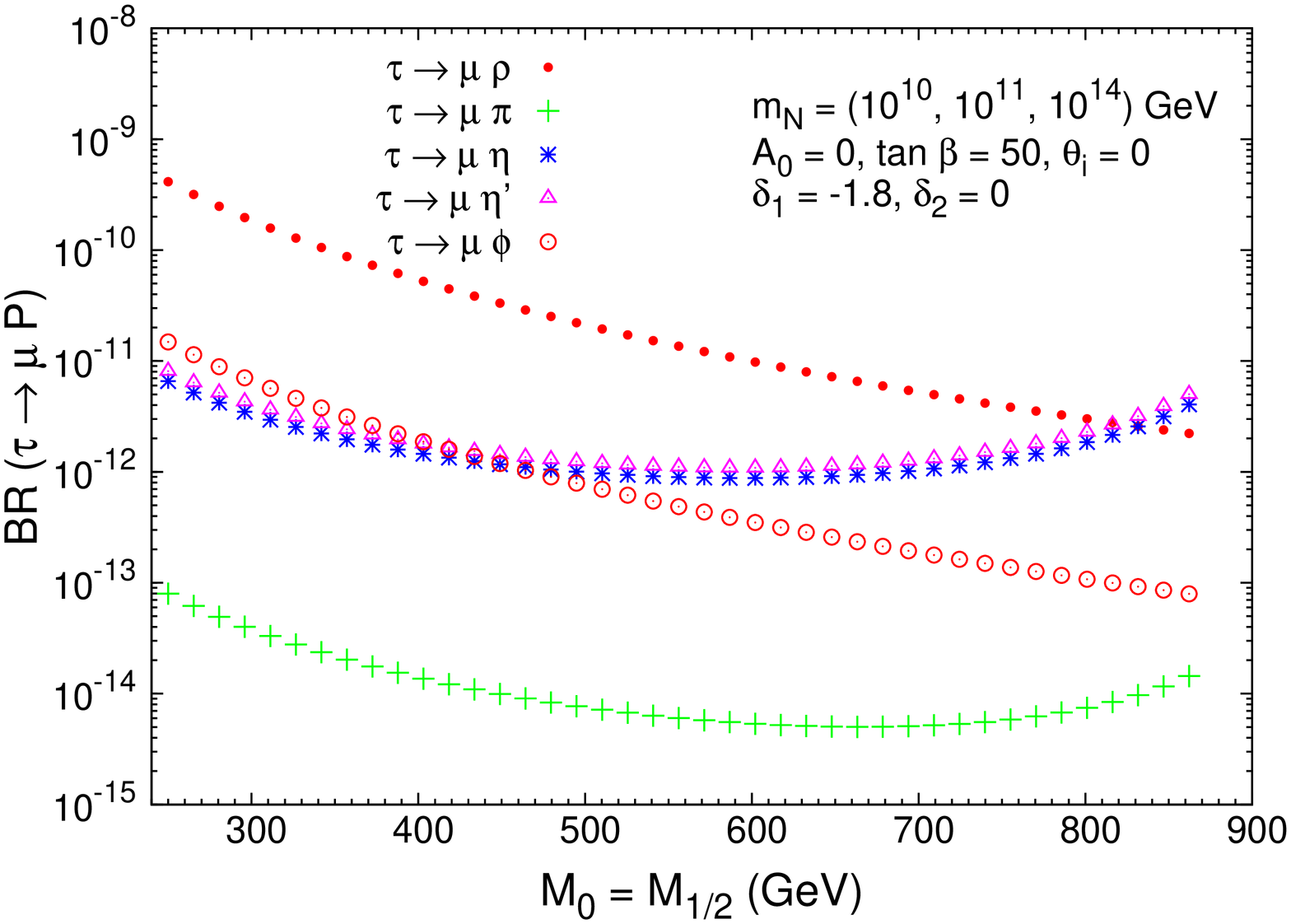,width=60mm,angle=270,clip=}   
     \end{tabular}
     \caption{BR($\tau \to \mu P$) for $P = \pi, \eta, \eta^\prime, \rho, \phi$ as a function of $M_\text{SUSY} = M_0
     = M_{1/2}$ in the contrained MSSM-seesaw scenarios: CMSSM (left
     panel) and NUHM (right panel).
     }\label{fig:tau_muP_R1} 
   \end{center}
 \end{figure}

The predictions of BR$(\tau \to \mu P)$, with $P$ being here a pseudoscalar meson $\pi, \eta,
\eta^\prime$ or a vector resonance $\rho, \phi$, as a function of $M_\text{SUSY}$ are displayed in
Fig.~\ref{fig:tau_muP_R1}. We also consider CMSSM (left panel) and
NUHM (right panel) scenarios. In the universal case we find the
following hierarchy, BR$(\tau \to \mu \rho)$ $>$ BR$(\tau \to \mu \phi)$ $>$
BR$(\tau \to \mu \eta^\prime)$
$\gtrsim$ BR$(\tau \to \mu \eta)$ $>$ BR$(\tau \to \mu \pi)$.
We obtain again the expected decoupling behaviour for large
$M_\text{SUSY}$ in this universal scenario, while in the NUHM scenario
the non-decoupling behaviour is clearly manifest for
$\tau \to \mu \eta$, $\tau \to \mu \eta^\prime$ and $\tau \to \mu\pi$.
The $\tau \to \mu \rho$ rates in the NUHM scenario are the largest
ones, except in the large $M_\text{SUSY}$ region, where $\tau \to \mu
\eta$ and $\tau \to \mu \eta^\prime$ rates exceed them. These two
channels are by far dominated by the Higgs-mediated contributions in
the full $M_\text{SUSY}$ explored interval, as can be seen for the
$\eta$ case in Fig~\ref{fig:tau_muhadrons_contributions}. The reason
for this Higgs dominance is because of the large Higgs couplings to
the strange components of the $\eta$ and $\eta^\prime$ mesons, which
result in large $A^0-\eta$ and $A^0-\eta^\prime$ ``mixings'' proportional
to $m_K^2$ as explicitely given in Eq.~(\ref{Bs}).

One of the most important outcomes from the previous analysis,
corresponding to the $\theta_i = 0$ choice,
is that for both scenarios and for the chosen input parameters,
the predicted rates for both $\tau \to \mu PP$ and $\tau \to \mu P$
channels do no reach their corresponding
experimental bounds, and even in the best cases of
$\tau \to \mu \pi^+ \pi^-$ and $\tau \to \mu \rho$ they are still two
orders of magnitude below their present
experimental sensitivities. In the following, we will therefore focus
on larger values of $\theta_2$.

 \begin{figure}[t!]
   \begin{center} 
     \begin{tabular}{cc} \hspace*{-12mm}
         \psfig{file=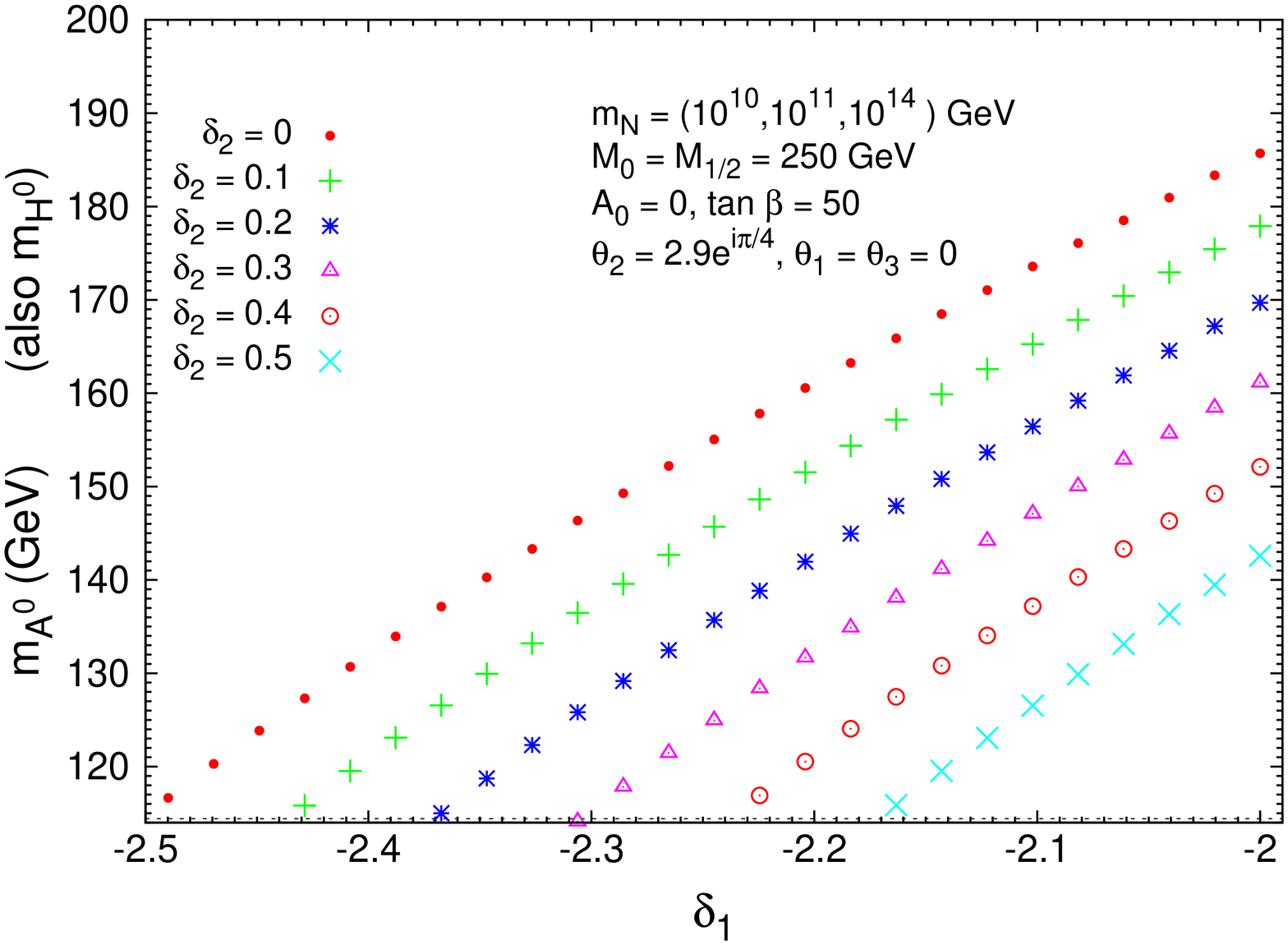,width=60mm,angle=270,clip=} 
 &
  	 \psfig{file=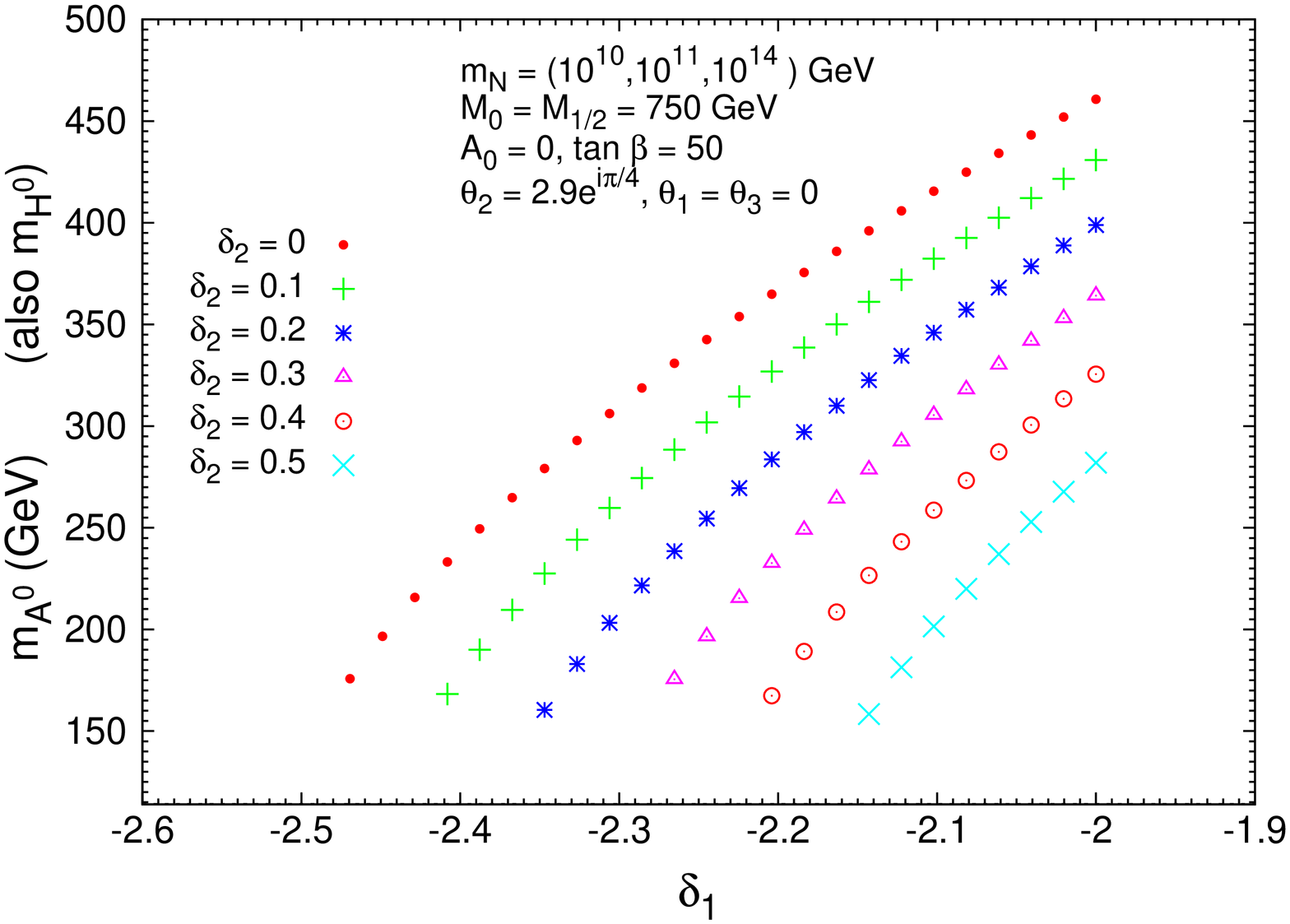,width=60mm,angle=270,clip=}
     \end{tabular}
     \caption{Light $m_{A^0}$ predictions as a function of
     non-universal parameters $\delta_1$
     and $\delta_2$ in the NUHM scenario. The predictions for $m_{H^0}$ are
     indistinguisable from those of $m_{A^0}$ in this figure.
     }\label{fig:mH_delta1} 
   \end{center}
 \end{figure}

In order to reach the larger rates as possible in the $\theta_i \neq
0$ case, one needs to explore first the optimal values of $\delta_1$
and $\delta_2$ which lead to light Higgs bosons. We summarise the
predictions for the relevant Higgs boson mass, $m_{A^0}$ (and
$m_{H^0}$), as a function 
of $\delta_1$ and $\delta_2$ in 
Fig.~\ref{fig:mH_delta1} for the extreme value of $\theta_2 = 2.9
e^{i\pi/4}$. The reason for this particular choice of $\theta_2$ is
due to the fact that it leads to the maximum value of $|\delta_{32}|$
which is compatible with our hypothesis of perturbativity, as shown in
Fig.~\ref{fig:theta2} and discussed in Section~\ref{th_framework}.    
We have chosen here 
$\tan \beta=50$ and two representative values of
$M_{\rm SUSY}=$ 250 and 750 GeV for moderate and heavy SUSY spectra,
respectively.  The other parameters are set to the values of
$m_{N_i} = (10^{10},10^{11},10^{14})$ GeV, 
$\theta_1=\theta_3=0$, $A_0=0$
and ${\rm sign} (\mu)=+1$. To ensure that our
results are indeed experimentally viable, 
we have included in this, and in the following figures, 
only the solutions where the three neutral 
Higgs boson masses are above the experimental bound for the lightest MSSM
Higgs boson, which at present is 110 GeV for
$\tan{\beta} > 5$ ($99.7\%$ C.L.)~\cite{Yao:2006px}.
The most interesting solutions with important phenomenological
implications are found for negative $\delta_1$
within the range $(-3, -2)$ and very small and positive $\delta_2$,
the choices selected for Fig.~\ref{fig:mH_delta1}.
In this figure, 
for all the explored values of $\delta_1$ and $\delta_2$,
we find a value of $m_{A^0}$ that is significantly 
smaller than what one would encounter in the universal
case (here represented by the choice $\delta_1=\delta_2=0$). 
This is truly remarkable in the case of large
soft breaking masses, as can be seen, for instance,
in the panel with $M_{\rm SUSY}=750$ GeV,
where low values of $m_{A^0}\sim 150$ GeV are still found.

 \begin{figure}[t!]
   \begin{center} 
     \begin{tabular}{cc} \hspace*{-12mm}
  	 \psfig{file=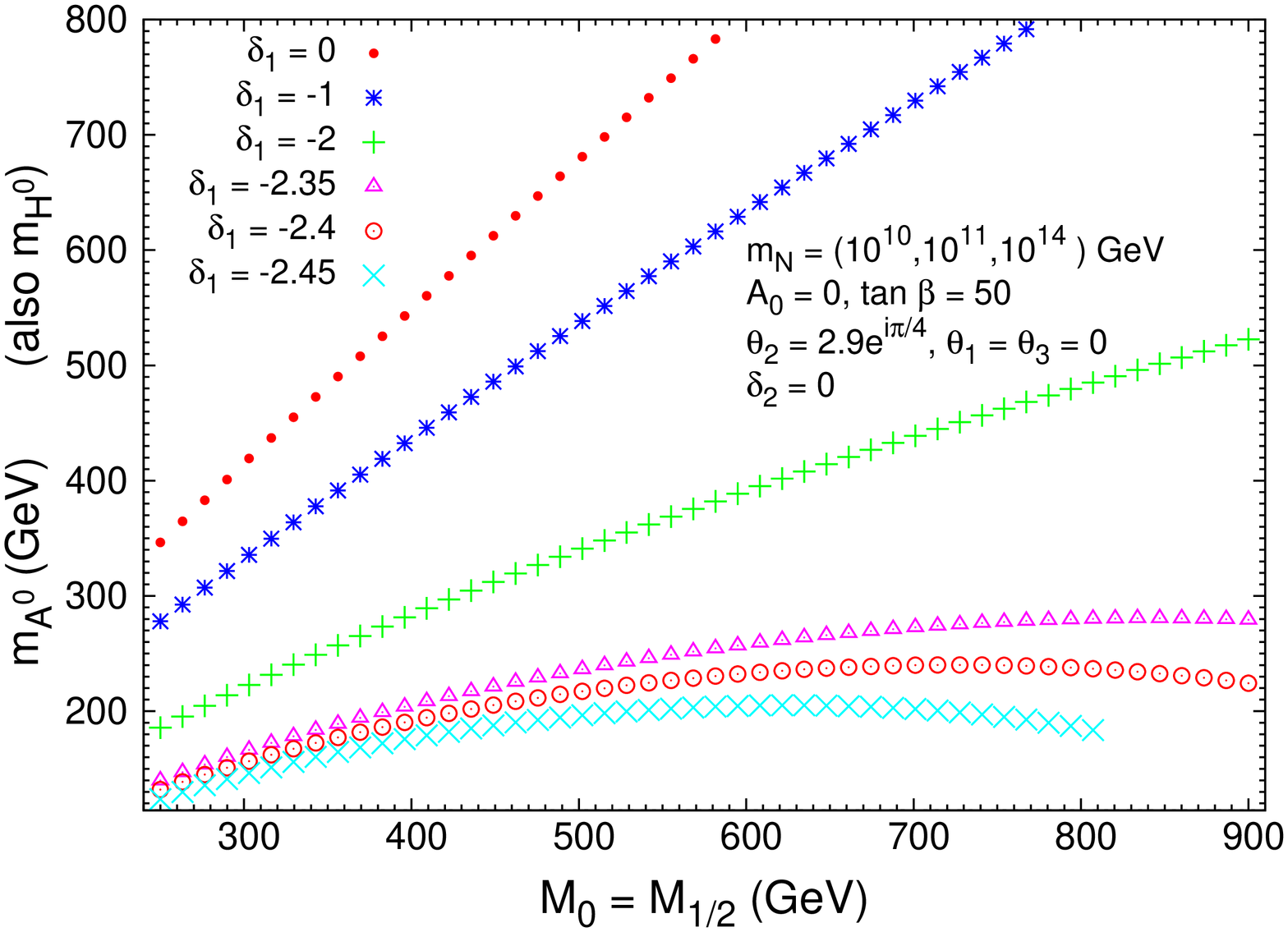,width=60mm,angle=270,clip=}
  &
  	 \psfig{file=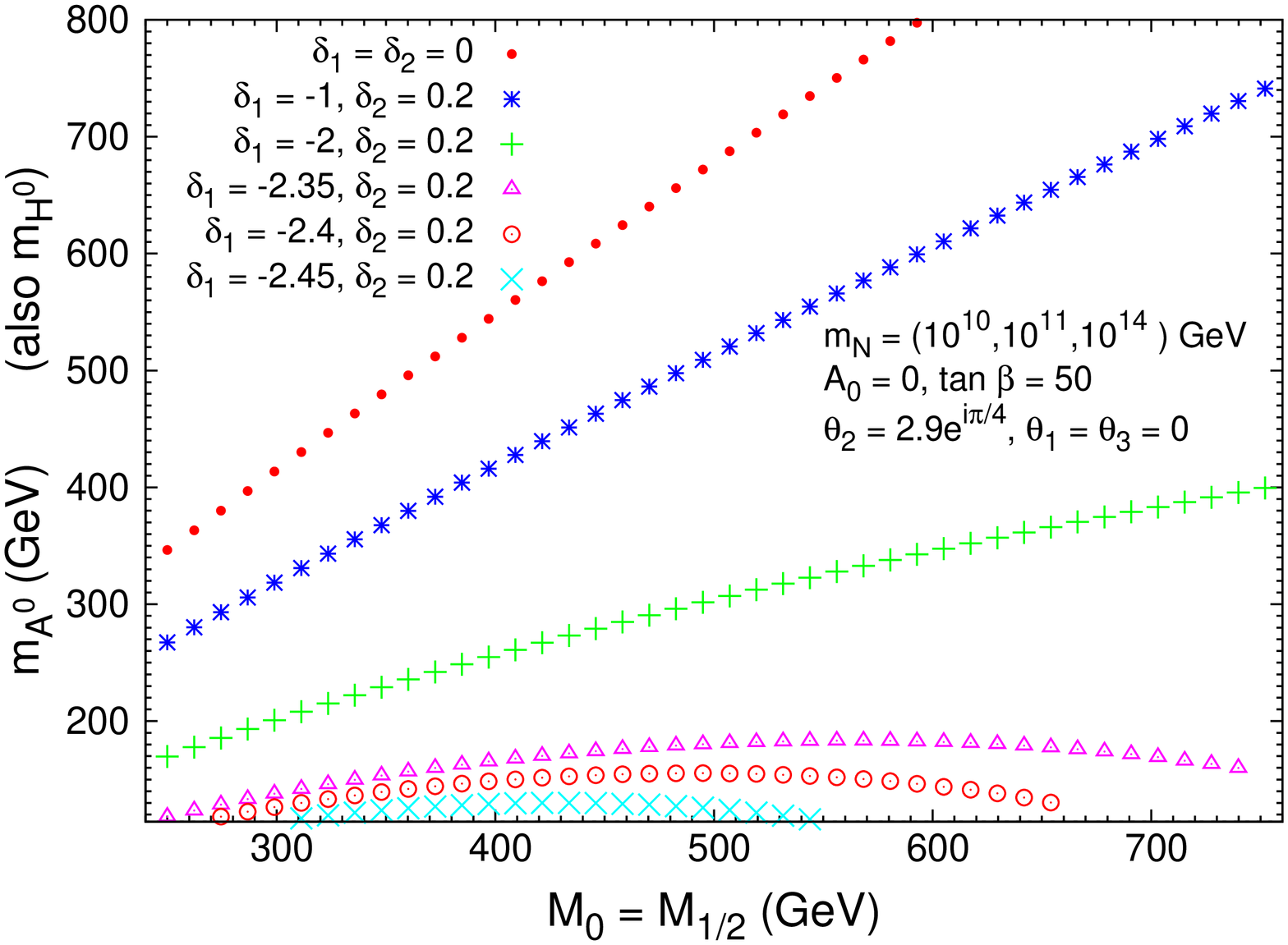,width=60mm,angle=270,clip=}
     \end{tabular}
     \caption{Light $m_{A^0}$ predictions as a function of
     $M_{\text{SUSY}} = M_0 = M_{1/2}$ in the NUHM scenario. The predictions for $m_{H^0}$ are
     indistinguisable from those of $m_{A^0}$ in this figure. The
     predictions for the CMSSM scenario ($\delta_1 = \delta_2 = 0$) are also included for
     comparison.  
     }\label{fig:mH_mSUSY} 
   \end{center}
 \end{figure}

The behaviour of the predicted $m_{A^0}$ as a function of 
$M_\text{SUSY}$ is depicted in
Fig.~\ref{fig:mH_mSUSY}. Here the
specific values of $\delta_1=\{-2.45,\,-2.4,\,-2.35,-2,\,-1,\,0\}$ 
and $\delta_2=0,\, 0.2$ have been considered. 
This figure illustrates again the interesting departure in NUHM scenarios from the linear
behaviour of $m_{A^0}$ with $M_\text{SUSY}$, which is 
generic in the universal case ($\delta_1=\delta_2=0$). The same
pattern with $M_\text{SUSY}$ was also found for the $\theta_i = 0$
case in~\cite{Arganda:2007jw}, but obviously for different choices of
$\delta_1$ and $\delta_2$.

 \begin{figure}[t!]
   \begin{center} 
     \begin{tabular}{cc} \hspace*{-12mm}
         \psfig{file=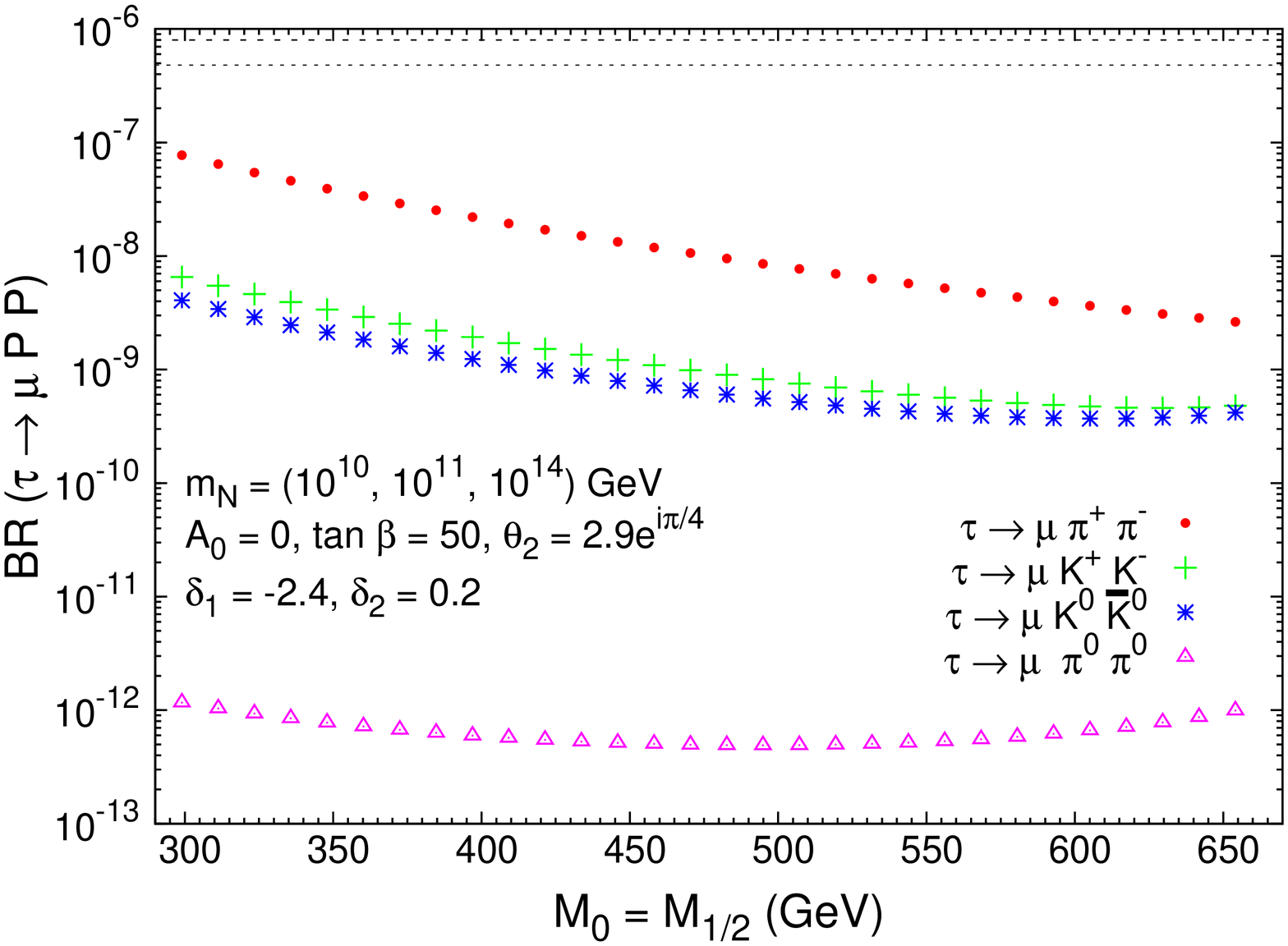,width=60mm,angle=270,clip=} 
 &
  	\psfig{file=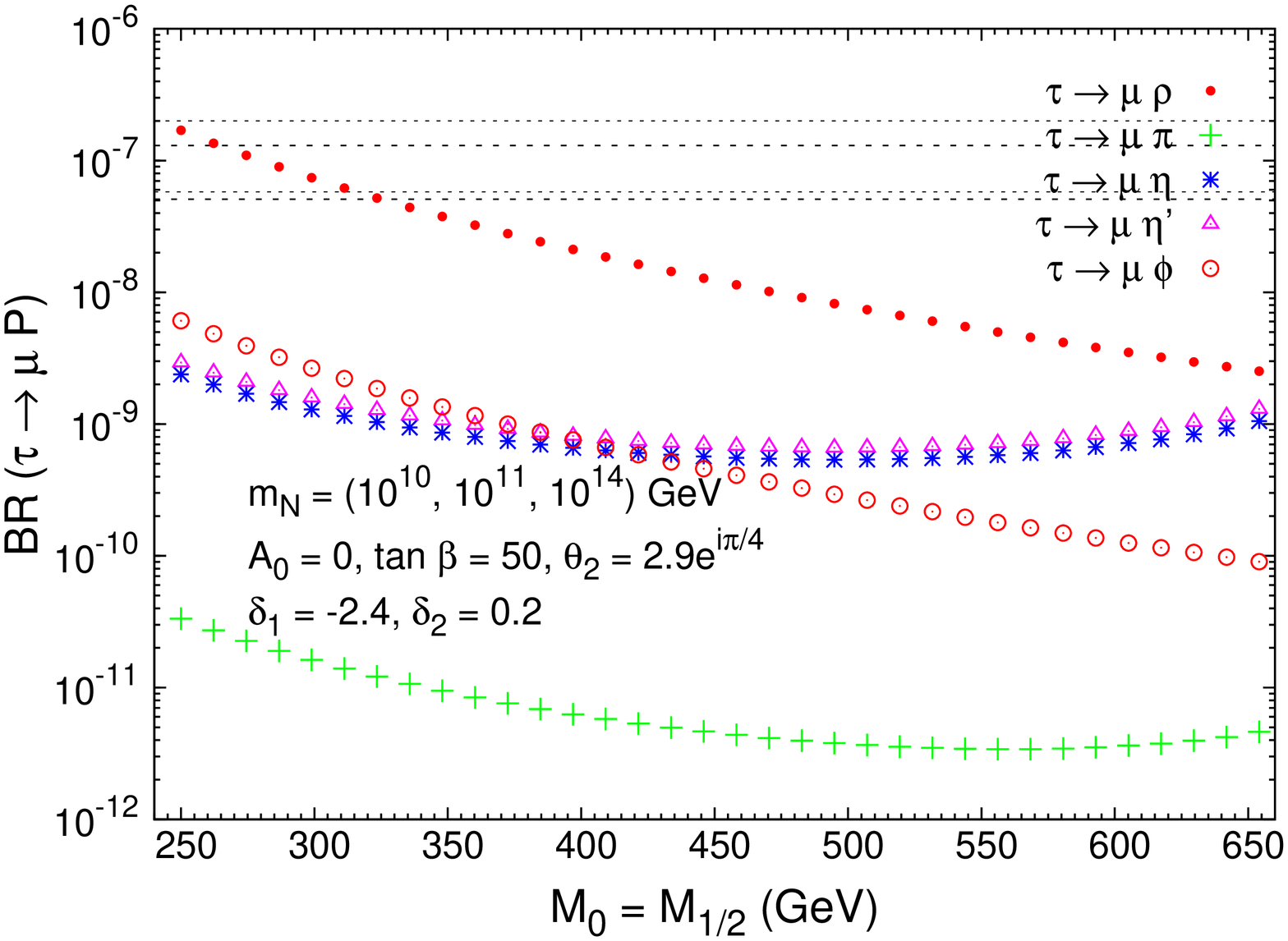,width=60mm,angle=270,clip=}   
     \end{tabular}
     \caption{Predictions of BR($\tau \to \mu PP$) and BR($\tau \to
     \mu P$) as a function of $M_{\text{SUSY}}$ in the NUHM scenario
     for a large $\tau - \mu$ mixing driven by $\theta_2 = 2.9 e^{i\pi/4}$.
     The horizontal lines are the present experimental bounds given in Table~\ref{LFVsemilep:bounds}.
     }\label{fig:tau_muhadrons_2.9eiPi4} 
   \end{center}
 \end{figure}

The corresponding predictions for $\theta_2 = 2.9 e^{i\pi/4}$ of the
nine LFV semileptonic $\tau$ decays studied in this work as a
function of $M_\text{SUSY}$ are shown in
Fig.~\ref{fig:tau_muhadrons_2.9eiPi4}. In this case,
we work with $\delta_1 = -2.4$ and $\delta_2 = 0.2$,
that drive us to Higgs boson masses around 150 GeV
even for heavy SUSY spectra, as can be seen in
Figs.~\ref{fig:mH_delta1} and~\ref{fig:mH_mSUSY}.
In this Fig.~\ref{fig:tau_muhadrons_2.9eiPi4} we can see that,
compared to predictions in Figs.~\ref{fig:tau_muPP_R1}
and~\ref{fig:tau_muP_R1}, the new choice of $\theta_2$
increase all the rates about two orders of magnitude.
All the rates exhibit the same hierarchy as
in the previous plots, being BR$(\tau \to \mu \pi^+ \pi^-)$
and BR$(\tau \to \mu \rho)$ the largest ones. Indeed, the predictions
of these two latter channels reach their present experimental
sensitivities at the low $M_\text{SUSY}$ region, below 200 GeV and 250 GeV
respectively, for this
particular choice of input parameters.

\subsection{Comparison between the full and approximate results}
It is interesting and useful to provide simple formulas that can approximate 
reasonably well our full predictions. The most popular approximation when
predicting LFV rates is to work with expressions that are valid only in the
large $\tan \beta$ region. The justification for this is obvious since all
these LFV rates
are known to grow with $\tan \beta$. It is specially important
in scenarios where the LFV rates are dominated by the Higgs mediated diagrams,
since these latter grow much faster with $\tan \beta$ than the photon or Z boson
mediated ones. Accordingly, we will pay more attention to the semileptonic 
$\tau \to \mu P$ and $\tau \to \mu PP$ channels that can be dominated by the Higgs bosons and whose present 
experimental sensitivities are the best ones. This leads us mainly to the 
$\tau \to \mu \eta$ and $\tau \to \mu K^+ K^-$ channels.
The other approximation which is used frequently in the
literature, due to its simplicity, is the mass insertion approximation, where 
the tau-muon
LFV is encoded in the dimesionless parameter $\delta_{32}$, already introduced
in Sec.\ref{th_framework}, and whose expression in the LLog
approximation is given in Eqs.~(\ref{delta23}) and (\ref{Y32:LLog}).

We start by considering the large $\tan\beta$ limit of 
the tau-muon-Higgs form factors that are the relevant ones for
the LFV Higgs-mediated processes. The full one-loop Higgs form factors were
 computed in~\cite{Arganda:2004bz} (see also~\cite{Arganda:2005ji}) and are collected in
Appendix A.2. At large $\tan\beta$, $H_L$ dominates $H_R$ by about a factor 
$m_\tau /m_\mu$. Moreover,
 $H_L^{A^0}$ and $H_L^{H^0}$ are by far the largest form
factors in this limit, and one can safely neglect $H_L^{h^0}$. 
More specifically, by using the mass insertion approximation, 
their chargino and neutralino contributions in the large $\tan\beta$ limit 
give, correspondingly, the following results \cite{Arganda:2004bz}, 
\begin{eqnarray}\label{HL}
H_{L,c}^{(A^0)}\,&= & \,iH_{L,c}^{(H^0)}\,=\, i \,\frac{g^3}{16\pi^2} \,
\frac{m_\tau}{12 m_W} \, \delta_{32} \, \tan^2{\beta} \, , \\[2mm]
H_{L,n}^{(A^0)}\,&= & \,iH_{L,n}^{(H^0)}\,=\, i \,\frac{g^3}{16\pi^2} \,
\frac{m_\tau}{24 m_W} \, (1-3 \tan^2{\theta_W}) \,\delta_{32} \, \tan^2{\beta}  \, .
\end{eqnarray}
One can further verify that $H_c$ dominates $H_n$ by about a factor 20, so
that in the following we will take $H_L \simeq H_{L,c}$.

On the other hand, we also consider the
large $\tan\beta$ limit of the functions that define the 
Higgs couplings to one meson, $B(P)$ in Eq. (\ref{Bs}), and to 
two mesons, $J(PP)$ in Eq. (\ref{Jotas}). It 
leads to the following results:
\begin{eqnarray}\label{Bs_Js}
B_L^{(A^0)}(\eta) &=& -B_R^{(A^0)}(\eta) \, = \, -i\frac{1}{4\sqrt{3}} \tan\beta 
\left[ (3 m_\pi^2 - 4 m_K^2) \cos\theta - 2 \sqrt{2} m_K^2 \sin\theta \right] \, , 
\nonumber \\
B_L^{(A^0)}(\eta') &=& -B_R^{(A^0)}(\eta') \, = \, -i\frac{1}{4\sqrt{3}} \tan\beta 
\left[ (3 m_\pi^2 - 4 m_K^2) \sin\theta + 2 \sqrt{2} m_K^2 \cos\theta \right] \, , 
\nonumber \\
B_L^{(A^0)}(\pi) &=&-B_R^{(A^0)}(\pi)\, = \, i\frac{1}{4} \tan\beta \, m_\pi^2 \, , 
\nonumber \\
J_L^{(H^0)}(K^+K^-)&=&J_R^{(H^0)}(K^+K^-) \, = \, -\frac{1}{4}\tan\beta \,(2m_K^2-m_\pi^2) \, , 
\nonumber \\ 
J_L^{(H^0)}(K^0{\bar K}^0)&=&J_R^{(H^0)}(K^0{\bar K}^0)\, = \,
-\frac{1}{2}\tan\beta \, m_\pi^2 \,, 
\nonumber \\  
J_L^{(H^0)}(\pi^+\pi^- )&=&J_R^{(H^0)}(\pi^+\pi^-) \,= \, J_L^{(H^0)}(\pi^0\pi^0) \,= \, J_R^{(H^0)}(\pi^0\pi^0) \, = \, -\frac{1}{4}\tan\beta \, m_\pi^2\, .
\end{eqnarray}
By using the above sequence of approximations and by neglecting the
muon mass, we finally get the following
simple results,  
\begin{eqnarray}
\text{BR}(\tau \to \mu \eta)_{H_\text{approx}}  &=& \frac{1}{8 \pi m_\tau^3} \left( m_\tau^2 -
m_\eta^2 \right)^2 \left| \frac{g}{2 m_W} \, \frac{F}{m_{A^0}^2} \, 
B_L^{(A^0)}(\eta) \, H_{L,c}^{(A^0)} \right|^2
\frac{1}{\Gamma_\tau } \nonumber \\[2mm]
&=& 1.2 \times 10^{-7} \left| \delta_{32} \right|^2 \left( \frac{100}
{m_{A^0}({\rm GeV})}
\right)^4 \left( \frac{\tan \beta}{60} \right)^6,
\label{taumueta_approx}
\end{eqnarray}
and
\begin{eqnarray}\label{taumuKK_Happrox}
\text{BR}(\tau \to \mu K^+K^-)_{H_\text{approx}} &=& 
\frac{1}{128m_\tau\pi^3} 
\left|\frac{g}{2m_W}\frac{1}{m_{H^0}^2}J_L^{(H^0)}(K^+K^-)H_{L,c}^{(H^0)}
\right|^2
\frac{1}{\Gamma_\tau} \nonumber \\[2mm]
&\times& \int^{s_{max}}_{s_{min}} 
ds\, (t_{max}-t_{min})(1-\frac{s}{m_\tau^2}) \nonumber \\[2mm]
&=& 2.8 \times 10^{-8}\left| \delta_{32} \right|^2 \left( \frac{100}
{m_{H^0}({\rm GeV})} \right)^4 \left(
\frac{\tan \beta}{60} \right)^6 \, ,
\end{eqnarray}
where $s_{max}$, $s_{min}$, $t_{max}$ and $t_{min}$ are given in Eq.~(\ref{lambda}).
The results for the other channels can be similarly obtained by using the
corresponding $B(P)$ or $J(PP)$ functions and the corresponding meson masses
(with
an additional 1/2 factor in the case of $BR(\tau \to \mu \pi^0 \pi^0)$ to account for
identical final state particles). We get,
\begin{eqnarray}
\text{BR}(\tau \to \mu \eta')_{H_\text{approx}}  &=& 1.5 \times 10^{-7} \left| \delta_{32} \right|^2 \left( \frac{100}
{m_{A^0}({\rm GeV})}
\right)^4 \left( \frac{\tan \beta}{60} \right)^6,\\
\text{BR}(\tau \to \mu \pi)_{H_\text{approx}}  &=& 3.6\times 10^{-10} \left| \delta_{32} \right|^2 \left( \frac{100}
{m_{A^0}({\rm GeV})}
\right)^4 \left( \frac{\tan \beta}{60} \right)^6,\\ 
\text{BR}(\tau \to \mu K^0{\bar K}^0)_{H_\text{approx}}  &=& 
3.0 \times 10^{-8}\left| \delta_{32} \right|^2 \left( \frac{100}
{m_{H^0}({\rm GeV})} \right)^4 \left(
\frac{\tan \beta}{60} \right)^6,\\
\text{BR}(\tau \to \mu \pi^+ \pi^-)_{H_\text{approx}}  &=& 
2.6 \times 10^{-10}\left| \delta_{32} \right|^2 \left( \frac{100}
{m_{H^0}({\rm GeV})} \right)^4 \left(
\frac{\tan \beta}{60} \right)^6\\
\text{BR}(\tau \to \mu \pi^0 \pi^0)_{H_\text{approx}}  &=& 
1.3 \times 10^{-10}\left| \delta_{32} \right|^2 \left( \frac{100}
{m_{H^0}({\rm GeV})} \right)^4 \left(
\frac{\tan \beta}{60} \right)^6.
\label{otherchannels_approx}
\end{eqnarray} 
In all the above approximate results of the LFV semileptonic tau decay rates we see 
explicitely the strong dependence 
with both $\tan\beta$ and the corresponding Higgs boson mass, being $(\tan\beta)^6$ 
and  $(1/m_H)^4$, respectively, which are characteristic of the Higgs mediated
processes. 

 \begin{figure}[t!]
   \begin{center} 
     \begin{tabular}{cc} \hspace*{-12mm}
     \psfig{file=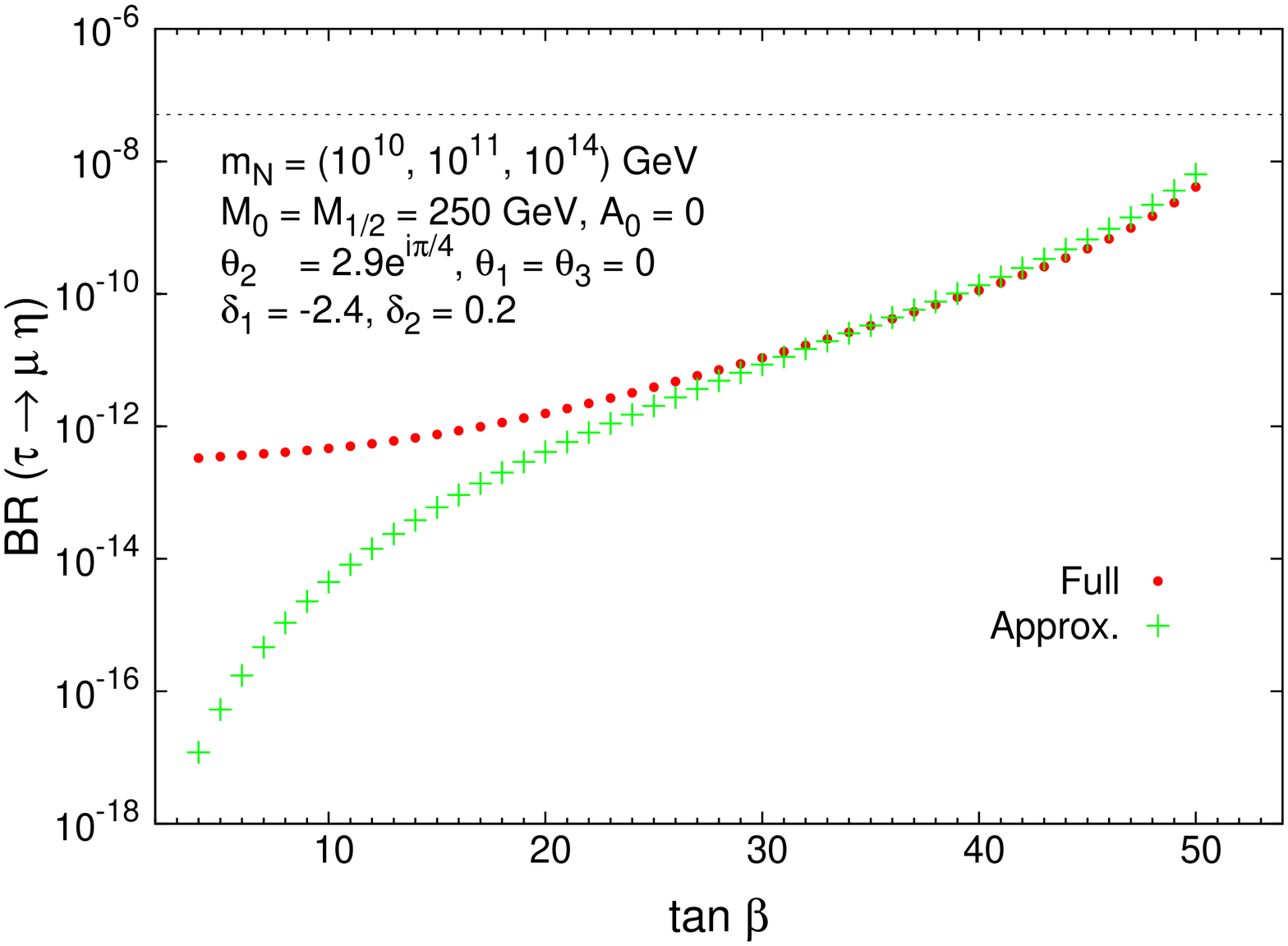,width=60mm,angle=270,clip=}
  &
     \psfig{file=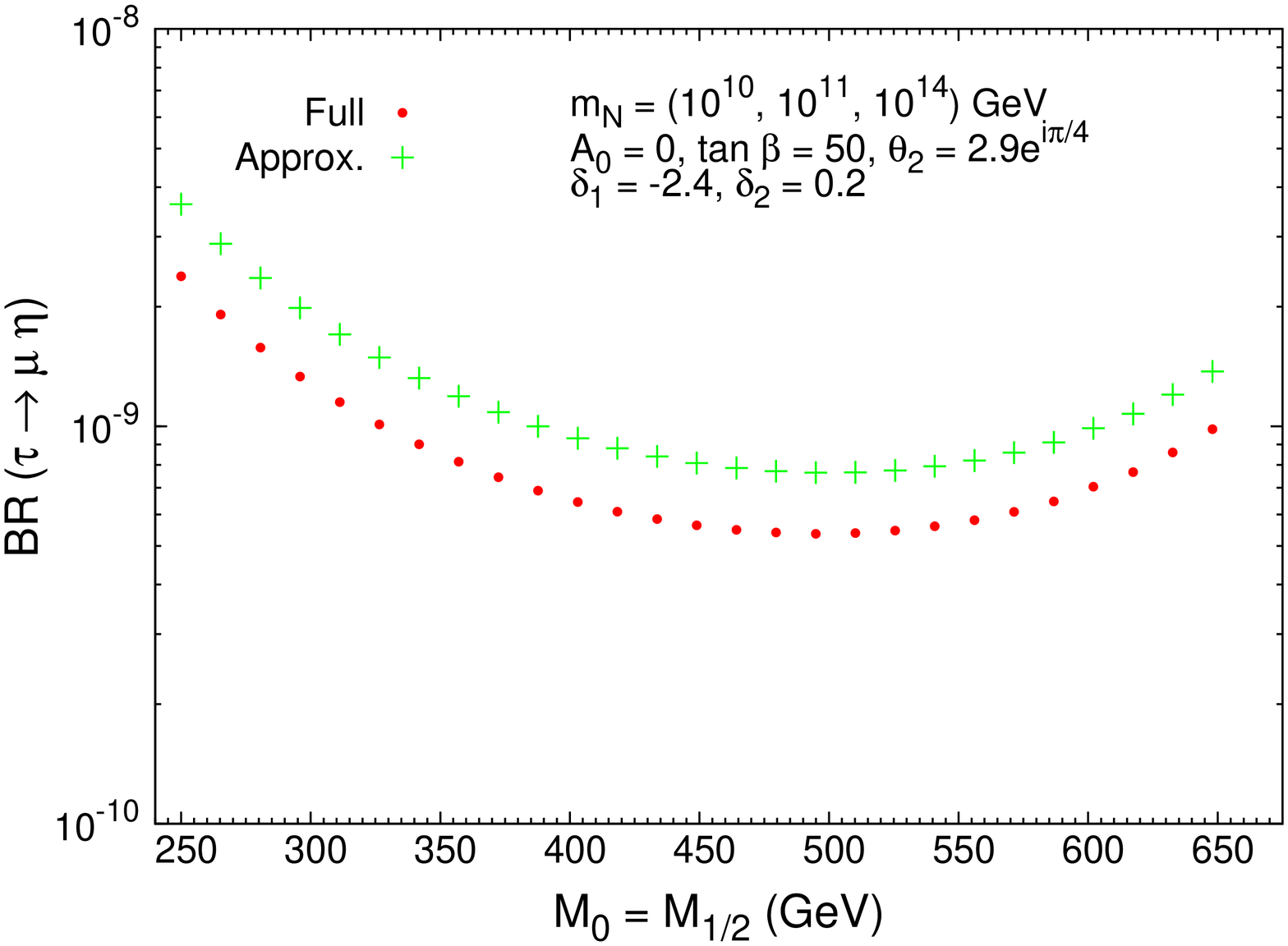,width=60mm,angle=270,clip=} 
     \end{tabular}
     \caption{Comparison between the predicted rates of BR$(\tau \to
     \mu \eta)$ in the NUHM from our full
     1-loop computation and from the approximate result of
     Eq. (\ref{taumueta_approx}) as a function of  $\tan \beta$ (left
     panel) and $M_\text{SUSY} = M_0 = M_{1/2}$ (right panel). The horizontal lines are the present experimental bounds given in Table~\ref{LFVsemilep:bounds}.
     }\label{fig:tau_mueta_approx} 
   \end{center}
 \end{figure}
 
Regarding the comparison with other works, first, we notice that our numerical prediction for BR$(\tau \to \mu \eta)$ in
Eq.~(\ref{taumueta_approx}) does not agree with the original estimate in~\cite{Sher:2002ew} that gives a decay rate a factor 7 larger than ours. We believe that 
the discrepancy comes from our different approaches to describe the hadronisation of quark
bilinears. Our numerical result is closer to that in~\cite{Brignole:2004ah} whose
prediction is larger than ours in a factor of 2. Notice, that the comparison
with this latter work must be done by switching off the bottom-loop induced
contributions and the higher order loop-effects enhanced by $\tan \beta$ factors
which were taken into account in~\cite{Brignole:2004ah} but we are not
including here. This means setting their $\xi_q$ parameters to $\xi_b=0$ and $\xi_s=1$ in 
their formulas. We believe that this small discrepancy is 
mainly due to the different approaches for hadronisation. In particular, they
neglect the $m_{u,d}$ masses whereas we are taking into account chiral symmetry 
breaking
effects via the explicit $m_\pi^2$ and $m_K^2$ dependences, which are well determined in the 
$\chi$PT approach. On the other hand, our prediction for 
$\text{BR}(\tau \to \mu \eta')$ is slightly above  
$\text{BR}(\tau \to \mu \eta)$, due basically to the larger Higgs coupling to 
$\eta'$, $|B_L(\eta')^{(A_0)}|> |B_L(\eta)^{(A_0)}|$. The prediction
in~\cite{Brignole:2004ah}
of $\text{BR}(\tau \to \mu \eta')$ is, however, a factor 100 smaller 
than ours. The prediction for  $\text{BR}(\tau \to \mu \pi)$ here and
in~\cite{Brignole:2004ah}
agree within a factor of 2. Finally, the prediction for $\text{BR}(\tau \to \mu K^+K^-)$ 
in~\cite{Chen:2006hp} is larger than our result in about a factor 50. 
 
The goodness of the above approximate result for $\tau \to \mu \eta$ in 
Eq.~(\ref{taumueta_approx}) 
can be seen in
Fig.~\ref{fig:tau_mueta_approx}, where it is compared with the full result as a function
of $\tan\beta$ and $M_{\rm SUSY}$. It is clear that, for $\tan \beta$ values
larger than about 30 the approximation is quite good, providing rates that 
are at most a factor of 2 above the full predictions.  
Moreover, the behaviour with 
$\tan\beta$ of the
full result at this region is well described by the 
$(\tan\beta)^6$ behaviour of the approximate one.
Regarding the behaviour with $M_{\rm SUSY}$, we see
again that the approximate and full results differ  by less than a factor of 2 
and they both follow the same pattern. The displayed dependence with $M_{\rm SUSY}$  
can be easily understood from the dependence of $m_{A^0}$ with this parameter, as
was shown in Fig.~\ref{fig:mH_mSUSY}. For the
studied range in this plot, $250\,<\,M_{\rm SUSY}\,({\rm GeV})\,<\,650$, this
leads to a relatively small variation in the rates of about $BR_{max}/BR_{min}\sim 5$.

The Higgs dominance approach, however, is not so good for other LFV tau decay
channels. In particular, it is clearly not a good approximation for
$\tau \to 3\mu$ because, in this case, there are other contributions from
$\gamma$-mediated, $Z$-mediated  and box diagrams that enter into the full
computation~\cite{Hisano:1995cp,Arganda:2005ji}. In the NUHM
scenarios that are considered here with small Higgs masses, one may guess 
that the Higgs mediated contribution could dominate the rates at large $\tan
\beta$, but it is not so as will be shown next. By performing a similar
analysis as we have done before, that is, by using the tau-muon-Higgs form factors 
in Eq.~\ref{HL} and plugging it into the exact formula for the
Higgs-contribution~\cite{Arganda:2005ji},
we get in the large $\tan \beta$ limit,
\begin{eqnarray} \!\!\!\!\!\!\!\!\!\!
\text{BR}(\tau \to 3 \mu)_{H_\text{approx}}  &=& 
\frac{G_F^2}{2048 \pi^3} \frac{m_\tau^7 m_\mu^2}{\Gamma_\tau} 
\left( \frac{1}{m_{H^0}^4} + \frac{1}{m_{A^0}^4} + \frac{2}{3m_{H^0}^2 m_{A^0}^2} \right) 
\left| \frac{g^2\delta_{32} }{96 \pi^2} \right|^2 (\tan\beta)^6 
\label{tau3mu_approx_formula}\\
 & = &1.2 \times 10^{-7} \left| \delta_{32} \right|^2 \left( \frac{100}{m_{A^0}({\rm GeV})}
 \right)^4 \left( \frac{\tan \beta}{60} \right)^6,
\label{tau3mu_approx}
\end{eqnarray}
which is in good agreement with the original result in~\cite{Babu:2002et} and
also with posterior estimates~\cite{Dedes:2002rh,Brignole:2004ah}.   
\begin{figure}[t!]
   \begin{center} 
     \begin{tabular}{cc} \hspace*{-12mm}
         \psfig{file=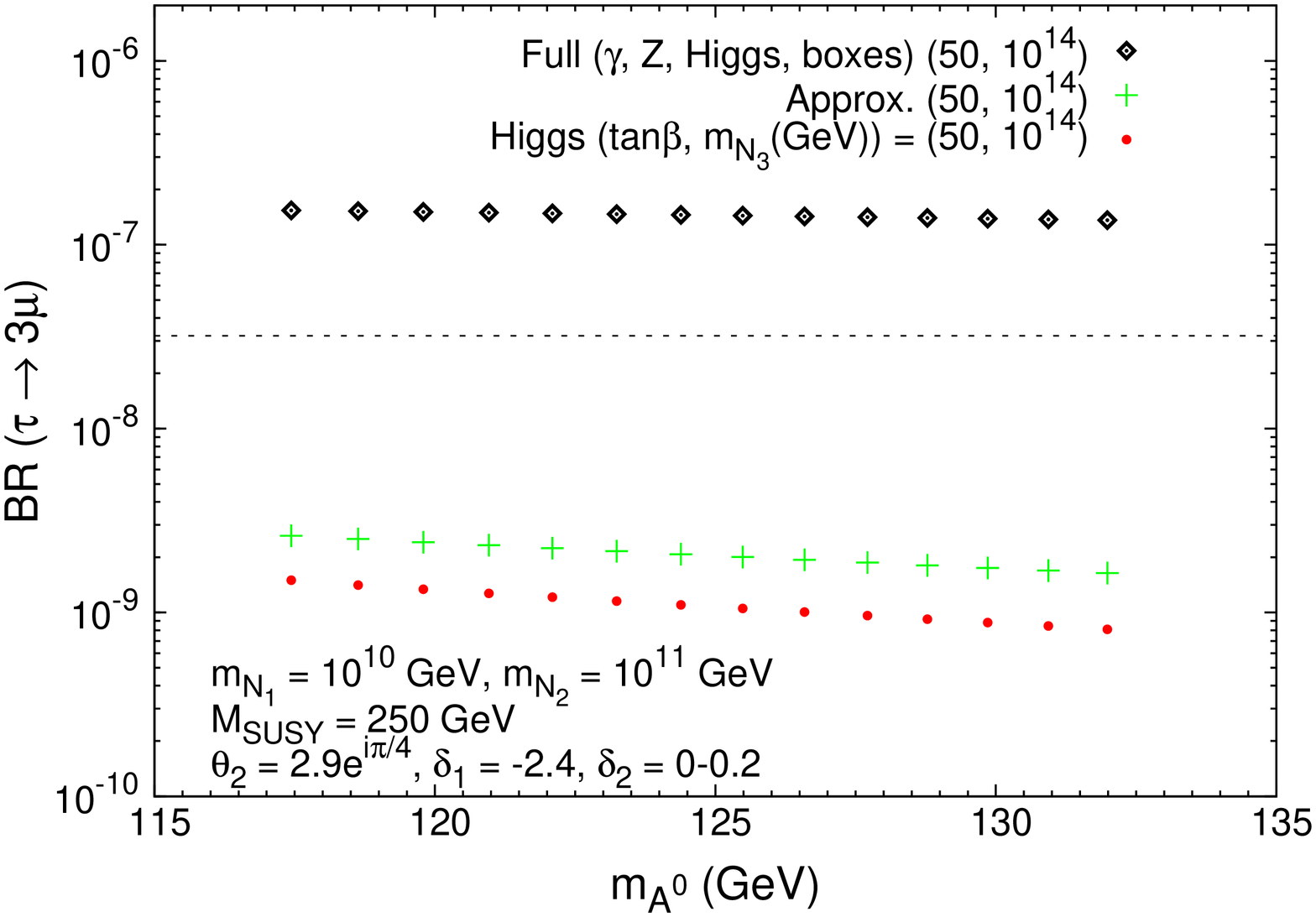,width=60mm,angle=270,clip=}
&	 
         \psfig{file=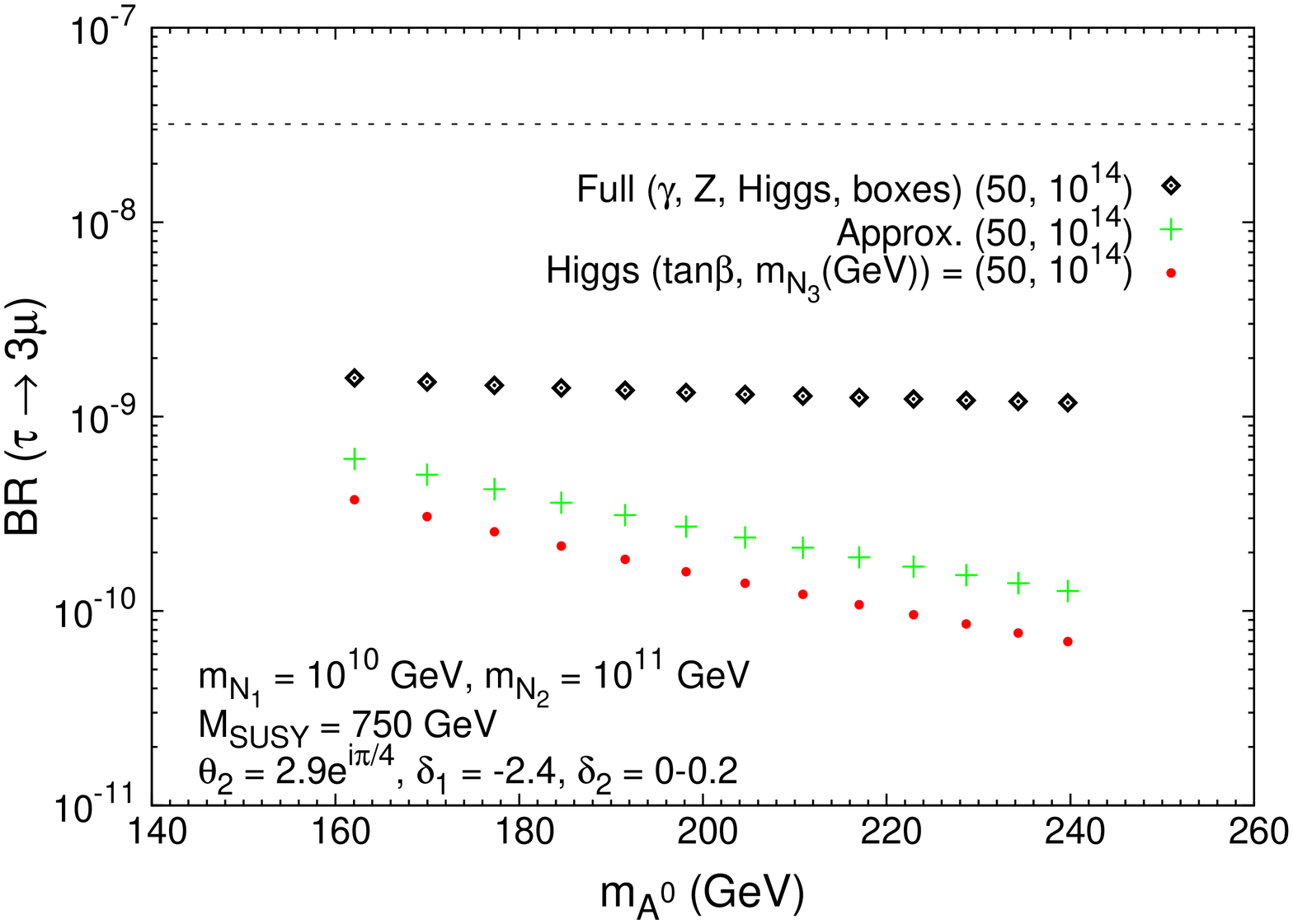,width=60mm,angle=270,clip=}	 
     \end{tabular}
     \caption{Comparison between the full and approximate results for 
     $\tau \to 3 \mu$ as a function of $m_{A^0}$ in the NUHM scenario,
     for $M_{\rm SUSY}=M_0=M_{1/2}=$ 250 GeV (left panel) and for 
    $M_{\rm SUSY}=M_0=M_{1/2}=$ 750 GeV (right panel). The dashed horizontal line is the
    present experimental upper bound~\cite{Miyazaki:2007zw}.
    }\label{fig:tau3mu}
   \end{center}
 \end{figure}
The comparison between the full (i.e. including one-loop SUSY diagrams mediated by 
$\gamma$, $Z$, $h_0$, $H_0$,
$A_0$ and box diagrams which are taken from~\cite{Arganda:2005ji}) and the approximate 
numerical results for this channel is
shown in Fig.~\ref{fig:tau3mu}.  
 We see that the formula in Eq.~(\ref{tau3mu_approx_formula}) 
 predicts rates that are about a factor of 2 larger than the exact Higgs-mediated 
 contribution. Therefore, for large $\tan \beta$ values, it provides a 
 good estimate of the Higgs
 contribution. 
  However, the total rates are much larger than the
 Higgs contribution, since the photon-mediated diagrams give by far the dominant 
 contribution in this channel. For instance, we see in Fig.~\ref{fig:tau3mu} that 
 the total and Higgs rates differ 
 in about two orders of 
 magnitude for 
 $M_{\rm SUSY}\sim$ 250 GeV  and in more than a factor 5 for 
 $M_{\rm SUSY}\sim$ 750 GeV. It is remarkable that, in this channel, 
 the photon dominance 
 holds largely even in scenarios with very heavy SUSY spectra, as for
 $M_{\rm SUSY}\sim$ 750 GeV, and Higgs bosons 
 as light as $m_H=160$ GeV. Therefore, the total rates for this channel 
 can be better approximated by the simplified formula of the photon-mediated
 contribution,
\begin{eqnarray}
\text{BR}(\tau \to 3\mu)_{\gamma_\text{approx}} &=& 
\frac{\alpha}{3\pi}
\left( \log \frac{m_\tau^2}{m_\mu^2}-\frac{11}{4}\right)\,
\text{BR}(\tau \to \mu \gamma)
\nonumber \\[2mm]
& =& 2.3 \times 10^{-3} \,\,\text{BR}(\tau \to \mu \gamma) 
\nonumber \\
& = & 3.4 \times 10^{-5} \left| \delta_{32} \right|^2 \left( \frac{100}
{M_{\rm SUSY}({\rm GeV})}\right)^4 \left( \frac{\tan \beta}{60} \right)^2,
\label{taumugamma_approx}
\end{eqnarray}
where the last line has been obtained by using the result of BR($\tau \to \mu
\gamma$) in the 
mass insertion approximation for equal SUSY mass scales and in the large
$\tan \beta$ limit. It is also interesting to compare this estimate with the present
experimental upper bound for this channel which is 
$3.2 \times 10^{-8}$~\cite{Aubert:2007pw,Miyazaki:2007zw}.
  We see in Fig.~\ref{fig:tau3mu} that, for the chosen
parameters in this plot, the predicted rates are above the present experimental 
bound for $M_{\rm SUSY}<$ 300 GeV.    

Similarly to the $\tau \to 3 \mu$ channel, the semileptonic $\tau \to \mu PP$ decays
(with the exception of $\tau \to \mu \pi^0 \pi^0$)
are clearly dominated by the photon contribution and therefore they can be better 
approximated by the corresponding simplified formulas of this contribution. By
neglecting the $\mu$ mass we have found the following approximate result,   
\begin{eqnarray}
\text{BR}(\tau \to \mu PP)_{\gamma_\text{approx}}
 &=& \int_{4 m_P^2}^{m_{\tau}^2}  ds  
\left( 1-\frac{s}{m_{\tau}^2} \right)^2 \left( 1+\frac{2 m_{\tau}^2}{s} \right) 
\left( 1- \frac{4 m_P^2}{s} \right)^{3/2} |F_V^{PP}(s)|^2 \nonumber \\[2mm] 
&&\times \frac{\alpha}{24 \, m_{\tau}^2} \text{BR}(\tau \to \mu \gamma). 
\end{eqnarray}
And from this we get, 
\begin{eqnarray}
\text{BR}(\tau \to \mu \pi^+ \pi^-)_{\gamma_\text{approx}}& = &2.5 \times  10^{-3}\,\, 
\text{BR}(\tau \to \mu \gamma) \nonumber \\
&=& 3.7 \times 10^{-5}  
\left| \delta_{32} \right|^2 \left( \frac{100}
{M_{\rm SUSY}({\rm GeV})}\right)^4 \left( \frac{\tan \beta}{60} \right)^2 \, ,\\
&& \nonumber \\
\text{BR}(\tau \to \mu K^+ K^-)_{\gamma_\text{approx}}& = &2.0 \times  10^{-4}\,\, 
\text{BR}(\tau \to \mu \gamma) \nonumber \\
&=& 3.0 \times 10^{-6}  
\left| \delta_{32} \right|^2 \left( \frac{100}
{M_{\rm SUSY}({\rm GeV})}\right)^4 \left( \frac{\tan \beta}{60} \right)^2 \,, \\
&& \nonumber \\
\text{BR}(\tau \to \mu K^0 \bar{K}^0)_{\gamma_\text{approx}}& = &1.2 \times  10^{-4}\,\, 
\text{BR}(\tau \to \mu \gamma) \nonumber \\
&=& 1.8\times 10^{-6}  
\left| \delta_{32} \right|^2 \left( \frac{100}
{M_{\rm SUSY}({\rm GeV})}\right)^4 \left( \frac{\tan \beta}{60} \right)^2 \, , \\
&& \nonumber \\
\text{BR}(\tau \to \mu \rho)_{\gamma_\text{approx}}& = &2.3 \times  10^{-3} \,\,
\text{BR}(\tau \to \mu \gamma) \nonumber \\
&=& 3.4\times 10^{-5}  
\left| \delta_{32} \right|^2 \left( \frac{100}
{M_{\rm SUSY}({\rm GeV})}\right)^4 \left( \frac{\tan \beta}{60} \right)^2 \, , \\
&& \nonumber \\
\text{BR}(\tau \to \mu \phi)_{\gamma_\text{approx}}& = &8.4 \times  10^{-5} \,\,
\text{BR}(\tau \to \mu \gamma) \nonumber \\
&=& 1.3\times 10^{-6}  
\left| \delta_{32} \right|^2 \left( \frac{100}
{M_{\rm SUSY}({\rm GeV})}\right)^4 \left( \frac{\tan \beta}{60} \right)^2 \, .
\end{eqnarray} 
\begin{figure}[t!]
   \begin{center} 
     \begin{tabular}{cc} \hspace*{-12mm}
         \psfig{file=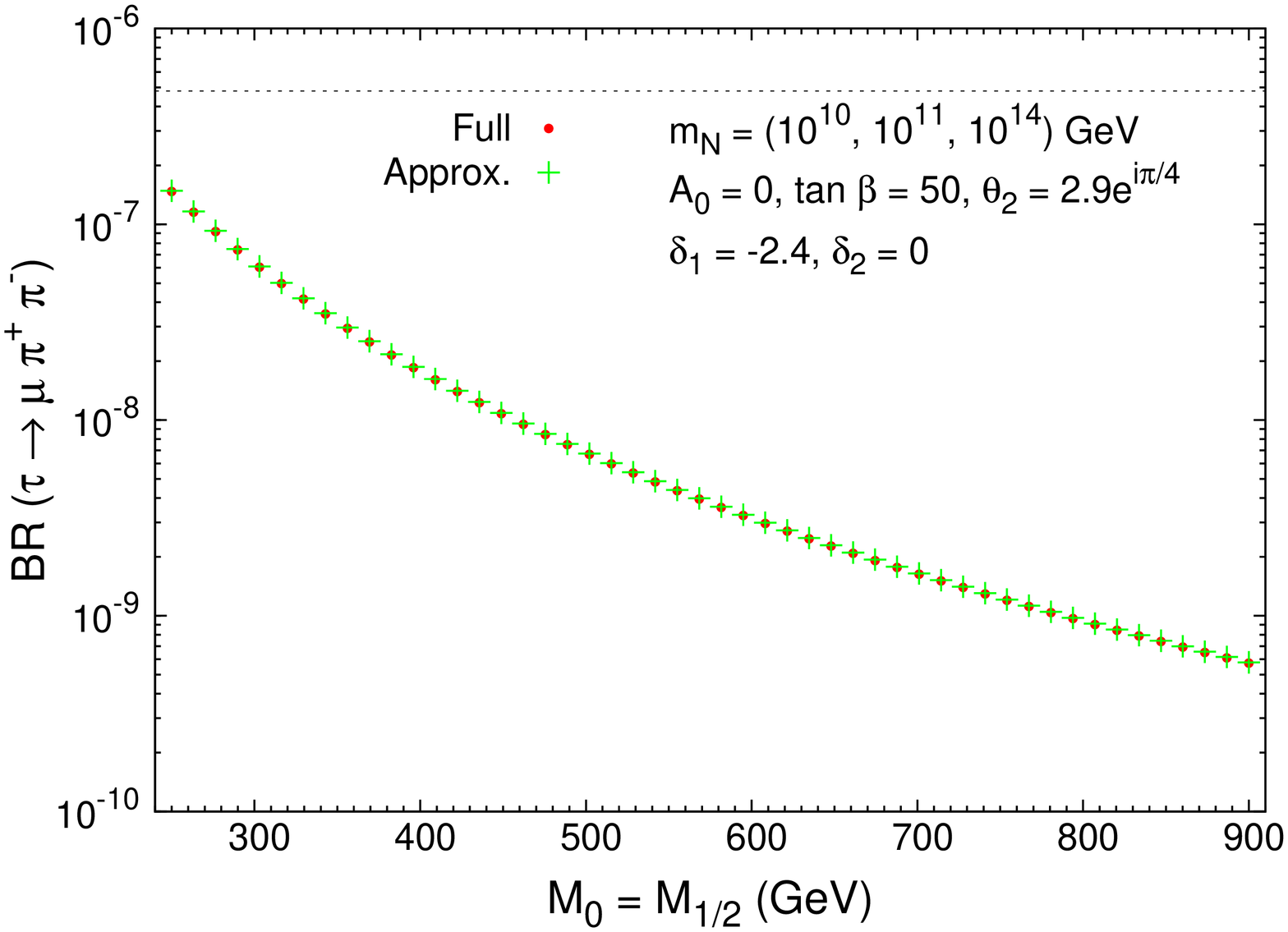,width=60mm,angle=270,clip=}
&	 
         \psfig{file=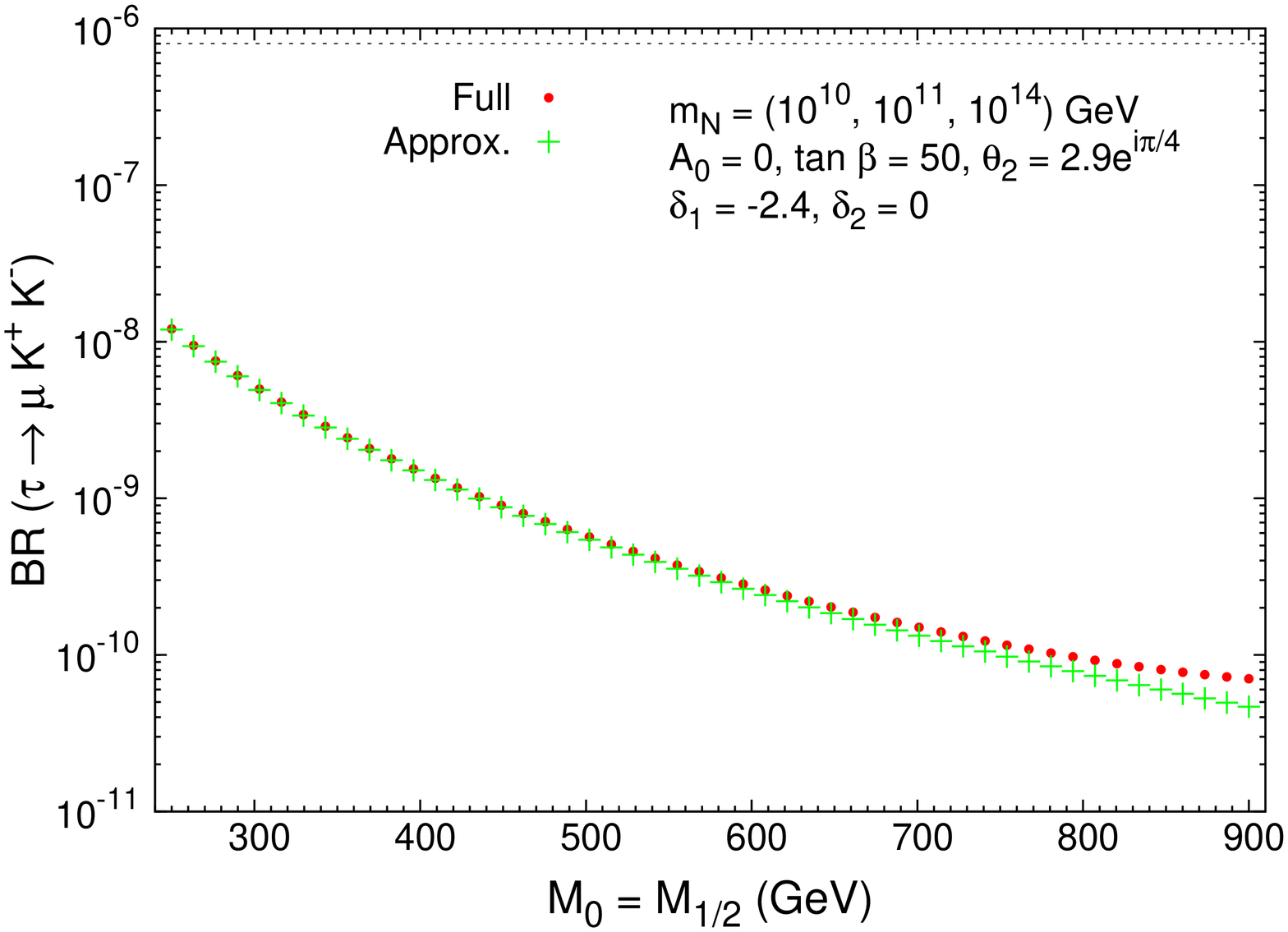,width=60mm,angle=270,clip=}	 
     \end{tabular}
     \caption{Comparison between the full rates and the approximate
     results in the NUHM scenario by
     considering just the photon-mediated contributions for 
     $\tau \to \mu \pi^+\pi^-$ (left panel) and $\tau \to \mu K^+K^-$ (right
     panel) as a function of $M_\text{SUSY} = M_0 = M_{1/2}$.  
     The dashed horizontal lines are the
    present experimental upper bounds~\cite{Yusa:2006qq}.
    }\label{fig:photondominance-approx}
   \end{center}
 \end{figure} 
As can be clearly seen in Fig.~\ref{fig:photondominance-approx} these
results approach pretty well the full rates for most of the $M_{\rm SUSY}$
studied region. For BR$(\tau \to \mu \pi^+\pi^-)$, they are indeed indistinguishable
in this plot. It is only at very large $M_{\rm SUSY} \geq 750 $  GeV that the
approximate result of BR$(\tau \to \mu K^+K^-)$ separates slightly from the full
result, due to the Higgs contribution which competes with the photon one in
this region.
 
\begin{figure}[t!]
   \begin{center} 
     \begin{tabular}{c} \hspace*{-12mm}
  	\psfig{file=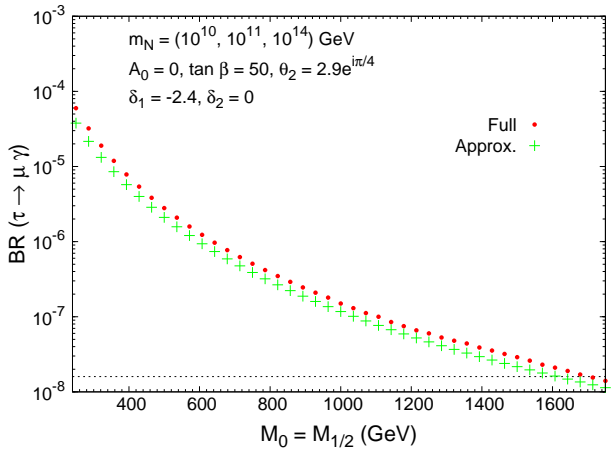,width=60mm,angle=270,clip=}   
     \end{tabular}
     \caption{Comparison between the full 1-loop
     prediction~\cite{Hisano:1995cp} and approximate results of Eq. (~\ref{taumugamma_approx}) for $\tau \to
     \mu \gamma$ as a function of $M_\text{SUSY} = M_0 =M_{1/2}$ in
     the NUHM scenario. The horizontal line is the present
     experimental upper bound~\cite{Banerjee:2007rj}.
     }\label{fig:taumugamma} 
   \end{center}
 \end{figure}

For completeness and comparison, we also include here the predictions for the leading 
LFV tau decay channel, $\tau \to \mu \gamma$.  Fig.~\ref{fig:taumugamma} displays
 the predictions of the
 full and approximate rates for this $\tau \to \mu \gamma$ channel.  
 The full rates
 are taken from~\cite{Arganda:2005ji} and the approximate ones are given by the
  result of the MI approach~\cite{Paradisi:2005tk,Hisano:1998fj},
which at large $\tan\beta$ and for 
  equal SUSY mass
  scales is,
 \begin{eqnarray}
\text{BR}(\tau \to\mu \gamma)_{\rm approx} &=& 
\frac{\alpha^3}{14400\pi^2} \frac{m_\tau^5}{\Gamma_\tau \sin^4\theta_W} \frac{\left|\delta_{32}\right|^2}{M_{\rm SUSY}^4}
(\tan \beta)^2 \nonumber \\[2mm]
& = & 1.5 \times 10^{-2} \left| \delta_{32} \right|^2 \left( \frac{100}
{M_{\rm SUSY}({\rm GeV})}\right)^4 \left( \frac{\tan \beta}{60} \right)^2  \, .
\label{taumugamma_approx}
\end{eqnarray}
In this case, and for the chosen parameters in this plot, 
the approximate and the full results agree to better than a factor 2. We have
verified, however,  that for other choices of $\delta_{1,2}$ the difference 
between them can be larger. Regarding this difference, we emphasise 
that in using the MI approach and LLog 
approximation one has
to be carefull because they are known to fail in some regions of the CMSSM 
parameter space. For instance, in~\cite{Paradisi:2005fk}, the departure of 
the MI from the
exact result is estimated to be up to $50\%$ for $|\delta_{32}| \sim
1$. In~\cite{Antusch:2006vw}
it has been found that the use of the MI and
LLog for large trilinear couplings, $A_0 \sim {\cal O}$ (1 TeV), can fail in several orders of magnitude.
      
The most 
evident conclusion from Fig.~\ref{fig:taumugamma} is   
 that for the chosen parameters in this
plot and for 
$M_{\rm SUSY}< 1600\,\,{\rm GeV}$, the $\tau \to \mu \gamma$ 
rates are above
the present experimental sensitivity, therefore this tau decay channel is at 
present the most  
competitive one in setting bounds on the tau-muon LFV. However, besides experimental
issues, the limitation 
of this channel is that it is
not sensitive at all to the Higgs sector. In this sense, the semileptonic 
channels are more interesting, and can be clearly competitive in the large $M_{\rm
SUSY} \sim {\cal O}(1-2 \, {\rm TeV})$ region.       

\begin{figure}[h!]
   \begin{center} 
     \begin{tabular}{c} \hspace*{-12mm}
         \psfig{file=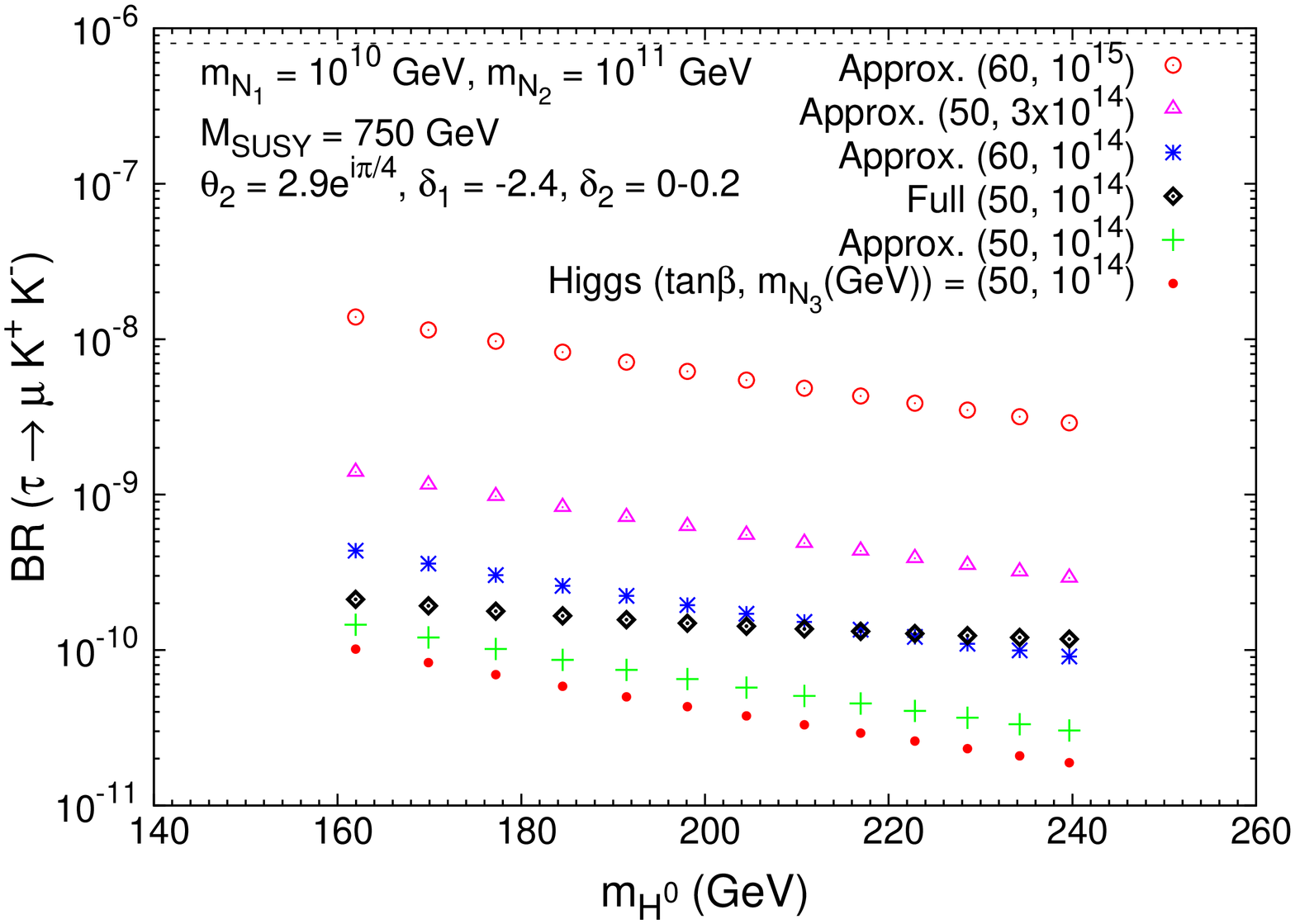,width=100mm,angle=270,clip=}
     \end{tabular}
     \caption{Predictions for BR($\tau \to \mu K^+K^-$) as a function
     of $m_{H^0}$ in the NUHM scenario. A comparison between the full
     1-loop computation and the approximation given by
     Eq. (~\ref{taumuKK_Happrox}) for various choices of large $\tan
     \beta$ and $m_{N_3}$ is included. The horizontal line is the
     present experimental upper bound~\cite{Yusa:2006qq}. 
     }\label{fig:final_taumuKK} 
   \end{center}
 \end{figure}

\begin{figure}[h!]
   \begin{center} 
     \begin{tabular}{c} \hspace*{-12mm}
         \psfig{file=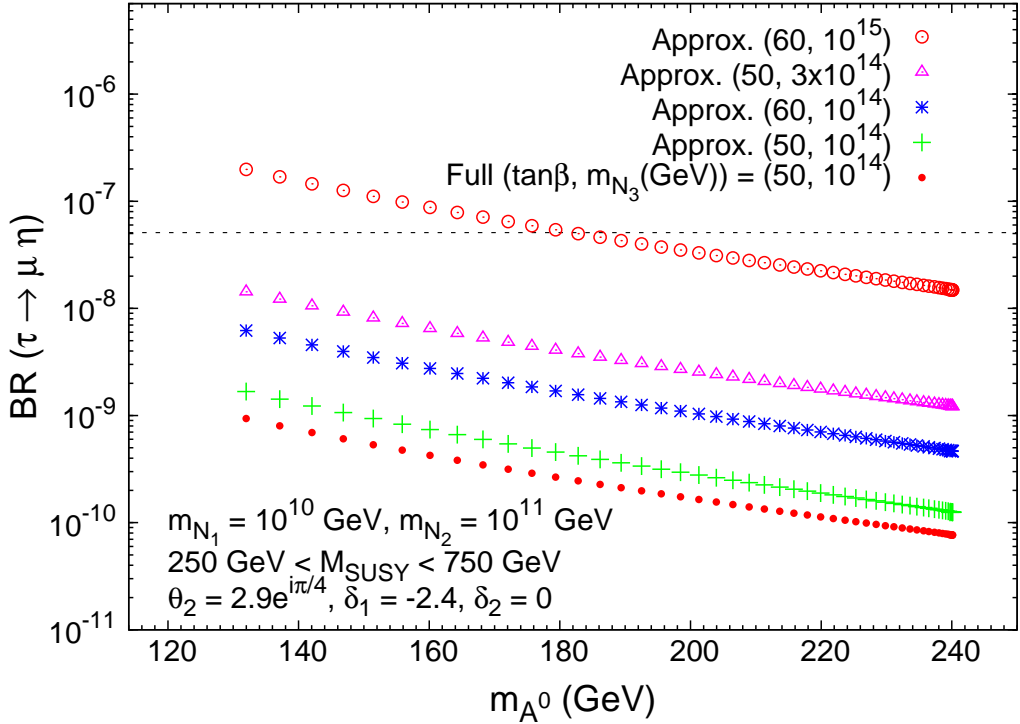,width=100mm,angle=270,clip=}
     \end{tabular}
     \caption{Predictions for BR($\tau \to \mu \eta$) as a function of
     $m_{A^0}$ in the NUHM scenario. A comparison between the full
     1-loop computation and the approximation given by
     Eq. (~\ref{taumuKK_Happrox}) for various choices of large $\tan
     \beta$ and $m_{N_3}$ is included. The horizontal line is the
     present experimental upper bound~\cite{Banerjee:2007rj}. 
     }\label{fig:final_taumueta} 
   \end{center}
 \end{figure}

In Figs.~\ref{fig:final_taumuKK} and \ref{fig:final_taumueta} we plot finally 
the predictions for $\text{BR}(\tau \to \mu K^+K^-)$ and $\text{BR}(\tau \to \mu
\eta)$ as a function of one the most relevant parameters for these Higgs-mediated processes
which is the corresponding Higgs boson mass.
Firstly, we see again that the approximate and exact results of the Higgs
contribution agree within a factor of two for both channels, but the agreement of
the full result with respect to the Higgs contribution is clearly worse in the case 
of $\tau \to \mu K^+ K^-$ than in $\tau \to \mu \eta$. In the latter, the
agreement is quite good because the $Z$-mediated contribution is negligible, and
this holds for all $M_{\rm SUSY}$ values in the studied interval,
250 GeV $<M_{\rm SUSY}<$ 750 GeV . In the first, it is
only for large $M_{\rm SUSY}$ that the $H$-mediated contribution competes with the 
$\gamma$-mediated one and the Higgs rates approach the total rates. For instance, 
Fig.~\ref{fig:final_taumuKK} shows that for $M_{\rm SUSY}=750$ GeV and 
$m_{H^0}=160$ GeV
the total rate is about a factor 2 above the Higgs rate, but for 
$m_{H^0}=240$ GeV it is already more than a factor 5 above.  

In these figures
we have also explored larger values of $m_{N_3}$ and $\tan \beta$, by using in those
cases the approximate formula, and in order to conclude about the values 
that predict rates comparable with the present experimental sensitivity.   
We can conclude then that, at present, it is certainly $\tau \to \mu \eta$ the most
competitive LFV semileptonic tau decay channel. The paremeter values that provide 
rates being comparable to the present sensitivities in this channel are $\tan
\beta = 60$ and $m_{N_3}= 10 ^{15}$ GeV which correspond to 
$|\delta_{32}| \simeq 2$. These large rates, however, should be taken with care
and be considered just as an
order of magnitude estimate since, as we have explained in Sec.~2.1, they  
correspond to neutrino Yukawa couplings which are clearly in the 
non-perturbative
regime. This is why we do not provide the corresponding full rates for them.

\section{Conclusions}\label{concs}
In this paper we have presented a complete one-loop computation of the branching
ratios for the LFV semileptonic $\tau$ decays within the context of two constrained 
MSSM-seesaw scenarios, the CMSSM and the NUHM. We have included both analytical and numerical
results for the particular channels: $\tau \to \mu PP$, with 
$PP= \pi^+\pi^-, \pi^0\pi^0, K^+K^-, K^0 {\bar K}^0$; 
$\tau \to \mu P$ with $P=\pi, \eta, \eta'$; and $\tau \to \mu \rho$,
$\tau \to \mu \phi$. The analysis of the channels  $\tau \to \mu PP$, with
$PP= \pi^+\pi^-, \pi^0\pi^0, K^0 {\bar K}^0$, and $\tau \to \mu \rho$,
$\tau \to \mu \phi$ are, to our knowledge, the first ones in the literature 
within the CMSSM-seesaw context. In addition, we have compared our predictions 
for $\tau \to 
\mu K^+K^-$ and for  $\tau \to \mu P$ with $P=\pi, \eta, \eta'$ with previous
predictions in the literature and found some discrepancies. 

Our treatment of hadronisation has involved two different procedures~:
for the $\gamma$ contribution we have employed state of the art form
factors (detailed in Appendix~\ref{ap:2}), as this amplitude is fairly dominated
by resonance states; for heavier intermediate contributions ($Z$ and
Higgses) we have a local (point-like) vertex driven by chiral symmetry.
It is difficult to estimate the errors of this procedure. For the
hadronization of the $\gamma$, large-$N_C$ inspired, the error should be
smaller than $30 \%$~\cite{Peris:1998nj} (at amplitude level), based on the fact that
subleading terms in the expansion have been included through the widths
of resonances. The hadronization of currents driven by the $Z$ or Higgses,
on the other side, is less known and it is not possible to give a reliable
error estimate.

Our results for $\tau \to \mu \pi^+\pi^-$ demonstrate that this channel is clearly dominated by
the photon-mediated contribution in all the studied region of 
100 GeV $\,< M_{\rm SUSY}\,<$ 1000 GeV.  In fact it is by far, the 
$\tau \to \mu PP$ channel with the highest rates, reaching values close to its present
experimental bound at $4.8 \times 10^{-7}$ for some input parameter values. 
Concretely, it happens for low 
$M_{\rm SUSY}\sim 100-200$ GeV, large $\tan\beta \sim 50-60$, large 
$m_{N_3} \sim 
10^{14}-10^{15}$ GeV and large ${\rm arg}(\theta_2) \sim \pi/4-\pi/2$ (these two 
latter parameters producing a large 
$\delta_{32}\sim {\cal O}(1)$). In contrast, $\tau \to \mu \pi^0\pi^0$ can
only be mediated by  $h^0$ and $H^0$ Higgs bosons and their rates are very 
small. Besides, they are not yet comparable with data, since there is no 
bound in this channel. 
The cases of 
$\tau \to \mu K^+K^-$ and $\tau \to \mu K^0 {\bar K}^0$ decays, are much more
interesting. In these two channels, the photon-mediated 
contribution dominates in most of the studied region of $M_{\rm SUSY}$, except at  
large, $M_{\rm SUSY} > $ 750 GeV values, where the Higgs-mediated and the
$\gamma$-mediated contributions can compete. This competition happens in 
specific constrained scenarios of NUHM 
type with low $m_{H^0} \sim 100-200$ GeV values and very heavy
SUSY spectrum with  $M_{\rm SUSY} > $ 750 GeV. This peculiar MSSM spectrum and
the fact that 
Higgs bosons couple stronger to $K^+K^-$ (and $K^0 {\bar K}^0$) than to $\pi^+\pi^-$
(and $\pi^0\pi^0$) is the reason why the $H$- and $\gamma$-mediated 
contributions 
can compete in $\tau \to \mu K^+K^-$
but not
in $\tau \to \mu \pi^+\pi^-$. Furthermore, due to the
fact that the photon diagram still dominates
BR($\tau \to \mu K^+K^-$) in a large region of the parameter space with 
100 GeV $\,<M_{\rm SUSY}\,<$ 750 GeV, the involved  hadronic form factors do 
play a crutial role in the final rates. Consequently, 
our results for this channel are in disagreement with those of~\cite{Chen:2006hp}
where they only included the Higgs-mediated contribution. 
We have also shown that the largest predicted
rates for BR$(\tau \to \mu K^+K^-)$ are, as in $\tau \to \mu \pi^+\pi^-$, 
at the region with
 low 
$M_{\rm SUSY}\sim 100-200$ GeV, large $\tan\beta \sim 50-60$, large 
$m_{N_3} \sim 
10^{14}-10^{15}$ GeV and large ${\rm arg}(\theta_2) \sim \pi/4-\pi/2$
values. However, the predicted rates do not
reach yet the present experimental sensitivity, which in this channel is at 
$8 \times 10^{-7}$.
 
Our results for  $\tau \to \mu \eta$ and  $\tau \to \mu \eta'$ demonstrate 
that these
two channels are largely dominated by the $A^0$-mediated contribution and their
predicted rates are very competitive in the case of NUHM scenarios with low  
$m_{A^0} \sim 100-200$ GeV values and large $\tan \beta \sim 50-60$. 
This is in qualitative agreement with previous
estimates in the literature. However, we have found some important 
numerical discrepancies 
with respect to the estimate in~\cite{Sher:2002ew}. Concretely, the predicted rates in
the present work are 
smaller than those in~\cite{Sher:2002ew} by a factor of about 7. We believe that these
discrepancies  are due to the different procedures of quark bilinear 
hadronisation. We claim that our results which are based on the well defined 
and more refined  
hadronisation prescription 
by $\chi$PT provide a better estimate. The rates for BR($\tau \to \mu \eta$) have 
also been compared with those in~\cite{Brignole:2004ah,Paradisi:2005tk}
which are within the different context of non-constrained MSSM and with input
$\delta_{32}$ not being connected to neutrino physics nor seesaw mechanism. We
have checked, that the predicted rates are in reasonable agreement with these
two works for, $\delta_{32} \sim {\cal O}(1)$ , which in our case is 
reached by input seesaw parameters of $m_{N_3} \sim 
10^{14}-10^{15}$ GeV and large ${\rm arg}(\theta_2) \sim \pi/4-\pi/2$. 
 
In addition, we have presented in this work a set of useful approximate formulas 
for  all
the semileptonic $\tau$ decays that we have compared with the full-one
loop results and concluded that they give reasonable good estimates, say 
differing
in less than a factor of two respect to the full result. We have also compared 
these results with those for the leptonic channel, 
$\tau \to 3\mu$, and the
radiative decay, $\tau \to \mu \gamma$. 

Our overall conclusion is that,
for the same Constrained MSSM-Seesaw input parameters,  $\tau \to \mu \gamma$ is 
the most competitive $\tau$ decay channel in testing the values of the LFV
parameter $\delta_{32}$, but it is not sensitive at all to the Higgs sector. 
Interestingly, the most competitive channels to explore simultaneously LFV and the
Higgs sector are $\tau \to \mu \eta$, $\tau \to \mu \eta'$ and also 
$\tau \to \mu K^+K^-$. The $\tau \to \mu K^+ K^-$ channel is certainly
more efficient than $\tau \to 3 \, \mu$ as far as the sensitivity to
the Higgs sector is concerned. Otherwise, the golden channels to tackle
the Higgs sector are undoubtly $\tau \to \mu \eta$ and $\tau \to \mu
\eta^\prime$. On the other hand, 
the rest of the studied semileptonic channels, 
$\tau \to \mu \pi^+\pi^-$, etc.,  will not provide additional information on LFV 
with respect to that provided by $\tau \to \mu \gamma$.
  
\section*{Acknowledgements}
We acknowledge P. Paradisi for clarifying us the results of
the $\tau \to \mu \gamma$ in the mass insertion approximation.
This work has been supported in part by the EU
   MRTN-CT-2006-035482 (FLAVIAnet), by MEC (Spain) under grants
   FPA2006-05423 and
   FPA2007-60323, by Generalitat Valenciana under grant
   GVACOMP2007-156, by Comunidad de Madrid under HEPHACOS project and
   by the Spanish Consolider-Ingenio 2010 Programme CPAN (CSD2007-00042).
\appendix

\label{apendices}
\renewcommand{\theequation}{\Alph{section}.\arabic{equation}}
\renewcommand{\thetable}{\Alph{section}.\arabic{table}}
\renewcommand{\thefigure}{\Alph{section}.\arabic{figure}}
\setcounter{section}{0}
\setcounter{equation}{0}
\setcounter{table}{0}
\setcounter{figure}{0}

\section{LFV form factors}
\label{apendice1}

In this Appendix we collect the main analytical formulae containing
the full 1-loop results of the SUSY contributions to the relevant $\tau
- \mu$ LFV form factors for the present work, corresponding to:
$\gamma \tau \mu$, $Z \tau \mu$ and $H \tau \mu$ vertices. All the
couplings and loop functions appearing in the following formulae are
defined in~\cite{Arganda:2005ji,Arganda:2007jw}. 

\subsection{Form factors for the $\gamma \tau \mu$ vertex}

\begin{figure}[hbtp]
  \begin{center} 
        \psfig{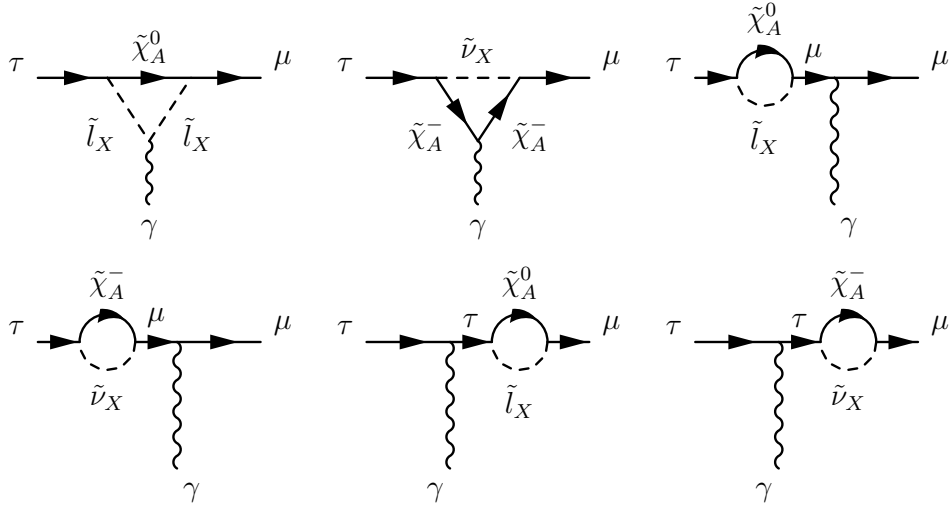}
    \caption{Relevant SUSY one-loop diagrams for the photon-mediated contributions to 
    LFV semileptonic $\tau$ decays.}
    \label{Photon_diagrams} 
  \end{center}
\end{figure}
Our convention for the form factors $A_{1,2}^{L,R}$ defining the $\gamma \tau \mu $ 
vertex is as follows:
\begin{equation}
i e \left[q^2 \gamma_\alpha (A_1^LP_L+A_1^RP_R)+
im_\tau\sigma_{\alpha\beta}q^\beta(A_a^LP_L+A_2^RP_R)\right],
\end{equation}
where $q$ is the off-shell photon momentum, $P_{L,R}=(1\mp\gamma_5)/2$, $e$ is the
electromagnetic positron charge and $m_\tau$ is the $\tau$ lepton mass. 

In the SUSY-seesaw context there are one-loop contributions to these form factors that come 
from the chargino and neutralino sectors respectively,
\begin{equation}
A_a^{L,R} = A_a^{(n)L.R} + A_a^{(c)L,R}, \quad a = 1, 2 \, .
\end{equation}
The neutralino contributions are given by,
\begin{eqnarray}
A_1^{(n)L} &=& \frac{1}{576 \pi^2} N_{\mu AX}^R N_{\tau AX}^{R \ast} \frac{1}{m_{\tilde{l}_X}^2} \frac{2 - 9 x_{AX} + 18 x_{AX}^2 - 11 x_{AX}^3 + 6 x_{AX}^3 \log{x_{AX}}}{\left( 1 - x_{AX} \right)^4} \nonumber \\
\\
A_2^{(n)L} &=& \frac{1}{32 \pi^2} \frac{1}{m_{\tilde{l}_X}^2} \left[ N_{\mu AX}^L N_{\tau AX}^{L \ast} \frac{1 - 6 x_{AX} + 3 x_{AX}^2 + 2 x_{AX}^3 - 6 x_{AX}^2 \log{x_{AX}}}{6 \left( 1 - x_{AX} \right)^4} \right. \nonumber \\
&+& N_{\mu AX}^R N_{\tau AX}^{R \ast} \frac{m_\mu}{m_\tau} \frac{1 - 6 x_{AX} + 3 x_{AX}^2 + 2 x_{AX}^3 - 6 x_{AX}^2 \log{x_{AX}}}{6 \left( 1 - x_{AX} \right)^4} \nonumber \\
&+& \left. N_{\mu AX}^L N_{\tau AX}^{R \ast} \frac{m_{\tilde{\chi}_A^0}}{m_\tau} \frac{1 -
x_{AX}^2 +2 x_{AX} \log{x_{AX}}}{\left( 1 - x_{AX} \right)^3} \right], \label{A2Lneut}\\
A_a^{(n)R} &=& \left. A_a^{(n)L} \right|_{L \leftrightarrow R},\label{ARneut}
\end{eqnarray}
where $x_{AX} = m_{\tilde{\chi}_A^0}^2/m_{\tilde{l}_X}^2$ and the indices are 
$A=1,..,4$, $X=1,..,6$.

The chargino contributions are given by
\begin{eqnarray}
A_1^{(c)L} &=& -\frac{1}{576 \pi^2} C_{\mu AX}^R C_{\tau AX}^{R \ast} \frac{1}{m_{\tilde{\nu}_X}^2} \frac{16 - 45 x_{AX} + 36 x_{AX}^2 - 7 x_{AX}^3 + 6 (2 - 3 x_{AX}) \log{x_{AX}}}{\left( 1 - x_{AX} \right)^4}, \nonumber \\
& & \\ \cr
A_2^{(c)L} &=& -\frac{1}{32 \pi^2} \frac{1}{m_{\tilde{\nu}_X}^2} \left[ C_{\mu AX}^L C_{\tau AX}^{L \ast} \frac{2 + 3 x_{AX} - 6 x_{AX}^2 + x_{AX}^3 + 6 x_{AX} \log{x_{AX}}}{6 \left( 1 - x_{AX} \right)^4} \right. \nonumber \\
&+& C_{\mu AX}^R C_{\tau AX}^{R \ast} \frac{m_\mu}{m_\tau} \frac{2 + 3 x_{AX} - 6 x_{AX}^2 + x_{AX}^3 + 6 x_{AX} \log{x_{AX}}}{6 \left( 1 - x_{AX} \right)^4} \nonumber \\
&+& \left. C_{\mu AX}^L C_{\tau AX}^{R \ast} \frac{m_{\tilde{\chi}_A^-}}{m_\tau} \frac{-3 +
4 x_{AX} - x_{AX}^2 - 2 \log{x_{AX}}}{\left( 1 - x_{AX} \right)^3} \right],
\label{A2Lchar} \\
A_a^{(c)R} &=& \left. A_a^{(c)L} \right|_{L \leftrightarrow R},\label{ARchar}
\end{eqnarray}
where in this case $x_{AX} = m_{\tilde{\chi}_A^-}^2/m_{\tilde{\nu}_X}^2$ and the indices are 
$A=1,2$, $X=1,2,3$. Notice that in 
both neutralino and chargino contributions a summation over the indices 
$A$ and $X$ is understood.

\subsection{Form factors for the $Z \tau \mu$ vertex}
Our convention for the form factors $F_{L,R}$ defining the $Z \tau \mu$ 
vertex is as follows:
\begin{equation}
-i\gamma_\mu\left[F_LP_L+F_RP_R\right].
\end{equation}
\begin{figure}[hbtp]
  \begin{center} 
        \psfig{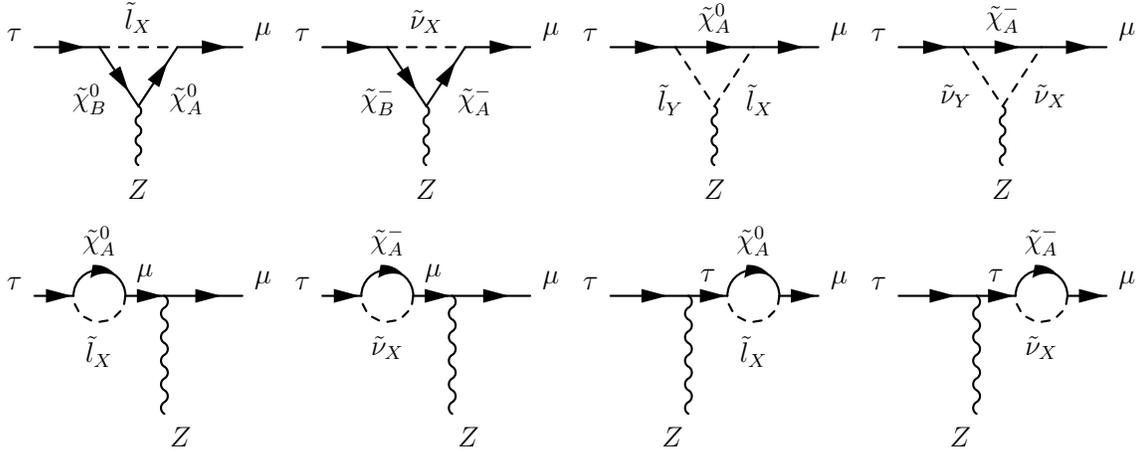}
    \caption{Relevant SUSY one-loop diagrams for the $Z$-mediated contributions to 
    LFV semileptonic $\tau$ decays.
    }\label{Z_diagrams} 
  \end{center}
\end{figure}

The $Z$-boson form factors have also the two kinds of contributions, from neutralinos
$(n)$ and charginos $(c)$, 
\begin{equation}\label{Zformfactors}
F_{L(R)} = F_{L(R)}^{(n)} + F_{L(R)}^{(c)} \, .
\end{equation}
The results for the corresponding form factors are the following:
\begin{eqnarray}\label{Zformafactor_expressions}
F_L^{(n)} &=& -\frac{1}{16 \pi^2} \left\{ N_{\mu BX}^R N_{\tau AX}^{R \ast} \left[ 2 E_{BA}^{R(n)} C_{24}(m_{\tilde{l}_X}^2, m_{\tilde{\chi}_A^0}^2, m_{\tilde{\chi}_B^0}^2) - E_{BA}^{L(n)} m_{\tilde{\chi}_A^0} m_{\tilde{\chi}_B^0} C_0(m_{\tilde{l}_X}^2, m_{\tilde{\chi}_A^0}^2, m_{\tilde{\chi}_B^0}^2) \right] \right. \nonumber \\
&+& \left. N_{\mu AX}^R N_{\tau AY}^{R \ast} \left[ 2 Q_{XY}^{\tilde{l}} C_{24}(m_{\tilde{\chi}_A^0}^2, m_{\tilde{l}_X}^2, m_{\tilde{l}_Y}^2) \right] + N_{\mu AX}^R N_{\tau AX}^{R \ast} \left[ Z_L^{(l)} B_1(m_{\tilde{\chi}_A^0}^2, m_{\tilde{l}_X}^2) \right] \right\}, \\
F_R^{(n)} &=& \left. F_L^{(n)} \right|_{L \leftrightarrow R}, \\
F_L^{(c)} &=& -\frac{1}{16 \pi^2} \left\{ C_{\mu BX}^R C_{\tau AX}^{R \ast} \left[ 2 E_{BA}^{R(c)} C_{24}(m_{\tilde{\nu}_X}^2, m_{\tilde{\chi}_A^-}^2, m_{\tilde{\chi}_B^-}^2) - E_{BA}^{L(c)} m_{\tilde{\chi}_A^-} m_{\tilde{\chi}_B^-} C_0(m_{\tilde{\nu}_X}^2, m_{\tilde{\chi}_A^-}^2, m_{\tilde{\chi}_B^-}^2) \right] \right. \nonumber \\
&+& \left. C_{\mu AX}^R C_{\tau AY}^{R \ast} \left[ 2 Q_{XY}^{\tilde{\nu}} C_{24}(m_{\tilde{\chi}_A^-}^2, m_{\tilde{\nu}_X}^2, m_{\tilde{\nu}_Y}^2) \right] + C_{\mu AX}^R C_{\tau AX}^{R \ast} \left[ Z_L^{(l)} B_1(m_{\tilde{\chi}_A^-}^2, m_{\tilde{\nu}_X}^2) \right] \right\}, \\
F_R^{(c)} &=& \left. F_L^{(c)} \right|_{L \leftrightarrow R},
\end{eqnarray}
where again the indices are $A,B=1,..,4$, $X,Y=1,..,6$ in the contributions from the
neutralino sector and $A,B=1,2$, $X,Y=1,2,3$ in the contributions from the chargino
sector, and a summation over the various indices is understood.

\subsection{Form factors for the $H \tau \mu$ vertex}
Our convention for the form factors $H^{(p)}_{L,R}$ defining the $H_p \tau \mu $ 
vertex is as follows:
\begin{equation}
i\left[H^{(p)}_LP_L+H^{(p)}_RP_R\right].
\end{equation}
\begin{figure}[hbtp]
  \begin{center} 
        \psfig{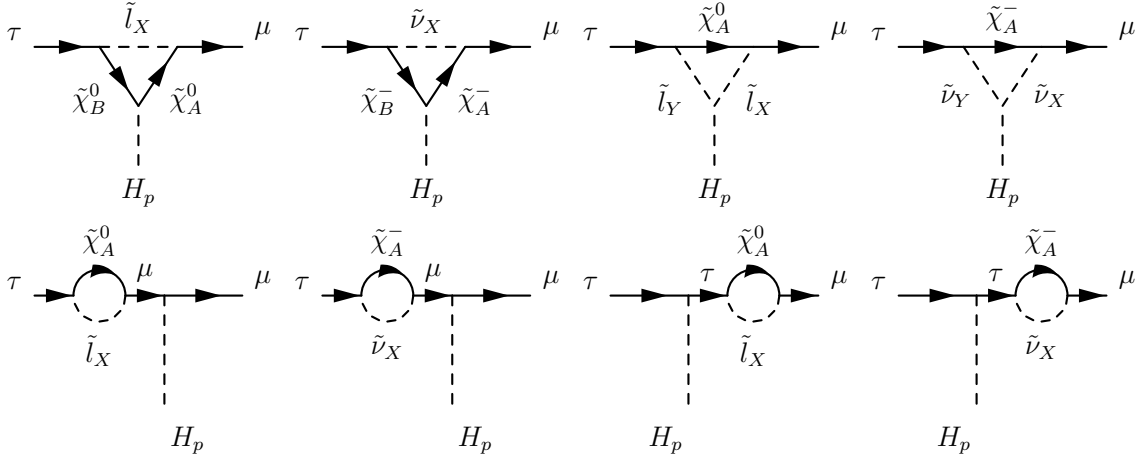}
    \caption{Relevant SUSY one-loop diagrams for the Higgs-mediated contributions to 
    LFV semileptonic $\tau$ decays.}\label{H_diagrams} 
  \end{center}
\end{figure}

As in the previous cases, we separate the contributions from the neutralino and
chargino sectors,
\begin{equation}
H^{(p)}_{L(R)} = H_{L(R),n}^{(p)} + H_{L(R),c}^{(p)}.
\end{equation}

The results for the form factors are the following,
\begin{eqnarray}
H_{L, n}^{(p)} &=& -\frac{1}{16 \pi^2} \left\{ \left[ B_0(m_{\tilde{\chi}_A^0}^2, m_{\tilde{\chi}_B^0}^2) + m_{\tilde{l}_X}^2 C_0(m_{\tilde{l}_X}^2, m_{\tilde{\chi}_A^0}^2, m_{\tilde{\chi}_B^0}^2) + m_\tau^2 C_{12}(m_{\tilde{l}_X}^2, m_{\tilde{\chi}_A^0}^2, m_{\tilde{\chi}_B^0}^2) \right. \right. \nonumber \\
&+& \left. m_\mu^2 (C_{11} - C_{12})(m_{\tilde{l}_X}^2, m_{\tilde{\chi}_A^0}^2, m_{\tilde{\chi}_B^0}^2) \right] N_{\mu AX}^L D_{R, AB}^{(p)} N_{\tau BX}^{R \ast} \nonumber \\
&+& m_\mu m_\tau (C_{11} + C_0)(m_{\tilde{l}_X}^2, m_{\tilde{\chi}_A^0}^2, m_{\tilde{\chi}_B^0}^2) N_{\mu AX}^R D_{L, AB}^{(p)} N_{\tau BX}^{L \ast} \nonumber \\
&+& m_\mu m_{\tilde{\chi}_B^0} (C_{11} - C_{12} + C_0)(m_{\tilde{l}_X}^2, m_{\tilde{\chi}_A^0}^2, m_{\tilde{\chi}_B^0}^2) N_{\mu AX}^R D_{L, AB}^{(p)} N_{\tau BX}^{R \ast} \nonumber \\
&+& m_\tau m_{\tilde{\chi}_B^0} C_{12}(m_{\tilde{l}_X}^2, m_{\tilde{\chi}_A^0}^2, m_{\tilde{\chi}_B^0}^2) N_{\mu AX}^L D_{R, AB}^{(p)} N_{\tau BX}^{L \ast} \nonumber \\
&+& m_\mu m_{\tilde{\chi}_A^0} (C_{11} - C_{12})(m_{\tilde{l}_X}^2, m_{\tilde{\chi}_A^0}^2, m_{\tilde{\chi}_B^0}^2) N_{\mu AX}^R D_{R, AB}^{(p)} N_{\tau BX}^{R \ast} \nonumber \\
&+& m_\tau m_{\tilde{\chi}_A^0} (C_{12} + C_0)(m_{\tilde{l}_X}^2, m_{\tilde{\chi}_A^0}^2, m_{\tilde{\chi}_B^0}^2) N_{\mu AX}^L D_{L, AB}^{(p)} N_{\tau BX}^{L \ast} \nonumber \\
&+& m_{\tilde{\chi}_A^0} m_{\tilde{\chi}_B^0} C_0(m_{\tilde{l}_X}^2, m_{\tilde{\chi}_A^0}^2, m_{\tilde{\chi}_B^0}^2) N_{\mu AX}^L D_{L, AB}^{(p)} N_{\tau BX}^{R \ast} \nonumber \\
&+& G_{XY}^{(p) \tilde{l}} \left[ - m_\mu (C_{11} - C_{12})(m_{\tilde{\chi}_A^0}^2, m_{\tilde{l}_X}^2, m_{\tilde{l}_Y}^2) N_{\mu AX}^R N_{\tau AY}^{R \ast} \right. \nonumber \\
&-& \left. m_\tau C_{12}(m_{\tilde{\chi}_A^0}^2, m_{\tilde{l}_X}^2, m_{\tilde{l}_Y}^2) N_{\mu AX}^L N_{\tau AY}^{L \ast} + m_{\tilde{\chi}_A^0} C_0(m_{\tilde{\chi}_A^0}^2, m_{\tilde{l}_X}^2, m_{\tilde{l}_Y}^2) N_{\mu AX}^L N_{\tau AY}^{R \ast} \right] \nonumber \\
&+& \frac{S_{L, \tau}^{(p)}}{m_\mu^2 - m_\tau^2} \left[ - m_\mu^2 B_1(m_{\tilde{\chi}_A^0}^2, m_{\tilde{l}_X}^2) N_{\mu AX}^L N_{\tau AX}^{L \ast} + m_\mu m_{\tilde{\chi}_A^0} B_0(m_{\tilde{\chi}_A^0}^2, m_{\tilde{l}_X}^2) N_{\mu AX}^R N_{\tau AX}^{L \ast} \right. \nonumber \\
&-& \left. m_\mu m_\tau B_1(m_{\tilde{\chi}_A^0}^2, m_{\tilde{l}_X}^2) N_{\mu AX}^R N_{\tau AX}^{R \ast} + m_\tau m_{\tilde{\chi}_A^0} B_0(m_{\tilde{\chi}_A^0}^2, m_{\tilde{l}_X}^2) N_{\mu AX}^L N_{\tau AX}^{R \ast} \right] \nonumber \\
&+& \frac{S_{L, \mu}^{(p)}}{m_\tau^2 - m_\mu^2} \left[ - m_\tau^2 B_1(m_{\tilde{\chi}_A^0}^2, m_{\tilde{l}_X}^2) N_{\mu AX}^R N_{\tau AX}^{R \ast} + m_\tau m_{\tilde{\chi}_A^0} B_0(m_{\tilde{\chi}_A^0}^2, m_{\tilde{l}_X}^2) N_{\mu AX}^R N_{\tau AX}^{L \ast} \right. \nonumber \\
&-& \left. \left. m_\mu m_\tau B_1(m_{\tilde{\chi}_A^0}^2, m_{\tilde{l}_X}^2) N_{\mu AX}^L N_{\tau AX}^{L \ast} + m_\mu m_{\tilde{\chi}_A^0} B_0(m_{\tilde{\chi}_A^0}^2, m_{\tilde{l}_X}^2) N_{\mu AX}^L N_{\tau AX}^{R \ast} \right] \right\}, \\
H_{R, n}^{(p)} &=& \left. H_{L, n}^{(p)} \right|_{L \leftrightarrow R} \quad p = 1, 2, 3.
\end{eqnarray}
Correspondingly, the result for the chargino contribution $H_{L (R), c}^{(p)}$ can be obtained from the previous $H_{L (R), n}^{(p)}$  by replacing everywhere,
\begin{eqnarray}
\tilde{l} &\to& \tilde{\nu} \nonumber \\
\tilde{\chi}^0 &\to& \tilde{\chi}^- \nonumber \\
N^{L(R)} &\to& C^{L(R)} \nonumber \\
D_{L(R)} &\to& W_{L(R)} \nonumber
\end{eqnarray}
In the previous formulae, the index $p$ refers to the each of the Higgs bosons.
Concretely,  $H_p = h^0, H^0, A^0$ for $p = 1, 2, 3$, respectively. The other indices
are again $A,B=1,..,4$, $X,Y=1,..,6$ in the contributions from the
neutralino sector and $A,B=1,2$ and $X,Y=1,2,3$ in the contributions from the chargino
sector. A summation over all the indices is also understood.
\renewcommand{\theequation}{\Alph{section}.\arabic{equation}}
\renewcommand{\thetable}{\Alph{section}.\arabic{table}}
\setcounter{section}{1}
\setcounter{equation}{0}
\setcounter{table}{0}

\section{Hadronic form factors}
\label{ap:2}

Our construction of the vector form factors $F_V^{PP}(s)$, defined by Eq.~(\ref{eq:ffem}), follows the idea put forward in~\cite{Ecker:1989yg} that lie on two key points~:
\begin{itemize}
\item[1/] At $s \ll M_R^2$ (being $M_R$ a generic resonance mass), the vector form factor should match
the ${\cal O}(p^4)$ result of $\chi$PT. Hence our form factors will satisfy the chiral constraint.
\item[2/] Form factors of QCD currents should behave softly at high transfer of momenta \cite{Lepage:1980fj}, i.e. they should vanish for $s \gg M_R^2$. Accordingly we will demand to our form factors that they
satisfy this asymptotic constraint.
\end{itemize}
In the $N_C \rightarrow \infty$ limit resonances have zero-width. However those present in the
relevant form factors in tau decays do indeed resonate due to the available phase space. As a consequence
we need to include energy-dependent widths for the wider resonances $\rho(770)$ and $\rho(1450)$, or constant
for the narrow ones~: $\omega(782)$ and $\phi(1020)$. For the $\rho(770)$ we take the definition
put forward in~\cite{GomezDumm:2000fz}~:
\begin{equation}
 \Gamma_{\rho}(s)  =  \frac{M_{\rho} s}{96 \pi F^2} \left[
\sigma_{\pi}^3(s) \, \theta(\, s \, - \, 4 m_{\pi}^2) \, + \,  \frac{1}{2} \, 
\sigma_K^3(s) \, \theta( \, s \, - \, 4 m_K^2) \right]
\, , 
\end{equation}
where $\sigma_P(s)  =  \sqrt{1-4 \frac{m_P^2}{s}}$, while for $\rho(1450)$ we employ a reasonable
parameterisation~:
\begin{equation}
 \Gamma_{\rho'}(s)  =  \Gamma_{\rho'}(M_{\rho'}^2) \, \frac{s}{M_{\rho'}^2}
\, \left( \frac{\sigma_{\pi}^3(s) \, \, + \,  \frac{1}{2} \, 
\sigma_K^3(s) \, \theta( \, s \, - \, 4 m_K^2)}{\sigma_{\pi}^3(M_{\rho'}^2) 
\,  + \,  \frac{1}{2} \, 
\sigma_K^3(M_{\rho'}^2) \, \theta( \, s \, - \, 4 m_K^2)} \right)
\theta(\, s \, - \, 4 m_{\pi}^2)  \, .
\end{equation}
The ${\cal O}(p^4)$ determination of the vector form factors was done in~\cite{Gasser:1984gg}. 
Requiring that our expressions match that result at small transfer of momentum we get the following
expressions~:
\begin{eqnarray} 
\label{eq:pp}
F^{\pi \pi}_V(s) & = & F(s) \, \, 
 \exp \left[ 2 \, Re \left(\tilde{H}_{\pi \pi}(s) \right) \, + \, 
Re \left(\tilde{H}_{KK} (s)
\right) \right]  \, \\[3.5mm]
F(s) & =  & 
\frac{M_{\rho}^2}{M_{\rho}^2-s-i M_{\rho} \Gamma_{\rho}(s)} \left[
1 + \left( \delta \, \frac{M_{\omega}^2}{M_{\rho}^2} \, - \, \gamma \, 
\frac{s}{M_{\rho}^2} 
\right) \, \frac{s}{M_{\omega}^2-s- i M_{\omega} \Gamma_{\omega}} \right] \nonumber \\
& &  - 
\frac{\gamma \, s}{M_{\rho'}^2-s-i M_{\rho'} \Gamma_{\rho'}(s)} 
\, \, ,
\nonumber \\[9mm]
F_V^{K^+K^-}(s) & = & \frac{1}{2} \, \frac{M_{\rho}^2}{M_{\rho}^2-s-i M_{\rho} \Gamma_{\rho}(s)} \, 
\exp \left[ 2 \, Re \left(\tilde{H}_{\pi \pi}(s) \right) \, + \, 
Re \left(\tilde{H}_{KK} (s)
\right) \right]  \, \nonumber \\
& &  +  \, \frac{1}{2} \, \left[ \sin^2 \theta_V \, \frac{M_{\omega}^2}{M_{\omega}^2-s-i M_{\omega} 
\Gamma_{\omega}} \, + \, 
\cos^2 \theta_V \, \frac{M_{\phi}^2}{M_{\phi}^2-s-i M_{\phi} \Gamma_{\phi}} \right] \,  \nonumber \\
&& \; \;\; \; \; \;\; \;\; \; \; \;\; \;\; \; \; \;\; \;\; \; \; \;\; \;\; \; \; \;  \; \; \;\; \;\; \; \; \;
 \; \; \;\; \;\; \; \; \; \; \; \;\; \;\; \; \; \; \; \; \;\; \;\; \; \; \;\times
\exp \left[ 3 \,  Re \left(\tilde{H}_{KK} (s) \,\right) \right]  \, ,
\nonumber \\[9mm]
F_V^{K^0\bar{K^0}}(s) & = & - \, \frac{1}{2} \, \frac{M_{\rho}^2}{M_{\rho}^2-s-i M_{\rho} \Gamma_{\rho}(s)} \, 
\exp \left[ 2 \, Re \left(\tilde{H}_{\pi \pi}(s) \right) \, + \, 
Re \left(\tilde{H}_{KK} (s)
\right) \right]  \, \nonumber \\
& &  +  \, \frac{1}{2} \, \left[ \sin^2 \theta_V \, \frac{M_{\omega}^2}{M_{\omega}^2-s-i M_{\omega} 
\Gamma_{\omega}} \, + \, 
\cos^2 \theta_V \, \frac{M_{\phi}^2}{M_{\phi}^2-s-i M_{\phi} \Gamma_{\phi}} \right] \,  \nonumber \\
&& \; \;\; \; \; \;\; \;\; \; \; \;\; \;\; \; \; \;\; \;\; \; \; \;\; \;\; \; \; \;  \; \; \;\; \;\; \; \; \;
 \; \; \;\; \;\; \; \; \; \; \; \;\; \;\; \; \; \; \; \; \;\; \;\; \; \; \;\times
\exp \left[ 3 \,  Re \left(\tilde{H}_{KK} (s) \,\right) \right]  \, , \nonumber
\end{eqnarray}
where we have used the definitions~:
\begin{eqnarray} \label{eq:functions}
\beta & = & \frac{\Theta_{\rho \omega}}{3 M_{\rho}^2} \; ,\nonumber \\
\gamma & = & \frac{F_V G_V}{F^2} \left( 1+ \beta \right) - 1 \; , \nonumber \\
\delta & = & \frac{F_V G_V}{F^2} - 1 \; , \nonumber \\
\tilde{H}_{PP}(s) & = & \frac{s}{F^2}  M_{P}(s)  \; , \nonumber \\
M_P(s) & = & \frac{1}{12} \left( 1 - 4 \frac{m_P^2}{s} \right) \, J_P(s) \,
- \, \frac{k_P(M_{\rho})}{6} \, + \, \frac{1}{288 \pi^2} \; , \nonumber \\
J_P(s) & = & \frac{1}{16 \pi^2} \left[ \sigma_P(s) \, \ln 
\frac{\sigma_P(s) -1}{\sigma_P(s) + 1}
+ 2 \right] \; , \nonumber \\
k_P(\mu) & = & \frac{1}{32 \pi^2} \left( \ln \frac{m_P^2}{\mu^2}+1 \right) \; .
\end{eqnarray}
Notice that the $\beta$ parameter includes the contribution of the isospin breaking $\rho - \omega$ 
mixing through $\Theta_{\rho \omega} = -3.3 \times 10^{-3} \, \mbox{GeV}^2$ \cite{Pich:2002ne}, 
and $F_V$ and $G_V$ are defined in Eq.~(\ref{eq:vectors}). Moreover the asymptotic constraint
on the $N_C \rightarrow \infty$ vector form factor indicates $F_V G_V \simeq F^2$ \cite{Ecker:1989yg}.
The mixing between the octet and singlet vector components employed in the construction
of the $I=0$ component of the kaon vector form factors is defined by~:
\begin{equation}
\left(  \begin{array}{c}
  \phi \\
  \omega 
 \end{array} \right)\, = \,  \left(
\begin{array}{cc}
\cos \theta_V & - \sin \theta_V \\
\sin \theta_V &  \cos \theta_V
\end{array}  \right) \;  \left(
\begin{array}{c}
 v_8 \\
v_0
\end{array} \right)\, ,
\end{equation}
and we will use ideal mixing, i.e. $\theta_V = 35^{\circ}$.




\begin{thebibliography}{99}

\bibitem{Hisano:1995nq}
  J.~Hisano, T.~Moroi, K.~Tobe, M.~Yamaguchi and T.~Yanagida,
  Phys.\ Lett.\  B {\bf 357} (1995) 579
  [arXiv:hep-ph/9501407];

\bibitem{Hisano:1995cp}
J.~Hisano, T.~Moroi, K.~Tobe and M.~Yamaguchi,
Phys.\ Rev.\  D {\bf 53} (1996) 2442
[arXiv:hep-ph/9510309].

\bibitem{Hisano:1998fj}
J.~Hisano and D.~Nomura,
Phys.\ Rev.\ D {\bf 59} (1999) 116005
[arXiv:hep-ph/9810479].

\bibitem{Kuno:1999jp}
  Y.~Kuno and Y.~Okada,
  Rev.\ Mod.\ Phys.\  {\bf 73} (2001) 151
  [arXiv:hep-ph/9909265].

\bibitem{seesaw:I}
P.~Minkowski,
Phys.\ Lett.\ B {\bf 67} (1977) 421;
%
M.~Gell-Mann, P.~Ramond and R.~Slansky, in {\it Complex Spinors and
Unified Theories} eds. P.~Van.~Nieuwenhuizen and D.~Z.~Freedman,
{\it Supergravity} (North-Holland, Amsterdam, 1979),
p.315 [Print-80-0576 (CERN)];
%
T.~Yanagida, in {\it Proceedings of the Workshop on the Unified Theory
and the Baryon Numebr in the Universe}, eds. O.~Sawada and
A.~Sugamoto (KEK, Tsukuba, 1979), p.95;
%
S.~L.~Glashow, in {\it Quarks and Leptons}, eds. M.~L\'evy et
al. (Plenum Press, New York, 1980), p.687;
%
R.~N.~Mohapatra and G.~Senjanovi\'c,
Phys.\ Rev.\ Lett.\  {\bf 44} (1980) 912.

\bibitem{seesaw:II}
%
R.~Barbieri, D.~V.~Nanopolous, G.~Morchio and F.~Strocchi,
Phys.\ Lett.\ B {\bf 90} (1980) 91;
%
R.~E.~Marshak and R.~N.~Mohapatra, {\it Invited talk given at Orbis
Scientiae, Coral Gables, Fla., Jan. 14-17, 1980}, VPI-HEP-80/02;
%
T.~P.~Cheng and L.~F.~Li,
Phys.\ Rev.\ D {\bf 22} (1980) 2860;
%
M.~Magg and C.~Wetterich,
Phys.\ Lett.\ B {\bf 94} (1980) 61;
%
J.~Schechter and J.~W.~F.~Valle,
Phys.\ Rev.\ D {\bf 22} (1980) 2227;
%
G.~Lazarides, Q.~Shafi and C.~Wetterich,
Nucl.\ Phys.\  B {\bf 181} (1981) 287;
%
R.~N.~Mohapatra and G.~Senjanovic,
Phys.\ Rev.\ D {\bf 23} (1981) 165.
%
E.~Ma and U.~Sarkar,
Phys.\ Rev.\ Lett.\  {\bf 80} (1998) 5716 [arXiv:hep-ph/9802445].

\bibitem{Yao:2006px}
W.~M.~Yao {\it et al.}  [Particle Data Group],
J.\ Phys.\ G {\bf 33} (2006) 1.

\bibitem{Borzumati:1986qx}
F.~Borzumati and A.~Masiero,
Phys.\ Rev.\ Lett.\  {\bf 57} (1986) 961.

\bibitem{Brooks:1999pu}
M.~L.~Brooks {\it et al.}  [MEGA Collaboration],
Phys.\ Rev.\ Lett.\  {\bf 83} (1999) 1521
[arXiv:hep-ex/9905013].

\bibitem{Ritt:2006cg}
S.~Ritt  [MEG Collaboration],
Nucl.\ Phys.\ Proc.\ Suppl.\  {\bf 162} (2006) 279.

\bibitem{PRIME}
The PRIME working group,
``Search for the $\mu-e$ Conversion Process at an Ultimate
Sensitivity of the Order of $10^{18}$ with PRISM'',
unpublished;
LOI to J-PARC 50-GeV PS, LOI-25,
{\tt http://psux1.kek.jp/jhf-np/LOIlist/LOIlist.html}

\bibitem{Aubert:2005ye}
 B.~Aubert {\it et al.}  [BABAR Collaboration],
 Phys.\ Rev.\ Lett.\  {\bf 95} (2005) 041802
 [arXiv:hep-ex/0502032].

\bibitem{Abe:2006sf}
 K.~Abe {\it et al.}  [Belle Collaboration],
 arXiv:hep-ex/0609049.

\bibitem{Hayasaka:2007vc}
K.~Hayasaka {\it et al.}  [Belle Collaboration],
arXiv:0705.0650 [hep-ex].

\bibitem{Banerjee:2007rj}
 S.~Banerjee,
 Nucl.\ Phys.\ Proc.\ Suppl.\  {\bf 169} (2007) 199
 [arXiv:hep-ex/0702017].

\bibitem{Miyazaki:2007zw}
  Y.~Miyazaki {\it et al.}  [Belle Collaboration],
  arXiv:0711.2189 [hep-ex].

\bibitem{Aubert:2007pw}
  B.~Aubert {\it et al.}  [BABAR Collaboration],
  Phys.\ Rev.\ Lett.\  {\bf 99} (2007) 251803
  [arXiv:0708.3650 [hep-ex]].

\bibitem{Babu:2002et}
K.~S.~Babu and C.~Kolda,
Phys.\ Rev.\ Lett.\  {\bf 89} (2002) 241802
[arXiv:hep-ph/0206310].

\bibitem{Dedes:2002rh}
A.~Dedes, J.~R.~Ellis and M.~Raidal,
Phys.\ Lett.\  B {\bf 549} (2002) 159
[arXiv:hep-ph/0209207].

\bibitem{Brignole:2003iv}
A.~Brignole and A.~Rossi,
Phys.\ Lett.\  B {\bf 566} (2003) 217
[arXiv:hep-ph/0304081];

\bibitem{Arganda:2005ji}
E.~Arganda and M.~J.~Herrero,
Phys.\ Rev.\  D {\bf 73} (2006) 055003
[arXiv:hep-ph/0510405].

\bibitem{Antusch:2006vw}
S.~Antusch, E.~Arganda, M.~J.~Herrero and A.~M.~Teixeira,
JHEP {\bf 0611} (2006) 090
[arXiv:hep-ph/0607263].

\bibitem{Kitano:2003wn}
  R.~Kitano, M.~Koike, S.~Komine and Y.~Okada,
  Phys.\ Lett.\  B {\bf 575} (2003) 300
  [arXiv:hep-ph/0308021].

\bibitem{Arganda:2007jw}
E.~Arganda, M.~J.~Herrero and A.~M.~Teixeira,
JHEP {\bf 0710} (2007) 104
[arXiv:0707.2955 [hep-ph]].

\bibitem{Yusa:2006qq}
Y.~Yusa {\it et al.}  [BELLE Collaboration],
Phys.\ Lett.\  B {\bf 640} (2006) 138
[arXiv:hep-ex/0603036].

\bibitem{Abe:2006qv}
K.~Abe {\it et al.}  [Belle Collaboration],
arXiv:hep-ex/0609013.

\bibitem{Aubert:2006cz}
B.~Aubert {\it et al.}  [BABAR Collaboration],
Phys.\ Rev.\ Lett.\  {\bf 98} (2007) 061803
[arXiv:hep-ex/0610067].

\bibitem{Miyazaki:2007jp}
Y.~Miyazaki {\it et al.}  [BELLE Collaboration],
Phys.\ Lett.\  B {\bf 648} (2007) 341
[arXiv:hep-ex/0703009].

\bibitem{Kane:1993td}
For a review see, for instance,\\
G.~L.~Kane, C.~F.~Kolda, L.~Roszkowski and J.~D.~Wells,
Phys.\ Rev.\  D {\bf 49} (1994) 6173
[arXiv:hep-ph/9312272].

\bibitem{Ellis:2002iu}
For a review see, for instance,\\
J.~R.~Ellis, T.~Falk, K.~A.~Olive and Y.~Santoso,
Nucl.\ Phys.\  B {\bf 652} (2003) 259
[arXiv:hep-ph/0210205];

\bibitem{Sher:2002ew}
M.~Sher,
Phys.\ Rev.\  D {\bf 66} (2002) 057301
[arXiv:hep-ph/0207136].

\bibitem{Brignole:2004ah}
A.~Brignole and A.~Rossi,
Nucl.\ Phys.\  B {\bf 701} (2004) 3
[arXiv:hep-ph/0404211].

\bibitem{Paradisi:2005tk}
P.~Paradisi,
JHEP {\bf 0602} (2006) 050
[arXiv:hep-ph/0508054].

\bibitem{Chen:2006hp}
C.~H.~Chen and C.~Q.~Geng,
Phys.\ Rev.\  D {\bf 74} (2006) 035010
[arXiv:hep-ph/0605299].

\bibitem{Fukuyama:2005bh}
  T.~Fukuyama, A.~Ilakovac and T.~Kikuchi,
  arXiv:hep-ph/0506295.

\bibitem{Blanke:2007db}
  M.~Blanke, A.~J.~Buras, B.~Duling, A.~Poschenrieder and C.~Tarantino,
  JHEP {\bf 0705} (2007) 013
  [arXiv:hep-ph/0702136].

\bibitem{Ecker:1988te}
G.~Ecker, J.~Gasser, A.~Pich and E.~de Rafael,
Nucl.\ Phys.\ B {\bf 321} (1989) 311;
J.~F.~Donoghue, C.~Ram\'{\i}rez and G.~Valencia,
Phys.\ Rev.\ D {\bf 39} (1989) 1947.

\bibitem{Lepage:1980fj}
 G.~P.~Lepage and S.~J.~Brodsky,
 Phys.\ Rev.\  D {\bf 22} (1980) 2157.

\bibitem{Abe:2007exa}
  K.~Abe {\it et al.}  [Belle Collaboration],
  arXiv:0708.3276 [hep-ex].

\bibitem{Porod:2003um}
W.~Porod,
Comput.\ Phys.\ Commun.\  {\bf 153} (2003) 275
[arXiv:hep-ph/0301101].

\bibitem{Casas:2001sr}
J.~A.~Casas and A.~Ibarra,
Nucl.\ Phys.\ B {\bf 618} (2001) 171
[arXiv:hep-ph/0103065].

\bibitem{Maki:1962mu}
Z.~Maki, M.~Nakagawa and S.~Sakata,
Prog.\ Theor.\ Phys.\  {\bf 28} (1962) 870.

\bibitem{Pontecorvo:1957cp}
B.~Pontecorvo,
Sov.\ Phys.\ JETP {\bf 6} (1957) 429
[Zh.\ Eksp.\ Teor.\ Fiz.\  {\bf 33} (1957) 549];
Sov.\ Phys.\ JETP {\bf 7} (1958) 172
[Zh.\ Eksp.\ Teor.\ Fiz.\  {\bf 34} (1957) 247].

\bibitem{neutrinodata_fits}
M.~C.~Gonz\'alez-Garcia and C.~Pe\~na-Garay,
Phys.\ Rev.\ D {\bf 68} (2003) 093003
[arXiv:hep-ph/0306001];
M.~Maltoni, T.~Schwetz, M.~A.~Tortola and J.~W.~F.~Valle,
New J.\ Phys.\  {\bf 6} (2004) 122
[arXiv:hep-ph/0405172];
G.~L.~Fogli, E.~Lisi, A.~Marrone and A.~Palazzo,
Prog.\ Part.\ Nucl.\ Phys.\  {\bf 57} (2006) 742
[arXiv:hep-ph/0506083].

\bibitem{'tHooft:1973jz}
G.~'t Hooft,
Nucl.\ Phys.\ B {\bf 72} (1974) 461;
G.~'t Hooft,
Nucl.\ Phys.\ B {\bf 75} (1974) 461;
E.~Witten,
Nucl.\ Phys.\ B {\bf 160} (1979) 57.

\bibitem{Peris:1998nj}
S.~Peris, M.~Perrottet and E.~de Rafael,
JHEP {\bf 9805} (1998) 011
[arXiv:hep-ph/9805442];
 M.~Knecht, S.~Peris, M.~Perrottet and E.~de Rafael,
 Phys.\ Rev.\ Lett.\  {\bf 83} (1999) 5230
 [arXiv:hep-ph/9908283];
 S.~Peris, B.~Phily and E.~de Rafael,
 Phys.\ Rev.\ Lett.\  {\bf 86} (2001) 14
 [arXiv:hep-ph/0007338];
 B.~Moussallam,
 Nucl.\ Phys.\  B {\bf 504} (1997) 381
 [arXiv:hep-ph/9701400];
 B.~Moussallam,
 Phys.\ Rev.\  D {\bf 51} (1995) 4939
 [arXiv:hep-ph/9407402];
M.~Knecht and A.~Nyffeler,
Eur.\ Phys.\ J.\ C {\bf 21} (2001) 659
[arXiv:hep-ph/0106034];
 P.~D.~Ruiz-Femen\'{\i}a, A.~Pich and J.~Portol\'es,
 JHEP {\bf 0307} (2003) 003
 [arXiv:hep-ph/0306157];
V.~Cirigliano, G.~Ecker, M.~Eidem\"uller, A.~Pich and J.~Portol\'es,
Phys.\ Lett.\ B {\bf 596} (2004) 96
[arXiv:hep-ph/0404004];
V.~Cirigliano, G.~Ecker, M.~Eidem\"uller, R.~Kaiser, A.~Pich and J.~Portol\'es,
JHEP {\bf 0504} (2005) 006
[arXiv:hep-ph/0503108];
 V.~Mateu and J.~Portol\'es,
 Eur.\ Phys.\ J.\  C {\bf 52} (2007) 325
 [arXiv:0706.1039 [hep-ph]].

\bibitem{Bijnens:2003rc}
 J.~Bijnens, E.~Gamiz, E.~Lipartia and J.~Prades,
 JHEP {\bf 0304} (2003) 055
 [arXiv:hep-ph/0304222];
 P.~Masjuan and S.~Peris,
 JHEP {\bf 0705} (2007) 040
 [arXiv:0704.1247 [hep-ph]].

\bibitem{Weinberg:1978kz}
S.~Weinberg,
PhysicaA {\bf 96} (1979) 327;
J.~Gasser and H.~Leutwyler,
Annals Phys.\  {\bf 158} (1984) 142.

\bibitem{Gasser:1984gg}
J.~Gasser and H.~Leutwyler,
Nucl.\ Phys.\ B {\bf 250} (1985) 465.

\bibitem{HerreraSiklody:1996pm}
 P.~Herrera-Sikl\'ody, J.~I.~Latorre, P.~Pascual and J.~Tar\'on,
 Nucl.\ Phys.\  B {\bf 497}, 345 (1997)
 [arXiv:hep-ph/9610549].

\bibitem{Kaiser:1998ds}
 R.~Kaiser and H.~Leutwyler,
 arXiv:hep-ph/9806336;
 F.~G.~Cao and A.~I.~Signal,
 Phys.\ Rev.\  D {\bf 60} (1999) 114012
 [arXiv:hep-ph/9908481].

\bibitem{Arganda:2004bz}
E.~Arganda, A.~M.~Curiel, M.~J.~Herrero and D.~Temes,
Phys.\ Rev.\  D {\bf 71} (2005) 035011
[arXiv:hep-ph/0407302].

\bibitem{Paradisi:2005fk}
  P.~Paradisi,
  JHEP {\bf 0510} (2005) 006
  [arXiv:hep-ph/0505046].

\bibitem{Ecker:1989yg}
G.~Ecker, J.~Gasser, H.~Leutwyler, A.~Pich and E.~de Rafael,
Phys.\ Lett.\ B {\bf 223} (1989) 425;
 F.~Guerrero and A.~Pich,
 Phys.\ Lett.\  B {\bf 412} (1997) 382
 [arXiv:hep-ph/9707347].

\bibitem{GomezDumm:2000fz}
 D.~G\'omez Dumm, A.~Pich and J.~Portol\'es,
 Phys.\ Rev.\  D {\bf 62} (2000) 054014
 [arXiv:hep-ph/0003320].

\bibitem{Pich:2002ne}
 A.~Pich and J.~Portol\'es,
 Nucl.\ Phys.\ Proc.\ Suppl.\  {\bf 121} (2003) 179
 [arXiv:hep-ph/0209224].





\end{thebibliography}
\end{document}